\documentclass[fleqn,usenatbib]{mnras}

\usepackage{newtxtext,newtxmath}
\usepackage{graphicx}
\usepackage{caption}     
\usepackage{subcaption}
\usepackage{wrapfig}
\usepackage{placeins}
\usepackage{float}
\usepackage[T1]{fontenc}

\DeclareRobustCommand{\VAN}[3]{#2}
\let\VANthebibliography\thebibliography
\def\thebibliography{\DeclareRobustCommand{\VAN}[3]{##3}\VANthebibliography}

\usepackage{graphicx}	% Including figure files
\usepackage{amsmath}	% Advanced maths commands
\title[4PCF of the Simulated ISM]{First Measurements of the 4-Point Correlation Function of Magnetohydrodynamic Turbulence as a Novel Probe of the Interstellar Medium}

\author[Williamson et al.]{
Victoria Williamson,$^{1}$\thanks{E-mail: williamson.v@ufl.edu }
James Sunseri,$^{2}$
Zachary Slepian, $^{1}$
Jiamin Hou $^{3}$
\& Alessandro Greco $^{1}$
\\
$^{1}$Department of Astronomy, University of Florida, 211 Bryant Space Science Center, Gainesville, FL 32611, USA\\
$^{2}$Department of Astrophysical Sciences, Princeton University, 4 Ivy Lane, Princeton, NJ 08544, USA \\
$^{3}$Max-Planck-Institut f\"ur extraterrestrische Physik, Gießenbachstraße 1, 85748 Garching, Germany \\
}

\date{Accepted XXX. Received YYY; in original form December 1 2024}

\pubyear{2024}

\begin{document}
\label{firstpage}
\pagerange{\pageref{firstpage}--\pageref{lastpage}}
\maketitle

\begin{abstract}
In the Interstellar Medium (ISM), gas and dust evolve under magnetohydrodynamic (MHD) turbulence. This produces dense, non-linear structures that then seed star formation. Observationally and theoretically, turbulence is quantified by summary statistics such as the 2-Point Correlation Function (2PCF) or its Fourier-space analog the power spectrum. These cannot capture the non-Gaussian correlations coming from turbulence's highly non-linear nature. We here for the first time apply the 4-Point Correlation Function (4PCF) to turbulence, measuring it on a large suite of MHD simulations that mirror, as well as currently possible, the conditions expected in the ISM. The 4PCF captures the dependence of correlations between quadruplets of density points on the geometry of the tetrahedron they form. Using a novel functionality added to the \textsc{sarabande} code specifically for this work, we isolate the purely non-Gaussian piece of the 4PCF. We then explore simulations with a range of pressures, $P$, and magnetic fields, $B$ (but without self-gravity); these are quantified by different sonic $(M_{\rm S})$ and Alfv\'enic $(M_{\rm A})$ Mach numbers. We show that the 4PCF has rich behavior that can in future be used as a diagnostic of ISM conditions. We also show that a large-scale coherent magnetic field leads to parity-odd modes of the 4PCF, a clean test of magnetic field coherence with observational ramifications. All our measurements of the 4PCF (10 $M_{\rm S}, M_{\rm A}$ combinations, 9 time-slices for each, 34 4PCF modes for each) are made public for the community to explore.
\end{abstract}


%%%%%%%%%%%%%%%%%%%%%%%%%%%%%%%%%%%%%%%%%%%%%%%%%%

%%%%%%%%%%%%%%%%% BODY OF PAPER %%%%%%%%%%%%%%%%%%

\section{Introduction}
Gas and dust in the interstellar medium (ISM) play a pivotal role in star formation, offering, respectively, the raw material, and the emission lines to cool efficiently so that the gas can form high-density cores and begin to fragment and collapse (\citealt{Elmegreen1987}, \citealt{Ferriere2001}). The ISM also acts as a medium for transferring matter, momentum, and energy produced by individual stars out into their host galaxies, a process known as ``feedback''(\citealt{Ceverino2009}, \citealt{Hayward2017}, \citealt{Grisdale2018}).

Turbulence in the ISM shapes the dynamics of star formation (\citealt{Elmegreen2001}, \citealt{MacLow2004}, \citealt{Vazquez-Semadeni2005}). It influences the gravitational collapse of material, the growth of a star's mass to the point that nuclear fusion can begin (\citealt{Williams_2005_dustgas}), and the formation of circumstellar disks (\textit{e.g.} \citealt{riaz_24_disks}). The turbulence also influences how, once stars have formed, stellar winds, jets, mass ejections, and supernovae then return material to the ISM (\citealt{Mckee_2007_starform}, \citealt{McDonough2007}, \citealt{Hayward2017}).

The ISM is highly turbulent and magnetized, rendering it challenging to fully characterize (\citealt{Elmegreen_2004_ISM}, \citealt{miller_1}). Turbulence refers to the irregular motion of gas and dust, and it is characterized by a power-law energy spectrum, suggesting that it is organized into eddies on different scales. Turbulence affects nearly all regions of the ISM and is the primary driver of diffusion of matter and dissipation of kinetic energy (\citealt{MacLow2004}, \citealt{Falceta-Goncalves_2014_turbulence}, \citealt{Girichidis2014}). Turbulence also produces localized density enhancements through compression, which can lead to gravitational collapse and star formation (\citealt{Federrath2009}, \citealt{Girichidis2014}). Hence, understanding turbulence is considered a crucial step in understanding star formation as a whole. 

The magnetized ISM interacts with charged particles as well as with the magnetic fields of other objects (\citealt{Ostriker2001}). Like turbulence, magnetic fields are also thought to play a significant role in star formation, through their impact on the dynamics of the medium (\citealt{Padoan1999}). They impact the behavior of charged particles and oppose forces of gravity, reducing fragmentation of massive dense cores and working against the contraction of these cores (\citealt{Motte_2021_highmass}). Therefore, a greater understanding of star formation processes and the ISM as a whole will require studies accounting for both turbulence and magnetic fields.

Magnetohydrodynamic (MHD) simulations offer a way to investigate these turbulent density structures in the ISM. The Catalogue for Astrophysical Turbulence Simulations (CATS) offers MHD simulations that can reproduce these important characteristics of the ISM (\citealt{Cho_2003_MHD}, \citealt{Burkhart_2009_MHDISM}, \citealt{Portillo_2018_3pcfISM}, \citealt{Burkhart2020}, \citealt{BialyBurkhart_2020_TurbulentDecorrelationScale}). However, these simulations could be made more accurate in the future with the help of measurements of non-Gaussian information on real data.  

In recent years, the field has turned to exploring higher-order statistics such as the 3-Point Correlation Function (3PCF) and its Fourier-space analog, the bispectrum. \cite{Slepian_2015_3pcfon^2}, \cite{Slepian_2020_Anisotropic3PCF}, \cite{Garcia_2021_LineOfSight3PCF}, and \cite{Friesen_2022_Galactos3PCF} present a set of algorithms that have greatly reduced the 3PCF's computational cost; these have been applied to large-scale structure in \cite{Slepian_2017_largescale3pcf}, \cite{Slepian_2017_BaryonAcousticOscillations}, \cite{Slepian_2017_LargeBOSS3PCF}, and \cite{Slepian_2018_BaryonDarkMatterVelocity}), as well as to simulated ISM turbulence \citealt{Portillo_2018_3pcfISM} using the method outlined in \cite{Slepian_3pcf_ft}. Notably, \cite{OBrien_2022_Bispectrum} also studies the bispectrum of MHD simulations, following earlier work by \cite{Burkhart_2009_MHDISM}. 

The current state-of-the-art 3PCF algorithm \citep{Slepian_2015_3pcfon^2} measures the statistic by combining spherical harmonics to form an isotropic basis (in this case, Legendre polynomials, up to normalization and phase, as in \citealt{Cahn_2022_IsotropicNPoint}; for a generating function for this basis, \citealt{gen_fn}). The Fourier implementation (\citealt{Slepian_3pcf_ft}, \citealt{Portillo_2018_3pcfISM}) defines the angular basis functions on a grid, radially bins them to generate kernels, and then convolves these (using an FT) with the original gridded data to obtain the expansion coefficients at each point simultaneously. The combination of these coefficients gives us the 3PCF's expansion in the isotopic basis (\citealt{Sunseri_2022_SARABANDE}). Unlike the 2PCF and power spectrum, the 3PCF captures some of the non-Gaussian information in the density field. 

 However, the 3PCF may well not capture all of the non-Gaussian information in the density field; indeed there no theoretical reason to expect that it would. Hence, the 4PCF is the next logical step. Recently, in cosmology, \cite{Philcox_2021_disconnected}, \cite{Hou_2022_oddparity}, and \cite{ortola} have all explored the 4PCF, with \cite{Cahn_2021_ParityViolation} proposing to use it as a test of parity violation. 
 
In this work, we present the first-ever use of the 4-Point Correlation Function on MHD simulations of the ISM; in fact to our knowledge it is the first-ever measurement of the 4PCF of any MHD turbulence. The recent release of the \texttt{python} package \textsc{sarabande} \citep{Sunseri_2022_SARABANDE} allows measurement of the 4PCF in $\mathcal{O}(N_{\mathrm{gm}} \log N_{\mathrm{gm}})
$ time using Fast Fourier Transforms (FFTs), with $N_{\mathrm{gm}}$ the number of grid points used for the FFT. \textsc{sarabande} requires analysis on regular gridded data sets, but, conveniently, the MHD simulations we analyze were done on such grids (as indeed generally simulations are). With \textsc{sarabande}, we are well-positioned to make the first exploration of non-Gaussian information in the density field's 4PCF.

% \begin{itemize}
%     \item Introduce why topic important (ISM/turbulence stuff)
%     \item Limits to simulations
%     \item Ppl have started using higher order statistics: 3PCF
%     \item problems faced by 3PCF
%     \item Introduce 4PCF, why so much better
%     \item first ppl to ever do this, have access to code, only ppl able to remove non Gaussian bits
%     \item paper is structured as follows...
% \end{itemize}
% \begin{itemize}
%     \item sell it, why should anyone care
%     \item why is 4PCF better than 3PCF for star formation: sensitive to parity, more sensitive to non Gaussianity, more sensitive to mag fields non-linear equations (might be sensitive to mag field directions)
    % \item Characterize gas in ISM and understand how stars form, problem b/c power spectrum only and 3PCF only one paper on it
    % \item PCF sensitive to nonGaussian in way 3PCF isnt
    % \item never measured the 4PCF, only ppl who can measure the connected (removes Gaussian, non-important info)
    % \item series of papers to develop NPCF recently, SARABANDE has been released fro 4PCF with FFT (talk abt nlogN scaling) using gridded data sets (perfect to this)
    
    % The 3PCF is a natural tool for turbulence studies as it encodes information about non-linear interactions that is absent in the 2-Point Correlation Function (2PCF), the configuration-space analog of the power spectrum.
% 3pcf not good because cant remove the non gaussian bits, 4pcf can

This paper is structured as follows: Section \ref{sec:Methods} describes the underlying mathematics behind the \textsc{sarabande} code, the connected 4PCF, angular and dependence of the function, the MHD simulations we use, and the supercomputers used to run these simulations. Section \ref{sec:Results} presents our measurements for the first time. Section \ref{sec:Discussion of Qualitative Trends} explains the patterns and trends we observed in the data. Section \ref{sec:Discussion & Conclusions} concludes and suggests some future directions for pursuing the path initiated in this work.
% The interstellar medium (ISM) is all of the matter and radiation existing between star systems, in the form of gas, dust, and cosmic rays. The three-phase model (Field, Goldsmith \& Habing (1969) and McKee \& Ostriker(1977)) describes the ISM in temperature-dependent phases: (1) a cold (T<300 K) dense phase consisting of clouds of hydrogen, (2) a warm (T $\simeq 10^4$ K) intercloud phase of neutral and ionized gas, and (3) very hot (T $\simeq 10^6$ K) gas shock heated by supernovae. 

% Magnetic fields and turbulence play the largest role in creating pressure in the ISM. 

% The turbulence of the ISM means it plays a crucial role in star formation and connecting stellar and galactic scales. Stars are born inside large molecular clouds, the densest regions of the ISM. Throughout their lifetimes, stars interact with the ISM by building up molecular clouds and replenishing the ISM with matter and energy, through planetary nebalue, stellar winds, and supernovae.

% The 4PCF has been known to be sensitive to parity-odd signals (\citealt{Hou_2022_oddparity})
\begin{figure*}
    \centering
    \includegraphics[width=\textwidth]{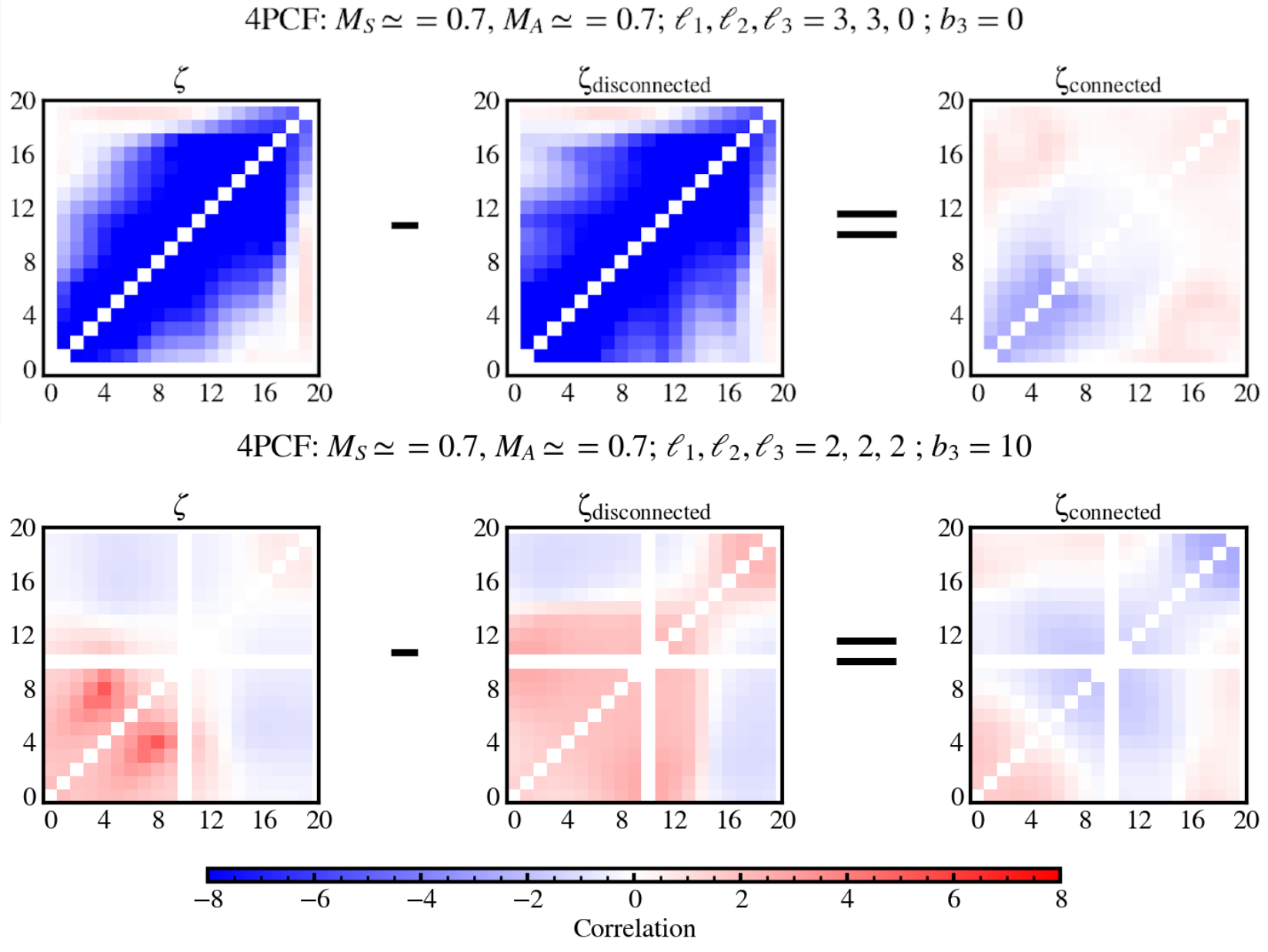}
    \caption{This depicts how we compute the ``connected'' 4PCF. The updated version of the \textsc{sarabande} \texttt{python} package, created specifically for this work, is able to separate the 4PCF into ``connected'' and ``disconnected'' pieces, allowing us to isolate the non-Gaussian information. The x- and y-axes correspond to bin indices in the configuration space, with the binning scheme described fully in Section~\ref{sec:connected_4pcf}. Each plot shows a 2D slice of the 4PCF, chosen to showcase specific combinations of the angular dependencies, $\ell_1, \ell_2$ and $\ell_3$. The color bar represents the correlation values, with blue shading indicating a deficit (-) and red shading indicating an excess (+) relative to random. This visualization provides an intuitive understanding of how the ``connected'' and ``disconnected'' terms contribute to the full 4PCF.}
.
    \label{fig:sarabande_processes.pdf}
\end{figure*}
\section{Methods}
\label{sec:Methods} % used for referring to this section from elsewhere
Here we outline the code we use to measure the 4PCF, explain the process by which the connected 4PCF is calculated, describe the angular dependence and parity, summarize the simulations on which we measured it, and report the computing cost of the work.
\subsection{SARABANDE \& The Isotropic Basis}
The code we run on these MHD simulations is the \texttt{python} package \textsc{sarabande}. This code implements the 4PCF, whose estimator is outlined in \cite{Philcox_2022_ENCORE} (with extension to higher spatial dimensions in \citealt{Philcox_2022_Npoint}), in the basis presented by \citealt{Cahn_2022_IsotropicNPoint}, generalizing work by \cite{Slepian_2015_3pcfon^2} on the 3PCF. 

The full 4PCF involves four distinct points, and the estimator uses the positions of these points to define three vectors. These vectors form a tetrahedron; this then is used to characterize the 4PCF, as we will now review.  

The full 4PCF, which we denote $\zeta$, is written as 
\begin{equation}
\label{eq:full_zeta_angle}
    \zeta = \langle \delta(\vec{x})\delta(\vec{x}+\vec{r}_1)\delta(\vec{x}+\vec{r}_2)\delta(\vec{x}+\vec{r}_3) \rangle_\mathcal{R},
\end{equation}
with $\delta$ the density fluctuation field. The left-hand side of the equation has arguments left unstated purposely due to the different ways to parameterize the full 4PCF. Here, the lengths of the tetrahedron sides extended from $\vec{x}$ are $r_1$, $r_2$, and $r_3$, while their directions are $\hat{r}_1$, $\hat{r}_2$, and $\hat{r}_3$. The subscript $\mathcal{R}$ in the equation denotes the region over which the 4PCF is averaged. 

Now, \cite{Cahn_2022_IsotropicNPoint} characterizes the angular dependence of the 4PCF using isotropic basis functions. The sum of radial coefficients $\zeta_{\Lambda}$ times these isotropic basis functions $\mathcal{P}_{\Lambda}$ represents the full 4PCF as 
\begin{equation}
\label{4pcf_basis}
    \zeta = \sum_{\Lambda} \hat{\zeta}_{\Lambda}(R) \mathcal{P}_{\Lambda}(\hat{R}).
\end{equation}
$\Lambda \equiv \{\ell_1, \ell_2, \ell_3\}$ is the set of angular momenta used to describe the dependence of the 4PCF on the direction vectors from one vertex (in this case, at $\vec{x}$) of the tetrahedron the four points form, and $R \equiv \{r_1, r_2, r_3\}$ while $\hat{R} \equiv \{\hat{r}_1, \hat{r}_2, \hat{r}_3\}$. 

In Eq. (\ref{4pcf_basis}), the isotropic basis functions $\mathcal{P}_{\Lambda}$ are defined as
\begin{equation}
    \mathcal{P}_{\Lambda}(\hat{R}) = \sum_{M} C^{\Lambda}_{M} Y_{\ell_1 m_1}(\hat{r}_1) Y_{\ell_2 m_2}(\hat{r}_2) Y_{\ell_3 m_3}(\hat{r}_3),
\end{equation}
% The lengths of said tetrahedron are defined by $r_1$, $r_2$, and $r_3$, while the angles between them are $\vec(r_1), \vec(r_2), \vec(r_3)$. 
and $C^{\Lambda}_{M}$ is given by the Wigner 3-$j$ symbol with an additional phase:
\begin{equation}
\label{eq:Wigner3-j}
    C^{\Lambda}_{M} = (-1)^{\ell_1 + \ell_2 + \ell_3} \begin{pmatrix}
\ell_1 & \ell_2 & \ell_3\\
m_1 & m_2 & m_3
\end{pmatrix}.
\end{equation}
 Above, we have defined $M \equiv(m_1,m_2,m_3)$. 
 
 The full 4PCF must be computed using the radial coefficients shown in Eq. (\ref{4pcf_basis}). This can be done in terms of binned convolution coefficients $a_{\ell m}^{\rm b}(\vec{x})$:
 \begin{align}
\label{eqn:almconv}
    a^{\rm b}_{\ell m}(\vec{x}) = \int d^3\vec{r}\; Y_{\ell m}^*(\hat{r})\Phi^{\rm b}(r)\delta(\vec{x} + \vec{r}).
\end{align}
The binning function $\Phi^{\rm b}$ is
\begin{align}
    \Phi^{\rm b}(r) = \frac{\Theta^{\rm b}(r)}{V_{\rm b}}.
\label{eq:bin_volume_unity}
\end{align}
The result is the binned radial coefficients about each point $\vec{x}$,
\begin{multline}
    \hat{\zeta}_{\ell_1 \ell_2 \ell_3}^{\; \rm b_1 \rm b_2 \rm b_3}(\vec{x}) = \delta(\vec{x})\sum_{M} \mathcal{E}(\Lambda) \; C^{\Lambda}_{M} \\ \times \; a_{\ell_1 m_1}^{\rm b_1}(\vec{x}) \; a_{\ell_2 m_2}^{\rm b_2}(\vec{x}) \; a_{\ell_3 m_3}^{\rm b_3}(\vec{x}).
\label{eqn:mu}
\end{multline}
$\mathcal{E}(\Lambda)$ is +1 for even parity, occurring when $\ell_1+\ell_2+\ell_3$ is an even number, and -1 for odd parity, occurring when this value is an odd number.

To ensure dimensionless results and consistent weighting, \textsc{sarabande} normalizes each radial bin by its volume. This accounts for variations in the number of grid cells within each bin. The bin volume is calculated as:
\begin{equation}
    V_{\text{bin}} = \left(\frac{L}{N_{\text{mesh}}}\right)^3 N_{\text{cells in bin}},
\end{equation}
where $L$ is the box size, $N_{\text{mesh}}$ is the grid resolution (number of grid points) per dimension (for our simulations, it will be 256), and $N_{\text{cells in bin}}$ is the number of grid cells in the bin. This normalization prevents artifacts' arising from varying bin sizes and ensures that the 4PCF reflects purely physical correlations.

The scales analyzed using \textsc{sarabande} range from the smallest measurable grid resolution to half the box size. Limiting the maximum scale to $L/2$ avoids redundant information due to periodic boundary conditions while covering the inertial range of turbulence, where critical non-Gaussian features emerge. These scale choices optimize the analysis for capturing the physical processes driving MHD turbulence.

\subsection{The Connected 4PCF} \label{sec:connected_4pcf}
While the full 4PCF is sensitive to non-Gaussian information, it is difficult to isolate the most meaningful information that can't be studied with a simple power spectrum or 2PCF. This is because, unlike the 2PCF and 3PCF, the 4PCF consists of two constituent parts: a connected piece and a disconnected piece. We have
\begin{equation}
    \hat{\zeta} = \hat{\zeta}^{\rm (connected)} + \hat{\zeta}^{(\rm disconnected)}.
\end{equation}
The disconnected piece is degenerate with the 2PCF, since it consists of products of 2PCFs, which are not sensitive to non-Gaussian information. 

In this work, we isolate the connected piece of the 4PCF, as first outlined in \citealt{Philcox_2021_disconnected}, to extract purely the non-Gaussian information the 4PCF has to offer. To accomplish this, we first refer back to Eq. (\ref{eq:full_zeta_angle}) and recognize that the averaging over realizations in the product of four density fluctuation fields can be decomposed as
\begin{multline}
\label{eq:disconnected_1}
    \langle \delta(\vec{x})\delta(\vec{x}+\vec{r}_1)\delta(\vec{x}+\vec{r}_2)\delta(\vec{x}+\vec{r}_3) \rangle_\mathcal{R} = \\ \langle \delta(\vec{x})\delta(\vec{x}+\vec{r}_1)\delta(\vec{x}+\vec{r}_2)\delta(\vec{x}+\vec{r}_3) \rangle_{\mathcal{R}}^{(c)} \\ + \underbrace{[\langle \delta(\vec{x})\delta(\vec{x}+\vec{r}_1) \rangle_\mathcal{R} \langle \delta(\vec{x}+\vec{r}_2)\delta(\vec{x}+\vec{r}_3) \rangle_\mathcal{R} + \rm 2 \; Perms.]}_{\rm Disconnected \; Piece}.
\end{multline}
Recognizing that the average over realizations and rotations of the product of two density fluctuation fields is the 2PCF, we can then rewrite as
\begin{equation}
    \zeta = \zeta^{\rm (c)} + \underbrace{[\xi(\vec{r}_1) \xi(\vec{r}_2 - \vec{r}_3) + \rm 2 \; Perms.]}_{\rm Disconnected \; Piece}.
\end{equation}
Following the derivation in \citealt{Philcox_2021_disconnected} we can write the resulting disconnected 4PCF as 
\begin{equation}
    \hat{\zeta}_{\ell_1 \ell_2 \ell_3}^{\; \rm b_1 \rm b_2 \rm b_3 (disc)}(\vec{x}) = C^{\Lambda}_{M} \left( \hat{\xi}_{\ell_1 m_1}^{\; \rm b_1} \hat{\xi}_{\ell_2 m_2 \ell_3 m_3}^{\; \rm b_2 b_3} + \rm 2 \; Perms. \right).
\end{equation}
With this disconnected piece in hand, we may now subtract it off from the full 4PCF to isolate the desired connected piece. Using an updated version of \textsc{sarabande} created specifically for this work, we isolate the connected piece in this way. The approach is depicted schematically in Fig. \ref{fig:sarabande_processes.pdf}.

\subsection{Angular Dependence and Parity}
\label{sec:angular_dependence&parity}
\begin{figure}
    \raggedright
    \includegraphics[width=0.5\textwidth]{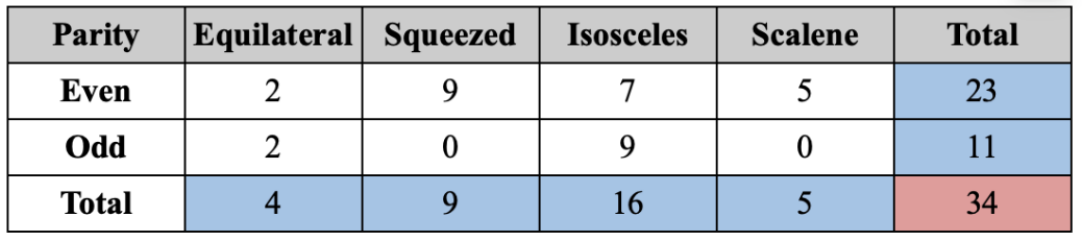}
    \caption{This table depicts the breakdown of unique $\ell$ combinations studied in this work. Of the 34 total combinations, 23 are parity even, while 11 are parity odd (see Sec. \ref{sec:angular_dependence&parity}). There are four equilateral $\ell$ combinations, which occur when all $\ell$s are the same. There are nine squeezed $\ell$ combinations, which is when only one $\ell = 0$; due to the 3-$j$ symbol, this causes the triangle formed by the three angular momentum vectors to be "squeezed" shut (see Eq. \ref{eq:Wigner3-j}). There are 16 isosceles and five scalene $\ell$ combinations, referring, respectively, to when two sides of the triangle are the same, and to when all sides of the triangle are different. The breakdown of these combination types into even or odd parity is shown as well.}
    \label{fig:paper_table}
\end{figure}

To fully capture the angular information in the 4PCF, we use Legendre polynomials and isotropic basis functions to expand the angular dependence between the tetrahedron's sides. This expansion introduces multipole moments, indexed by $\ell_1, \ell_2, \ell_3$, which represent the angular structure of the 4PCF. Parity-even components, where $\ell_1$ + $\ell_2$ + $\ell_3$ is even, dominate the signal, with 23 of the 34 unique $\ell$ combinations being even (Fig. \ref{fig:paper_table}). This reflects the inherent symmetries we find in turbulence. The remaining 11 parity-odd components, where $\ell_1 + \ell_2 + \ell_3$ is odd, are weaker but offer insight into asymmetries in the density field from magnetic fields, anisotropic turbulence driving, and other interesting factors. For the sake of computational efficiency, the expansion is halted at $\ell_{\text{max}} = 3$. 

\graphicspath{{figures/figures/color_sim_slices.pdf}}
 
\begin{figure*}
    \centering
    \includegraphics[width=\textwidth]{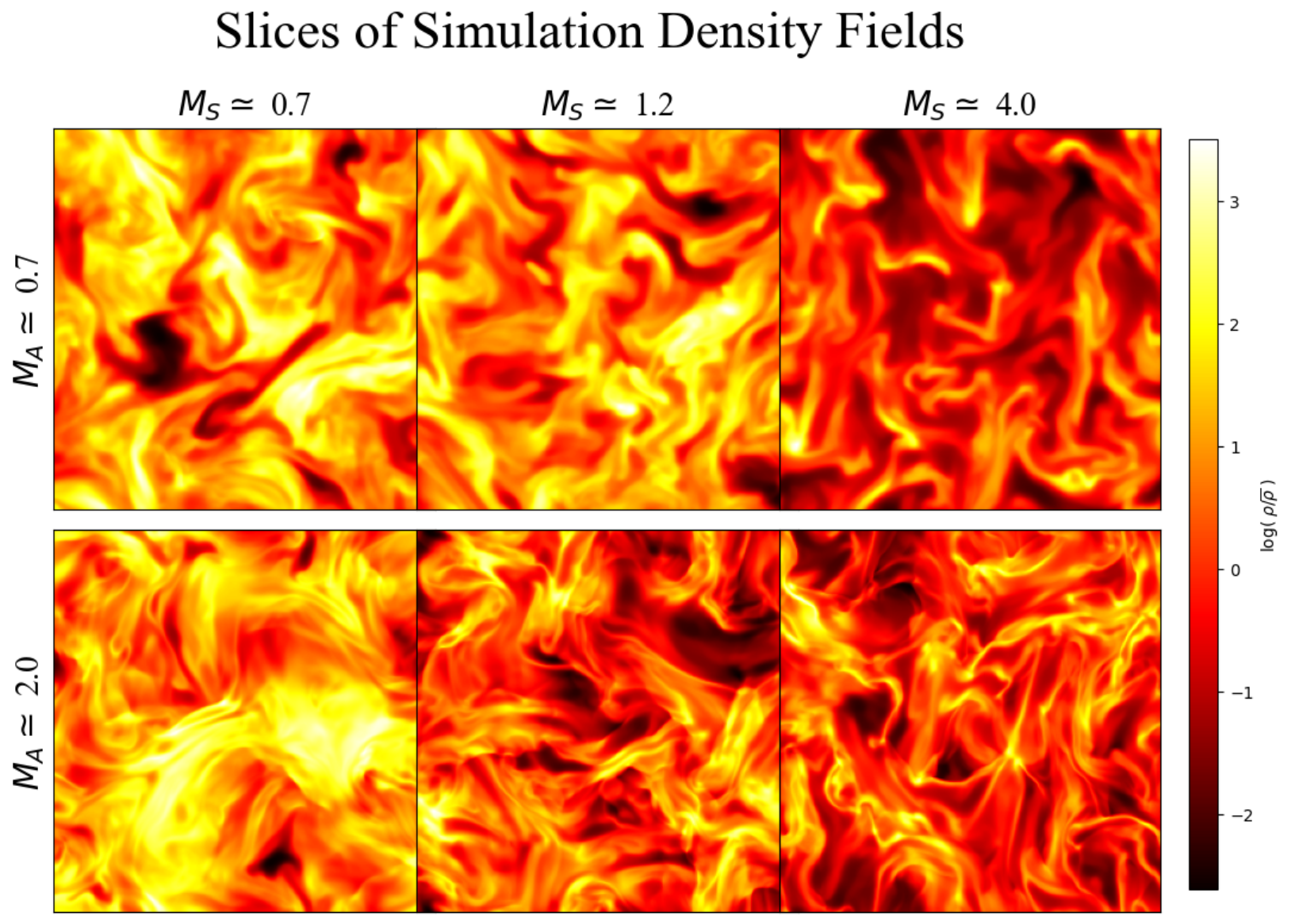}
    \caption{These are sample slices of density fields from the MHD simulation sets used in this work, with a logarithmic color map, similar to Fig. 2 in \citealt{Portillo_2018_3pcfISM}. This demonstrates the correlation between higher sonic and Mach-Alfv\'enic numbers and small-scale filaments of high density (Sec. \ref{sec:MHD_sims}). Slower movement (lower $M_{\rm A}$ or $M_{\rm S}$) allows time for the small-scale power in the fluid to diminish and smooth out existing filaments.}
    \label{fig:color_sim_slices}
\end{figure*}

\subsection{Simulations of MHD Turbulence}
\label{sec:MHD_sims}
The MHD simulations we used to obtain the density profiles for sub- and supersonic turbulent gas with varying magnetic field strengths were isothermal and non-self-gravitating. To exclude outside factors, we employ the ideal MHD equations solved in other works (\citealt{Portillo_2018_3pcfISM}),
\begin{equation}
    \frac{\partial \rho}{\partial t} + \nabla \cdot (\rho \vec{\varv}) = 0,
 \label{eq:MHD_1st}
\end{equation}
\begin{equation}
    \frac{\partial (\rho \vec{\varv})}{\partial t} + \nabla \cdot [\rho \vec{\varv} \vec{\varv} + (p + \frac{B^2}{8\pi})\vec{I} - \frac{1}{4\pi} \vec{B}\vec{B}] = \vec{f},
 \label{eq:MHD_2nd}
\end{equation}
\begin{equation}
    \frac{\partial \vec{B}}{\partial t} - \nabla \times (\vec{v} \ \times \vec{B}) = 0 \label{eq:MHD_3rd}
\end{equation}
 where $\rho$ is the density, $\vec{B}$, is the magnetic field, \textit{p} is the gas pressure, $\vec{I}$ is the identity matrix, and $\vec{B}\vec{B}$ is the magnetic stress matrix. Eq. (\ref{eq:MHD_1st}) is the continuity equation which represents conservation of mass demonstrates how the change in mass over time plus the divergence of mass flow rate over the unit area must be zero. Eq. (\ref{eq:MHD_2nd}) represents conservation of momentum, stating that the change in momentum over time plus the divergence of momentum flex, plus the gradient of pressure and magnetic force must equal the external forces acting on the fluid. Eq. (\ref{eq:MHD_3rd}) represents conservation of magnetic flux by demonstrating that the rate of change of the magnetic field over time minus the curl of the velocity-induced magnetic field must be zero. We assume zero-divergence condition, periodic boundary conditions, and an isothermal equation of state. We assume a wave number of $k$ = 2.5, where $k$ is expressed in terms of the number of cells in the simulation grid. The wave number is 1/2.5 the box size and is driving a large-scale solenoidal for the source term $\vec{f}$. The simulations runs used are available as part of the Catalog for Astrophysical Turbulence Simulations (CATS; \citealt{Cho_2003_MHD}, \citealt{Burkart_Lazarian_2016}, \citealt{Burkhart2020}).

Our simulations explore varying parameters of sonic Mach numbers and Alfv\'enic Mach numbers, with $v$ as the velocity, $c_{s}$ as sound speed, and $v_A$ as the Alfv\'en speed. Our simulations run with $M_S$ $\simeq$ 0.7, 1.2, and 7.0, and $M_A$ $\simeq$ 0.7 and 2.0.

Any desired physical scale can be assigned for the box size, since the simulations are non-self-gravitating. We use $\approx$ when stating the sonic and Alfv\'enic Mach numbers because they are averages which fluctuate slightly over time.

The resulting density fields from these simulations are shown in Fig. \ref{fig:color_sim_slices}, where we present sample slices using a logarithmic color map. These slices illustrate how varying Mach numbers influence the density structure. Higher sonic and Alfvénic Mach numbers are associated with small-scale filaments of high density, while lower Mach numbers show smoother structures with less prominent filamentation. The slower motion of the gas in subsonic regimes allows time for the small-scale power in the fluid to diminish, smoothing out existing filaments, as evidenced by the smoother density field. These visualizations help contextualize the turbulent structures present in the simulations and set the stage for our subsequent analysis of the 4PCF.

\subsection{Normalization and Scaling}
Before applying \textsc{sarabande} to compute the 4PCF, the density fields from our MHD simulations undergo a preliminary processing step to standardize the data and emphasize non-Gaussian structures. The density fields of turbulent media are often log-normal, meaning the logarithm of the density fluctuations follows a Gaussian distribution (\citealt{vazquez_1994_hierarchical}, \citealt{federrath_2008_density}, 
\citealt{burkhart_2011)_stattools}). To enable meaningful statistical analysis, we construct logarithmic density fluctuations as follows:
\begin{equation}
    \delta(\vec{x}) \equiv \frac{\log \rho(\vec{x}) - \langle \log \rho \rangle}{\sigma(\log \rho)},
 \label{eq:log_scale}
\end{equation}
where $\rho(\vec{x})$ is the density field, $\langle \log \rho \rangle$ is the mean logarithm of the density, and $\sigma(\log \rho)$ is its standard deviation (\citealt{Portillo_2018_3pcfISM}). Taking the logarithm of the density field serves to reduce the asymmetry typically observed in turbulent density fields. By transforming the density field into logarithmic space, the distribution becomes closer to a Gaussian distribution. As explained in \citealt{Portillo_2018_3pcfISM}, this transformation ensures that the density fluctuation field within each simulation box has zero mean and unit variance. We then compute the 4PCF of the normalized density fluctuation field. This normalization guarantees consistent comparisons across simulations with varying Mach and Alfv\'enic numbers. This provides a standardized framework for \textsc{sarabande} to isolate complex non-Gaussian features in the data, such as higher-order correlations, which are crucial for detecting subtle structures in turbulence.

\subsection{Summary of Computing Cost}
To measure the connected 4PCF on all 50+ simulations,
we used high-performance computing clusters at both Princeton University (Della) and University of Florida (HiPerGator).The main computational runs were performed on Della, while HiPerGator was primarily used for testing, debugging, and optimization of the workflow. Both the Della cluster at Princeton University and the HiPerGator cluster at the University of Florida are heterogeneous clusters supporting several different computer architectures. When using Della for our analysis, we exclusively used nodes with 2.4 GHz Intel Broadwell Central Processing Units (CPUs). Each node contains 28 CPUs with a total shared memory of 128 GBs. The max instruction set for cores is AVX2. With HiPerGator, we used nodes equipped with 2x AMD Epyc 7742 (Rome) 64-core processors, providing a total of 256 cores per node. Each node has 2TB of RAM and supports the AVX-512 instruction set. Additionally, these nodes also include 28TB of NVMe local storage for high-speed data access.

The \textsc{sarabande} package uses an OpenMP parallelization paradigm to unravel nested for loops necessary for the combining binned convolution coefficients $ a_{\ell m}^{\rm b}$ into the radial 4PCF coefficients ${\zeta}_{\ell_1 \ell_2 \ell_3}^{\; \rm b_1 \rm b_2 \rm b_3}$. Using an entire node (28 Broadwell CPUs) on the Della Cluster, \textsc{sarabande} measured the connected 4PCF
with an $\ell_{\text{max}} = 3$ at a radial resolution of 20 bins for a single $256^3$ data cube in 45,000+ CPU hours, which was $\sim$ 24 Wall Hours. 

\graphicspath{{figures/figures/fullpowerlaw.pdf}}
 
\begin{figure*}
    \centering
    \includegraphics[width=\textwidth, height=0.49\textheight]
    {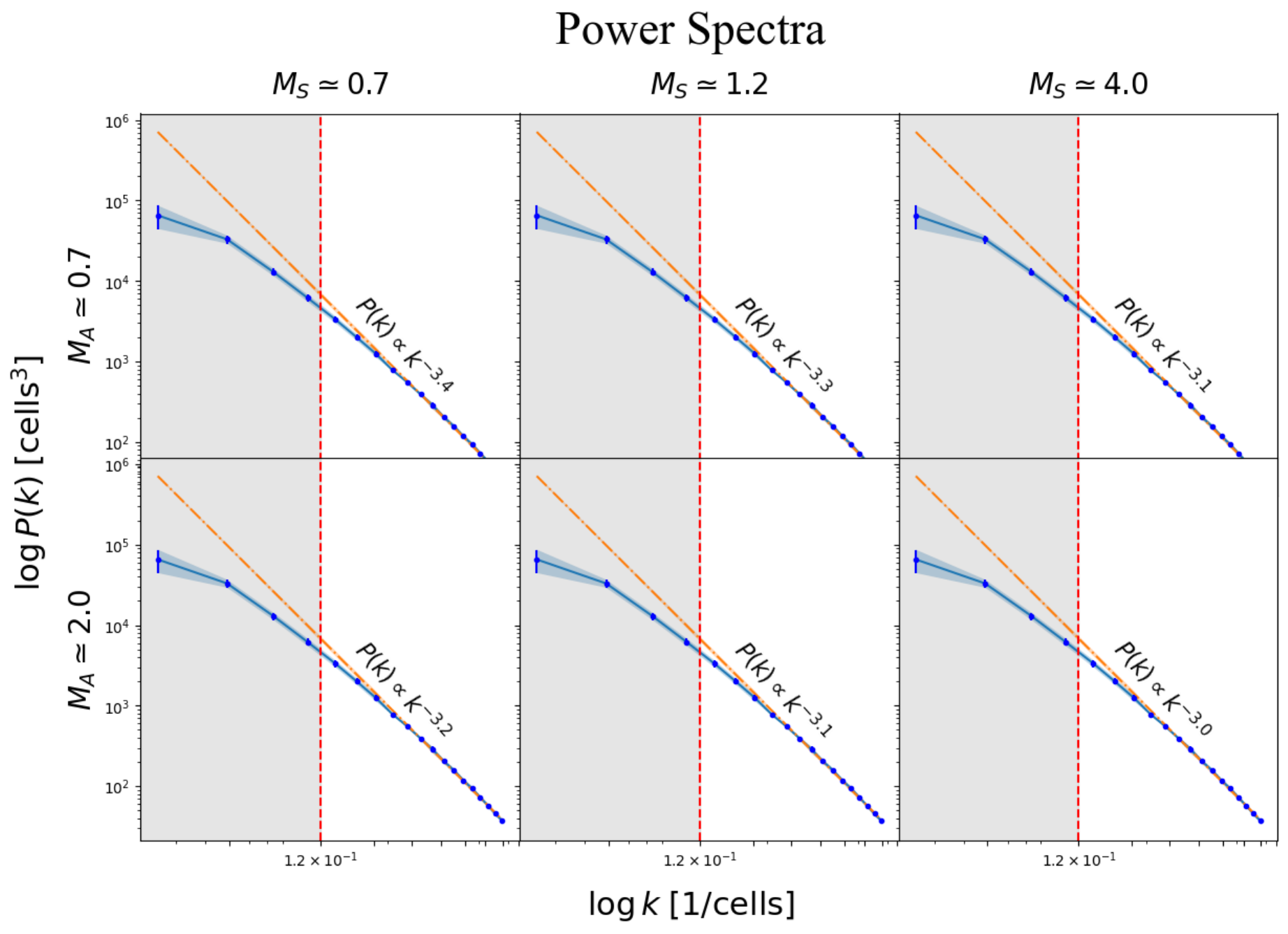}
    \caption{Here we show the power spectra of our simulations' density fields. The solid blue line represents the power spectrum averaged over all time slices (9, at each $M_{\rm A}, M_{\rm S}$ set), while the orange dot-dashed line represents our power-law fit. The slope of each power-law fit is indicated within the panel. The shaded region to the left of the vertical dotted red line, situated at 1.2 x $10^-1$, corresponds to large spatial scales (low $k$) comparable to the box size. These regions are dominated by box-scale effects, so we exclude the corresponding position-space scales from our analysis since they lie outside the inertial range, where the power spectrum follows a power law.}
    \label{fig:fullpowerlaw}
\end{figure*}

\section{Overview of Results}
\label{sec:Results}
Here, we give an overview of the general results we present, and then in Section \ref{sec:Discussion of Qualitative Trends} qualitatively discuss the trends we find.

In this work, we present the first-ever measurements of the connected 4PCF on simulations of MHD turbulence in the ISM, as seen in Figs. \ref{fig:equil_start.pdf} - \ref{fig:(3,2,1)_end.pdf}. These plots illustrate significant excess and deficit correlation in certain regions. Dark red regions indicate a correlation excess, where tetrahedra in the simulations are significantly more common than would be expected in a random distribution. Dark blue regions represent regions of correlation deficit, where tetrahedra are less common. The darkest regions on the colorbar, corresponding to values approaching +8 and -8 , correspond to the strongest non-Gaussian correlations, which are more likely to indicate intrinsic features of the density field. Regions of lighter coloring, closer to zero on the color bar, reflect minimal to no detection of correlation and are more likely consistent with noise. 

\begin{figure}
    \raggedright
    \includegraphics[width=0.5\textwidth]{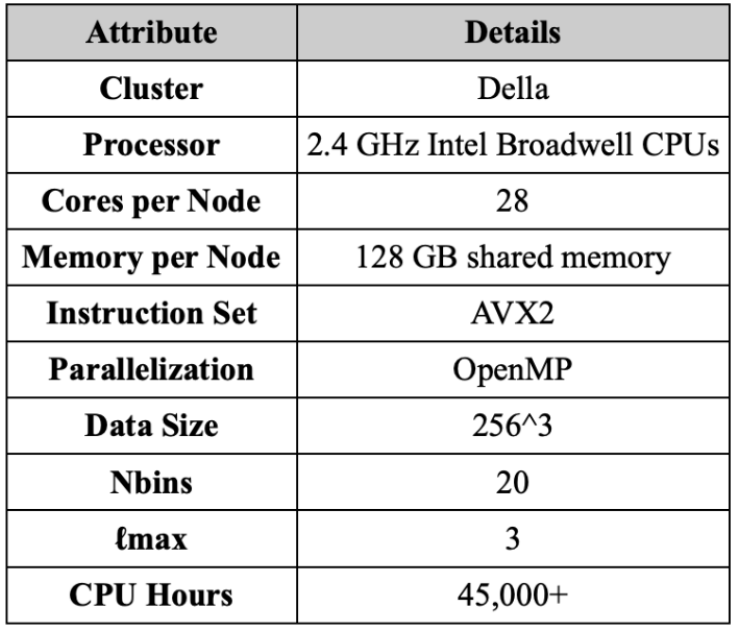}
    \caption{This is the computing cost summary and configuration of the Della cluster used for the main 4PCF measurements in this work. Specifications include node architecture, parallelization methods, and resource usage for a $256^3$ data cube with $\ell_{\text{max}} = 3$ and 20 radial bins.}
    \label{fig:computing_table}
\end{figure}

\section{Discussion of Qualitative Trends}
\label{sec:Discussion of Qualitative Trends}
Here we outline the qualitative trends observed in the results, separated by the parameters $M_A$, $M_S$, $b$, parity modes, and $\ell$ configurations. For each section we analyze how behavior changes with that specific parameter, if all other variables were held constant.

\begin{figure*}
    \centering
    \includegraphics[width=0.95\textwidth]{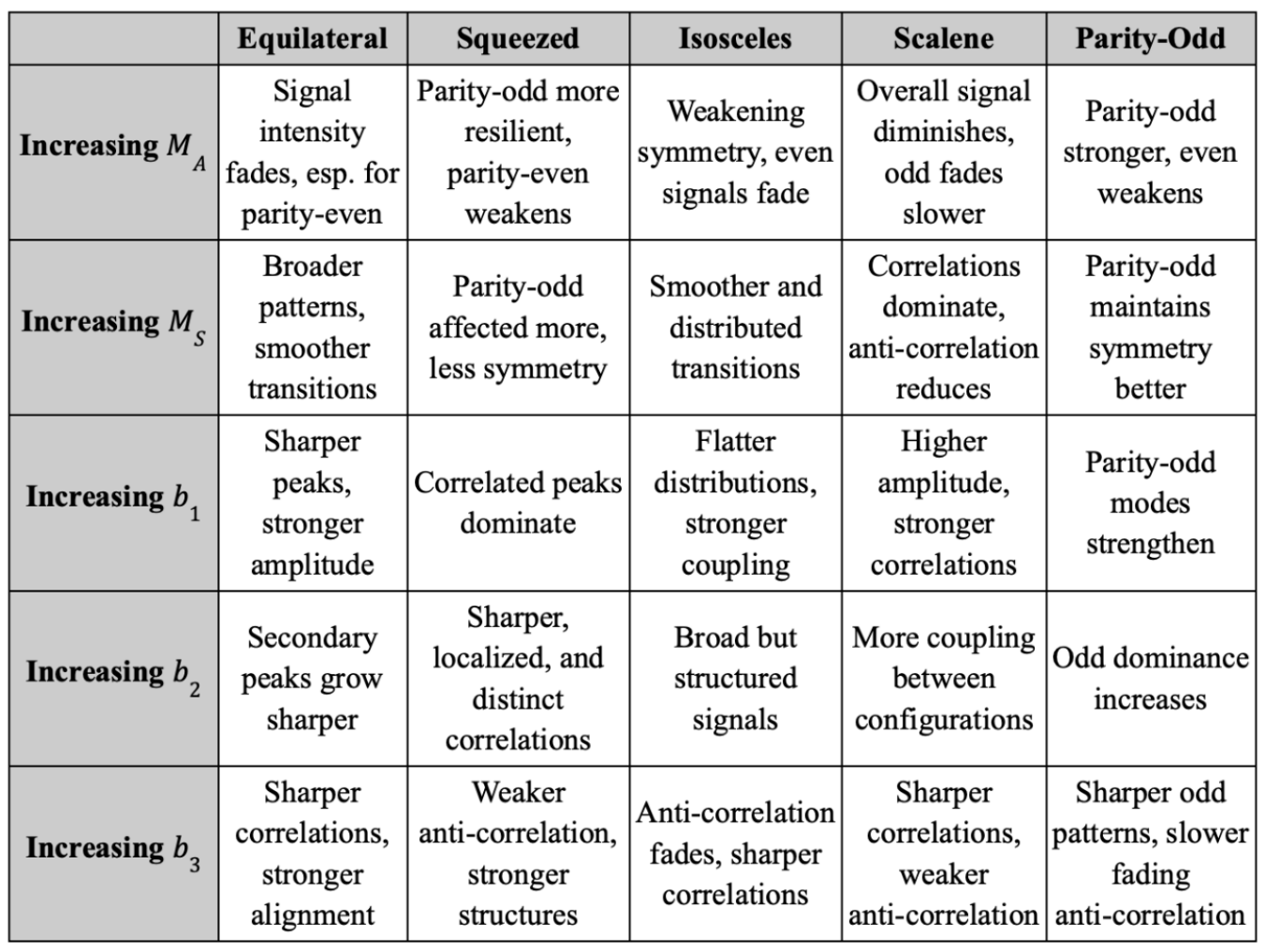}
    \caption{This table summarizes the observed trends for different 4PCF configurations (equilateral, squeezed, isosceles, scalene, and parity-odd) as the independent variables $M_A, M_S, b_1, b_2$ and $b_3$ are increased. While parity-odd modes are treated separately here, they can also be categorized into equilateral, squeezed, isosceles, and scalene configurations. For more detailed explanations and discussions of these qualitative trends, refer to Section \ref{sec:Discussion of Qualitative Trends}.}
    \label{fig:patterns_table}
\end{figure*}

% Exploration of these findings in later papers could attempt to further explain the role the ISM plays in star formation. There is a possibility for \textsc{sarabande} and the methodology employed in this work to be utilized on small regions of the intergalactic medium, likely only within parameters in which turbulence could dominate over the gravitational pull of galaxies. 

\subsection{Behavior with Increasing $M_A$}
As $M_A$ increases, we see a general reduction of signal intensity for both correlation and anti-correlation features. In equilateral configurations particularly, such as Fig. \ref{fig:(3,2,1)_end.pdf}, higher $M_A$ panels exhibit weaker colors compared to lower $M_A$. This trend is strongest for the parity-even modes, with signals gradually fading to near zero for the $M_A = 2.0$. Overall, this suggest that higher $M_A$ values lead to a suppression of the 4PCF signal. 
We also see shifts in high intensity regions over bins as $M_A$ increases. At smaller $M_A$, peaks are more widely distributed across bins, while at higher $M_A$ these peaks tend to cluster closer to smaller $b_2$ or $b_3$ values, dependent on configuration. This suggests that increasing $M_A$ causes a localization effect in the 4PCF signal distribution, as visible in Fig. \ref{fig:isosceles1_end.pdf}.
In terms of symmetry changes, we see that parity-even and parity-odd configurations are affected by $M_A$ as well. For parity-even configurations, symmetry between correlated (red) and anti-correlated (blue) signals weakens as $M_A$ increases. Lower $M_A$ values tend to exhibit more evenly distributed red-blue patterns, while higher $M_A$ panels tend to favor one polarity. For parity-odd configurations, increasing $M_A$ consistently diminishes the intensity but the symmetry structure is able to be maintained better.
Based on $\ell$ configurations, we see mode specific trends as well. As a whole lower $\ell$ modes appear less affected by increases in $M_A$, while modes with higher $\ell_1, \ell_2, \ell_3$ combinations have more drastic decreases in intensity. Equilateral configurations see a more pronounced suppression of intensity compared to isosceles or squeezed configurations as $M_A$ increases. With equilateral configurations, peaks become almost nonexistent in some $M_A = 2.0$ panels. In isosceles and squeezed configurations, the localization effect is much more apparent.
\subsection{Behavior with Increasing $M_S$}
As $M_S$ increases, the signal intensity trend is a bit more unclear, with both red and blue regions intensifying in some configurations but diminishing in others. This varies depending on the bins and mode type. We see parity-even modes exhibit consistent increasing intensity in certain bins, while parity-odd modes show localized suppression for those same bins. Signal intensity does tend to concentrate in specific regions of the bin space ($b_2, b_3$) for higher $M_S$. This localization effect becomes more prominent with larger $M_S$, which suggests a narrowing of the 4PCF signal distribution over this change, as evident in Fig. \ref{fig:i&s1_start.pdf}.
For parity-even configurations, correlated and anti-correlated regions become less symmetric, which one polarity beginning to dominate as $M_S$ increases. This is particularly prominent in equilateral and squeezed configurations, as seen in Fig. \ref{fig:squeezed2_start.pdf}. With parity-odd configurations, symmetry is better preserved but does weaken slightly at the highest $M_S$ value, when $M_S = 4.0$.
When $M_S$ values are lower, peaks in signal are more fairly distributed across bins but this changes as $M_S$ increases. The signal peaks shift to regions with higher $b_2$ or $b_3$, which indicates that higher $M_S$ values influence the clustering of the 4PCF in higher-bin regions.
In terms of mode-specific trends, we see that higher $\ell$ modes, such as $\ell = (3, 3, 0)$ (Fig. \ref{fig:(3,3,0)_start.pdf}, \ref{fig:(3,3,0)_end.pdf}) or $\ell = (3, 2, 1)$ (Fig. \ref{fig:(3,2,1)_start.pdf}, \ref{fig:(3,2,1)_end.pdf}), show a noticeable increase in signal amplitude as $M_S$ increase, especially in isosceles and squeezed configurations. Lower $\ell$, like $\ell = (0, 0, 0)$, show more uniform trends through the same $M_S$ changes. 
In equilateral configurations, the signal intensity grows significantly with increasing $M_S$, while the signal also becomes highly localized at specific bins. Isosceles and scalene configurations tend to have more equally distributed increases in intensity across a broader range of bins. Squeezed configurations often show a similar increase in signal intensity but these are concentrated in specific regions, commonly intermediate $b_2$ and low $b_3$ bins.
\subsection{Behavior with Increasing $b_1$}
As $b_1$ increases, there is a significant decrease in the overall signal intensity in most configurations. This is clear from the weakening of red and blue regions in the higher $b_1$ panels. Signals are strongest for smaller $b_1$ values, such as $b_1 = 0$ and $b_1 = 5$, and particularly near the equilateral regions of the bin space. The opposite happens for $b_1 = 10$ and $b_1 = 15$ when the signal becomes weaker and more fragmented, with the red and blue regions being significantly reduced in size and strength.
As $b_1$ grows we see the strongest signals shift towards larger $b_2$ and $b_3$ bins, with the signal shifting from being distributed across a broad range of bins to being concentrated in higher bins. This suggests that larger $b_1$ values correspond to suppressed 4PCF signal in outlying bins.
As $b_1$ grows, parity symmetry weakens and one polarity (positive or negative signal) dominates at certain configurations.
Higher $\ell$ modes, such as $\ell = (3, 2, 1)$ (Fig. \ref{fig:(3,2,1)_start.pdf}, \ref{fig:(3,2,1)_end.pdf}), display faster suppression of signal as $b_1$ increases, especially in equilateral configurations. At these equilateral configurations, we see almost no signal visible at the largest $b_1$. In isosceles configurations, they retain higher signals compared to equilateral configurations but still have significantly reduced intensity. In these modes, signal peaks tend to shift to higher bins as $b_1$ grows. Lower $\ell$ modes, such as $\ell =  (0, 0, 0)$ (Fig. \ref{fig:equil_start.pdf}), will retain some residual signal even at the largest $b_1$, particularly in squeezed configurations. However, in these squeezed configurations the signal becomes highly localized, with peaks near specific $b_2$ and $b_3$ values. We see a similar trend in scalene configurations but with slightly weaker signals (see Fig. \ref{fig:isosceles3_end.pdf}).
\subsection{Behavior with Increasing $b_2$}
As $b_2$ increases, we observed a gradual decrease in signal intensity, in both correlation and anti-correlation signals. The strongest signals occur at lower $b_2$ values, in particular when $b_1$ is lower as well. At larger $b_2$, the signal becomes much weaker and more fragmented, with a reduction in both red and blue regions regions.
As $b_2$ increases, we observe signals becoming more concentrated in specific $b_3$ regions, as opposed to being distributed across broader regions of $b_3$ prior to this change. This likely indicates a narrowing of the 4PCF signal distribution. We also see large regions of the bin space with no significant signal (white areas), which indicates suppression in off-peak regions as well.
As $b_2$ increases, we also see parity symmetry weaken, similar to with $b_1$, but in this case parity-odd modes dominate more configurations. The parity-odd signals tend to retain more structure and intensity at higher $b_2$, compared to parity-even modes, which exhibit faster suppression.
Lower $\ell$ modes, like $\ell = (1, 1, 1)$ (Fig. \ref{fig:equil_start.pdf}, \ref{fig:equil_end.pdf}), are more resilient to $b_2$ increases, with residual signals even at the largest $b_2$ values. Higher $\ell$ modes show faster suppression of signal intensity comparatively (see Fig. \ref{fig:isosceles1_end.pdf}). 
In equilateral configurations, increasing $b_2$ corresponds to a steep reduction in intensity, particularly at larger $b_1$ and $M_S$ values. Isosceles configurations retain signals relatively strong at smaller $b_2$ but show gradual decrease in signal intensity as $b_2$ increases. Signal peaks also become more localized in specific $b_3$ regions in these configurations, as shown in Fig. \ref{fig:i&s2_end.pdf}. Squeezed configurations retain significant signal intensity at higher $b_2$, but the signal becomes highly localized. Scalene configurations show a more moderate reduction in signal intensity, with parity-odd modes retaining more structure compared to parity-even modes.
\subsection{Behavior with Increasing $b_3$}
As $b_3$ increases, there is a general suppression of signal intensity, particularly in configurations with higher $b_1$ or $b_2$. At larger $b_3$, signals become weaker and more fragmented, with noticeable gaps in signal regions, as seen in Fig. \ref{fig:equil_end.pdf}.
At these larger $b_3$, signals also tend to concentrate in specific bin regions, often near smaller $b_1$ or $b_2$ values, shifting from a broader distribution at small $b_3$ values. We observe significant suppression of signals in off-peak bin regions, which leaves behind only localized peaks.
As $b_3$ increases, we also see parity symmetry weaken, in a trend similar to what we observe with increasing $b_2$. Parity-odd modes dominate most configurations, as they tend to retain residual structure and intensity even at the highest $b_3$.
Lower $\ell$ modes like $\ell = (1, 1, 1)$ (Fig. \ref{fig:equil_end.pdf}) are more resilient to $b_3$ increases, with residual signals even at the largest $b_3$ values. Higher $\ell$ modes, such as $\ell = (3, 2, 1)$ in Fig. \ref{fig:(3,2,1)_end.pdf}, are more sensitive to $b_3$, showing faster suppression of signal intensity and structure. Equilateral configurations display rapid suppression of signal as $b_3$ increases, with only small residuals signals remaining at the largest $b_3$. Isosceles configurations gradually weaken at higher $b_3$ and experience a shift in peaks towards smaller $b_1$ and $b_2$ (see Fig. \ref{fig:isosceles2_end.pdf}). In squeezed configurations, signals are relatively resilient even at higher $b_3$ but become highly localized and parity asymmetric. In scalene configurations, suppression of signal is moderate as $b_3$ grows and parity-odd modes retain more structure than parity-even modes.
\subsection{Difference Between Parity-Odd and Parity-Even Modes}
\subsubsection{Parity-Even}
As a whole, parity-even modes exhibit stronger signal intensity than parity-odd modes, as expected. This is consistent with theoretical predictions, as parity-even configurations align with the isotropic nature of turbulence. Parity-even modes show significant signal suppression as parameters, such as $M_A, M_S, b_1, b_2,$ and $b_3$, increase, with suppression most evident in equilateral and isosceles configurations, where the signal almost disappears at the highest parameter values (see Fig. \ref{fig:(3,2,1)_end.pdf}). Signals tend to concentrate in specific bin regions at higher parameter values, displaying localization of peaks under these conditions.
The hallmark of parity-even modes is a symmetric signal distribution across the bin space. Positive (red) and negative (blue) signals are balanced at lower parameter values. This symmetry weakens as parameters increase, with one polarity (often red) dominating in most configurations. Higher parity-even $\ell$ modes, like $\ell = (3, 2, 1)$, experience the fastest suppression and most significant fragmentation as parameters increase, as seen in Fig. \ref{fig:isosceles1_end.pdf}. Parity-even lower $\ell$ modes, like $\ell = (1, 1, 1)$ (Fig. \ref{fig:equil_start.pdf}), are less sensitive to suppression but still exhibit significant weakening at higher parameters.
In equilateral configurations at higher parameter values, parity-even modes experience rapid suppression and structural fragmentation. For isosceles configurations, parity-even modes see more moderate suppression with localized peaks remaining at higher parameters. In squeezed configurations, parity-even modes are suppressed but are able to maintain some strength in localized bins. As scalene configurations approach higher parameters, parity-even modes exhibit weaker signals than parity-odd modes.
\subsubsection{Parity-Odd}
\begin{figure*}
    \centering
    \includegraphics[width=0.95\textwidth, height=0.84\textheight]{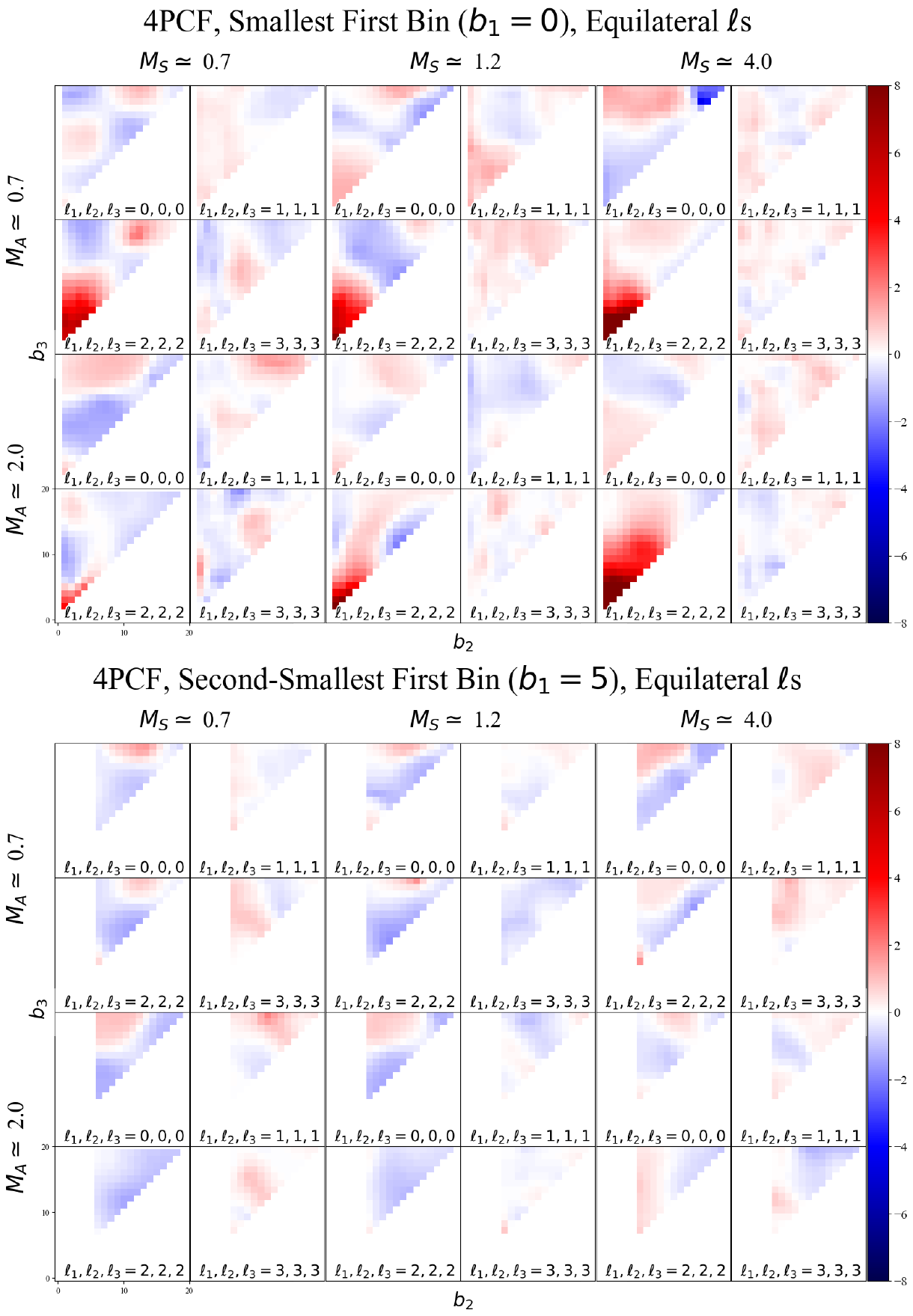}
    \caption{The 3D connected 4PCF signal-to-noise ratio, normalized by the standard deviation across time slices, computed in the equilateral multipole basis for $\ell_1, \ell_2, \ell_3 = 0, 1, 2,$ and $3$. The plot showcases results from simulations with varying Mach and Alfv\'enic numbers, highlighting how these physical parameters impact higher-order clustering. The signal is shown as a function of binned radial distances $b_1$, $b_2$, and $b_3$, capturing the clustering dependence on angular and radial configurations. The top panel represents when $b_1 = 0$, where the clustering amplitude is suppressed, highlighting isotropic and low-anisotropy configurations. The bottom panel corresponds to when $b_1 = 5$, emphasizing the enhanced clustering signal from higher-amplitude correlations. The lower triangular region of the plot is excluded because it corresponds to unphysical configurations that violate the triangle inequality for radial distances. The vertical column on the left-hand side is whited out because the analysis requires $b_2 > b_1$, ensuring that the ordering of distances is consistent with the convention used in the 4PCF calculation. Dark red regions correspond to excess tetrahedra relative to a random distribution, while dark blue regions indicate a deficit. The darkest regions, where values approach +8 and -8 on the color bar, correspond to the strongest detections, where non-Gaussian correlations intrinsic to the density field are most likely to be represented. Regions of lighter coloring, closer to 0 on the color bar, reflect minimal to no detection of correlation and are more likely consistent with noise.}
    \label{fig:equil_start.pdf}
\end{figure*}

\begin{figure*}
    \centering
    \includegraphics[width=0.95\textwidth, height=0.89\textheight]{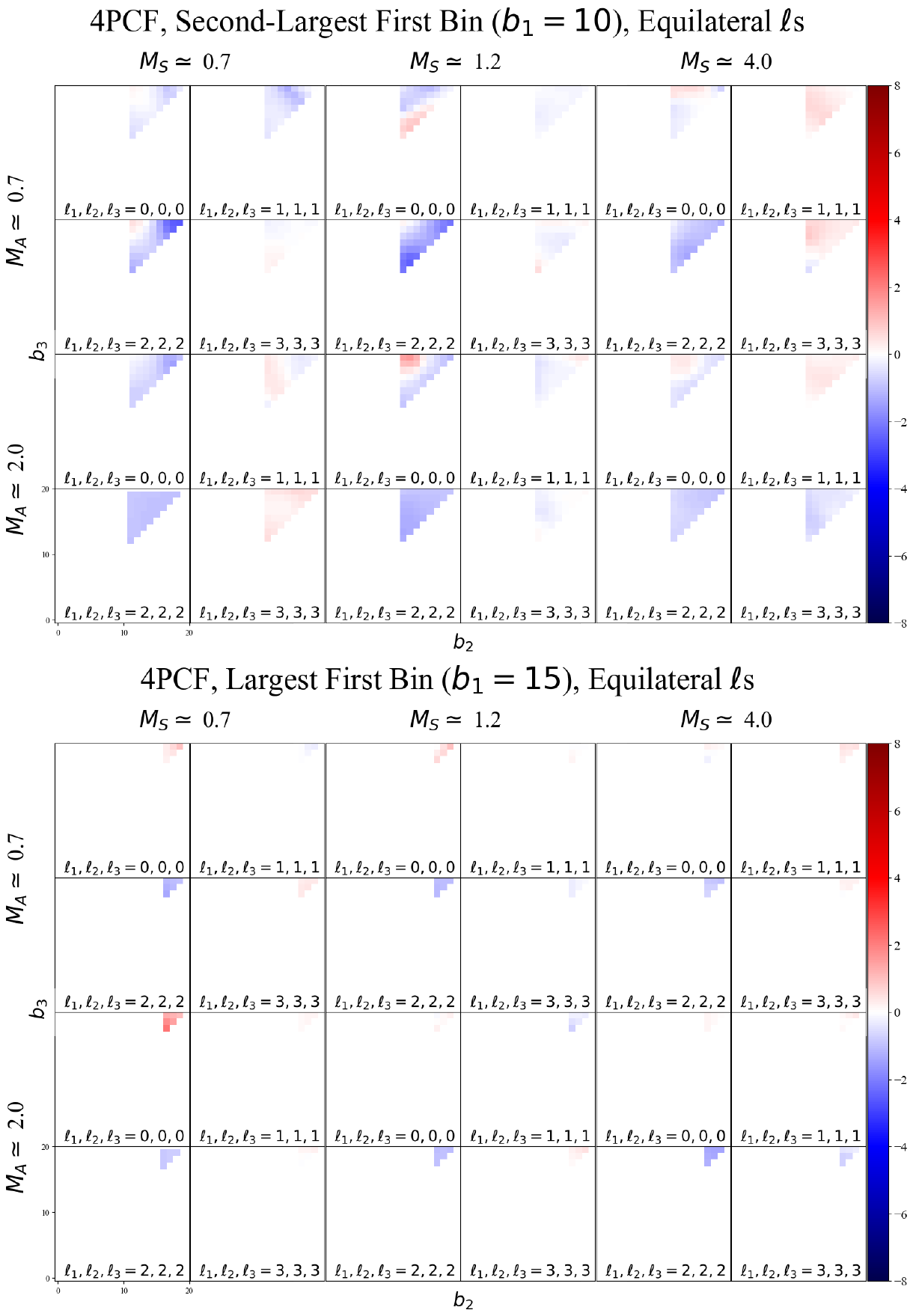}
    \caption{The 3D connected 4PCF signal-to-noise ratio, normalized by the standard deviation across time slices, computed in the equilateral multipole basis for $\ell_1$, $\ell_2$, $\ell_3 = 0, 1, 2, $ and $3$. The top panel represents when $b_1 = 10$ and the bottom panel corresponds to when $b_1 = 15$. These plots adhere to the same $b_1, b_2$ limitations and can be interpreted by the same method described for Fig. \ref{fig:equil_start.pdf}. At higher $M_A$, anti-correlations diminish rapidly, leaving correlated features more prominent. Parity-even signals weaken significantly, while parity-odd signals retain more structure. Further analysis is offered in Section \ref{sec:Discussion of Qualitative Trends}.}
    \label{fig:equil_end.pdf}
\end{figure*}

\begin{figure*}
    \centering
    \includegraphics[width=0.95\textwidth, height=0.89\textheight]{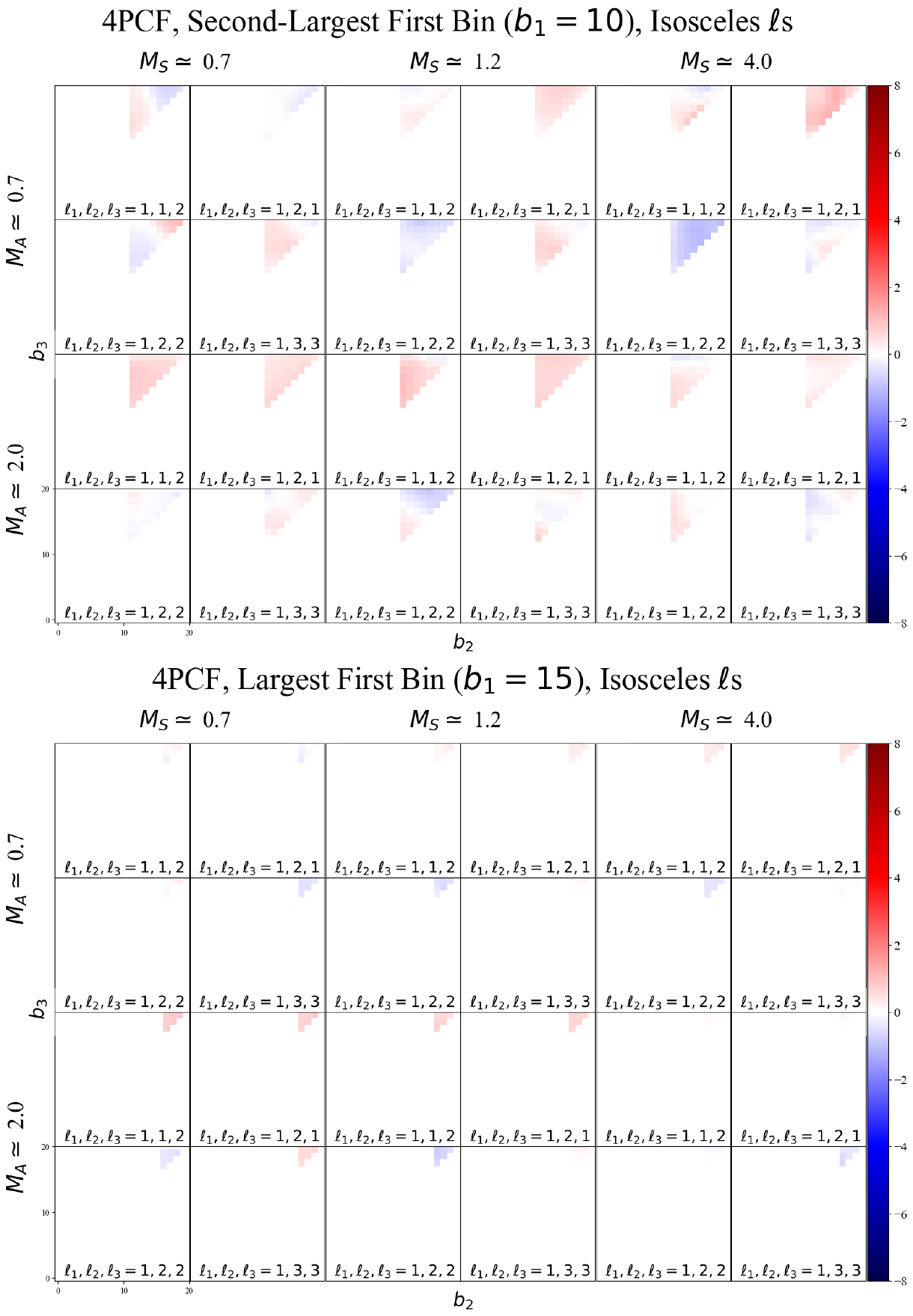}
    \caption{The 3D connected 4PCF signal-to-noise ratio, normalized by the standard deviation across time slices, computed in the isosceles multipole basis for $\ell_1, \ell_2, \ell_3 = (1, 1, 2)$, $\ell_1, \ell_2, \ell_3 = (1, 2, 1)$, $\ell_1, \ell_2, \ell_3 = (1, 2, 2)$, and $\ell_1, \ell_2, \ell_3 = (1, 3, 3)$. The top panel represents when $b_1 = 10$ and the bottom panel corresponds to when $b_1 = 15$. These plots adhere to the same $b_1, b_2$ limitations and can be interpreted by the same method described for Fig. \ref{fig:equil_start.pdf}. At higher $M_A$, peaks shift toward smaller bins, highlighting localization of the signal. Anti-correlations diminish, leaving correlated features dominant in the configuration. Further analysis is offered in Section \ref{sec:Discussion of Qualitative Trends}.}
	\label{fig:isosceles1_end.pdf}
\end{figure*}

\begin{figure*}
    \centering
    \includegraphics[width=0.95\textwidth, height=0.89\textheight]{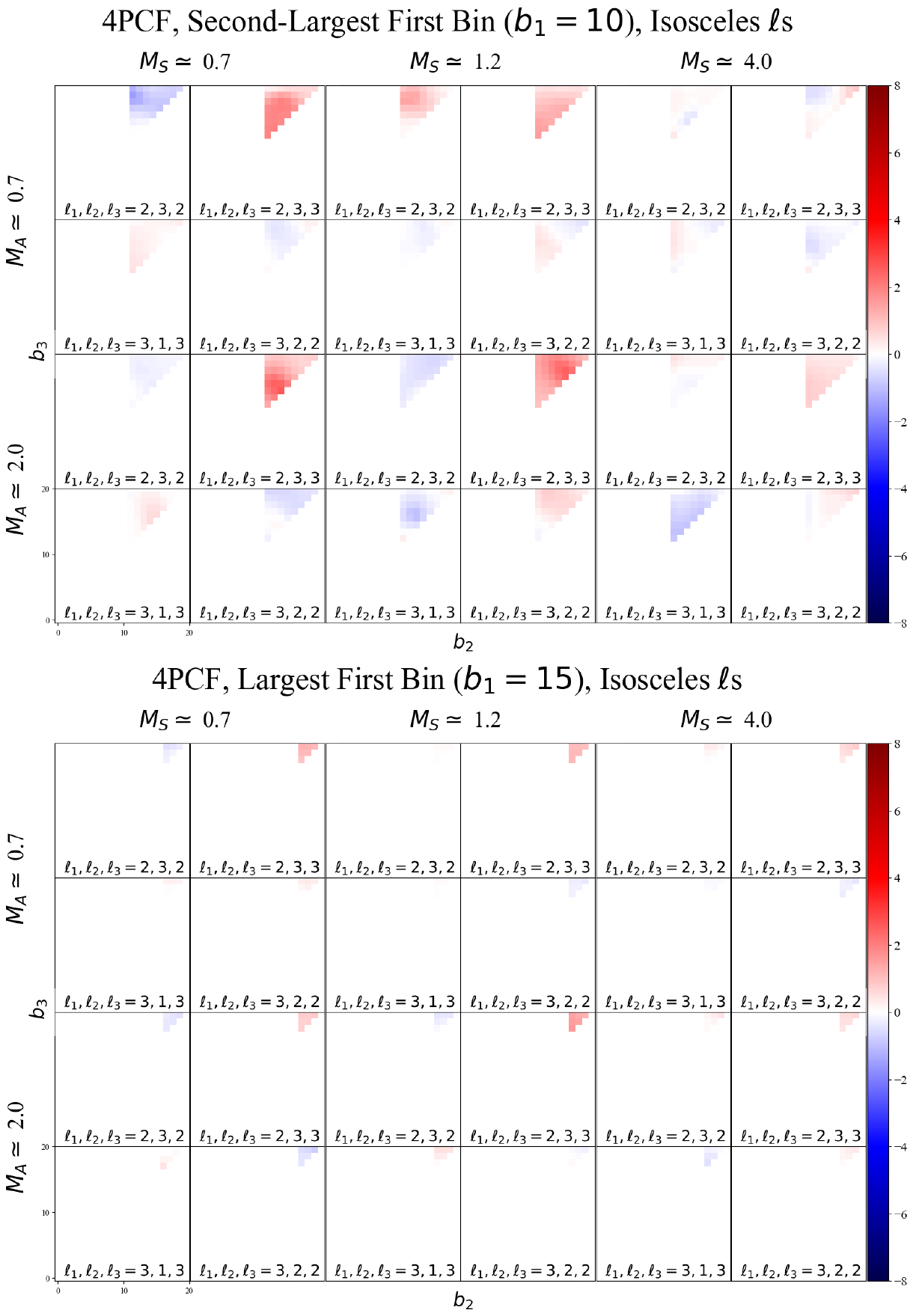}
    \caption{Same as Fig. \ref{fig:isosceles1_end.pdf} but for $\ell_1, \ell_2, \ell_3 = (2, 3, 2)$, $\ell_1, \ell_2, \ell_3 = (2,3,3)$, $\ell_1, \ell_2, \ell_3 = (3, 1, 3)$, and $\ell_1, \ell_2, \ell_3 = (3, 2, 2)$. The first image shows these $\ell$ combinations when $b_1 = 10$  and the second image shows when $b_1 = 15$. At higher $M_A$, correlated features sharpen as anti-correlations fade. Parity-odd modes retain structure, while parity-even modes weaken significantly. Further analysis is offered in Section \ref{sec:Discussion of Qualitative Trends}.}
	\label{fig:isosceles3_end.pdf}
\end{figure*}

Parity-odd components are fainter but remain detectable. This suggests that subtle asymmetries are present, which could be linked to a variety of factors such as anisotropic driving forces or magnetic field effects. The detection of these modes indicates that even in turbulent environments, small-scale asymmetries are preserved, which may be critical in understanding underlying processes. Despite parity-odd modes being generally weaker in strength, they seem to be far less sensitive to suppression as parameters, such as $M_A, M_S, b_1, b_2,$ and $b_3$, increase (see Fig. \ref{fig:equil_start.pdf}).
Parity-odd modes exhibit slower suppression, with residual signals surviving even at high parameter values. These modes often maintain asymmetry in their structure throughout, with stronger signals in more localized regions. In squeezed and scalene configurations, parity-odd modes are the most robust, able to retain signal strength across a broad bin range. Parity-odd modes inherently exhibit asymmetric signal patterns. This asymmetry persists even as parameters increase, which leads to more distinct localized peaks, as seen in Fig. \ref{fig:i&s2_end.pdf}. The structure of parity-odd signals is also less fragmented than parity-even signals at high parameter values.
Parity-odd higher $\ell$ modes retain structure better than their parity-even counterparts, with signals persisting in localized regions. Parity-odd lower $\ell$ modes show the greatest resilience, retaining strong, localized signals across all configurations and parameter values.
In equilateral configurations at higher parameter values, parity-odd modes retain more structure, even as intensity diminishes. For isosceles configurations, parity-odd modes are more robust, showing less fragmentation and stronger signals overall. In squeezed configurations, parity-odd modes dominate significantly at higher parameter values, being able to retain both structure and signal well. As scalene configurations approach higher parameters, parity-odd modes exhibit relatively stronger signals than parity-even modes.
\subsection{Comparison of Equilateral, Squeezed, Isosceles, and Scalene Modes}
\subsubsection{Equilateral Modes}
Equilateral configurations are generally more sensitive to parameter increases, with rapid suppression observed across all modes. At lower parameter values, signals are broadly distributed with balanced red and blue regions, but these weaken significantly as $M_A, M_S, b_1, b_2,$ and $b_3$ increase. We observe the fastest suppression at higher $\ell$ modes, with minimal residual signals at higher parameters, as seen in Fig \ref{fig:(3,3,0)_end.pdf}. In equilateral configurations, parity-even modes exhibit strong symmetry at small parameters but fragment quickly as they increase. As this occurs, signal peaks cluster near smaller $b_1, b_2, b_3$ values, leaving these off-equilateral regions significantly weaker in signals.
\subsubsection{Squeezed Modes}
\begin{figure*}
    \centering
    \includegraphics[width=0.95\textwidth, height=0.89\textheight]{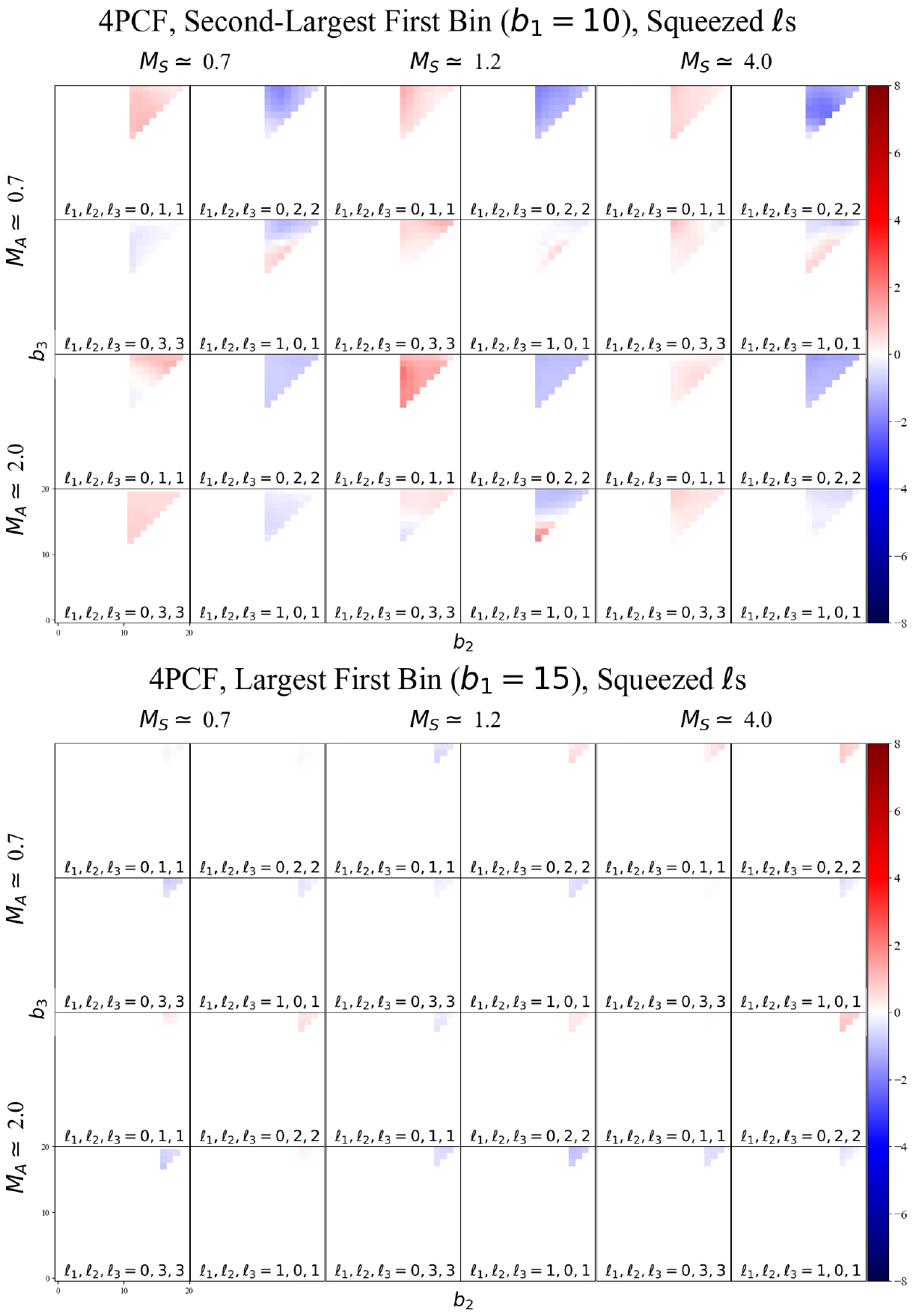}
    \caption{The 3D connected 4PCF signal-to-noise ratio, normalized by the standard deviation across time slices, computed in the squeezed multipole basis for $\ell_1, \ell_2, \ell_3 = (0, 1, 1)$, $\ell_1, \ell_2, \ell_3 = (0, 2, 2)$, $\ell_1, \ell_2, \ell_3 = (0, 3, 3)$, and $\ell_1, \ell_2, \ell_3 = (1, 0, 1)$. The top panel represents when $b_1 = 10$ and the bottom panel corresponds to when $b_1 = 15$. These plots adhere to the same $b_1, b_2$ limitations and can be interpreted by the same method described for Fig.  \ref{fig:equil_start.pdf}. Higher $M_a$ results in localization of peaks, with rapidly diminishing anti-correlations. Parity-odd modes retain more structure, while parity-even modes weaken faster. Further analysis is offered in Section \ref{sec:Discussion of Qualitative Trends}.}
    \label{fig:squeezed1_end.pdf}
\end{figure*}

Squeezed configurations are the most resilient to parameter increases. Signals remain strong and structured even at the largest parameter values, in particular for parity-odd modes. Peaks are more concentrated but retain significant strength across bin space, making squeezed configurations stand out as the most robust of all the configurations (see Fig. \ref{fig:squeezed1_end.pdf}). Parity-odd mode signals in squeezed configurations retain strength and structure better than any other configuration. At higher parameter, signals are highly localized, with clear red and blue regions near smaller $b_1$ and $b_3$ bins. Compared to other configurations, squeezed modes exhibit minimal suppression of signal intensity, even for higher $M_A, M_S$ and $b_3$.
\subsubsection{Isosceles Modes}
\begin{figure*}
    \centering
    \includegraphics[width=0.95\textwidth, height=0.89\textheight]{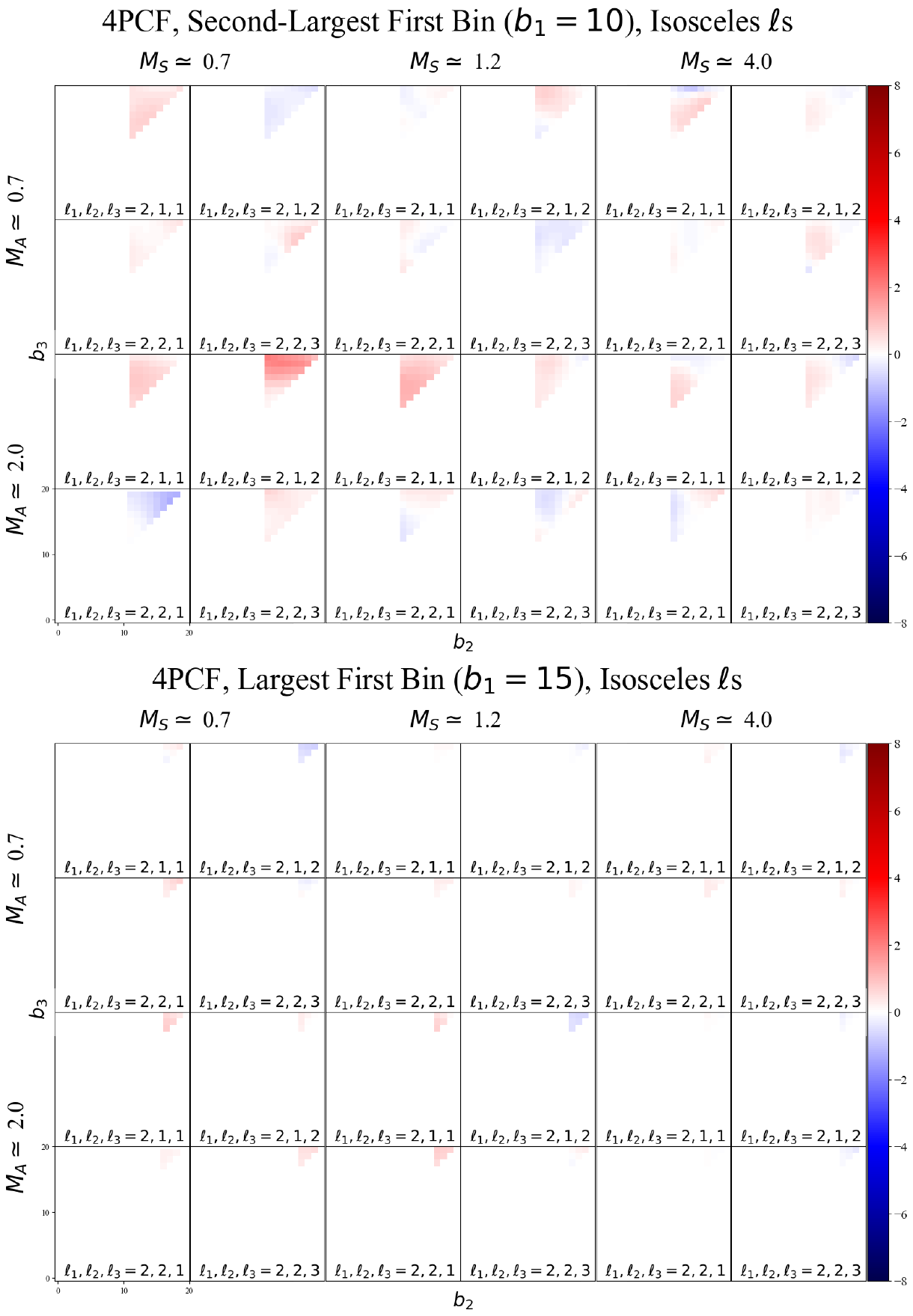}
    \caption{Same as Fig. \ref{fig:isosceles1_end.pdf} but for $\ell_1, \ell_2, \ell_3 = (2,1,1)$, $\ell_1, \ell_2, \ell_3 = (2,1,2)$, $\ell_1, \ell_2, \ell_3 = (2,2,1)$, and $\ell_1, \ell_2, \ell_3 = (2,2,3)$. The first image shows these $\ell$ combinations when $b_1 = 10$  and the second image shows when $b_1 = 15$. As $M_A$ increases, anti-correlations fade rapidly, leaving correlated regions  more dominant. Signal intensity decreases overall, with a particular reduction in intensity for parity-even modes. Further analysis is offered in Section \ref{sec:Discussion of Qualitative Trends}.}
	\label{fig:isosceles2_end.pdf}
\end{figure*}

Isosceles configurations exhibit moderate sensitivity to parameter increases. Signals weaken at these parameters but these configurations are able to retain localized peaks in certain bin regions. Parity-odd modes tend to show more resilience to parity-even modes, which suppress faster. Parity-odd modes in isosceles configurations, like in Fig. \ref{fig:isosceles2_end.pdf}, maintain structure better than equilateral configurations, even at higher $b_1, b_2, b_3$. Signal peaks shift to smaller $b_3$ at higher parameters, which indicates narrowing activity regions. In isosceles configurations, parity-even signals remain symmetric for smaller parameter values but exhibit fragmentation and asymmetry as parameters increase.
\subsubsection{Scalene Modes}
\begin{figure*}
    \centering
    \includegraphics[width=0.95\textwidth, height=0.89\textheight]{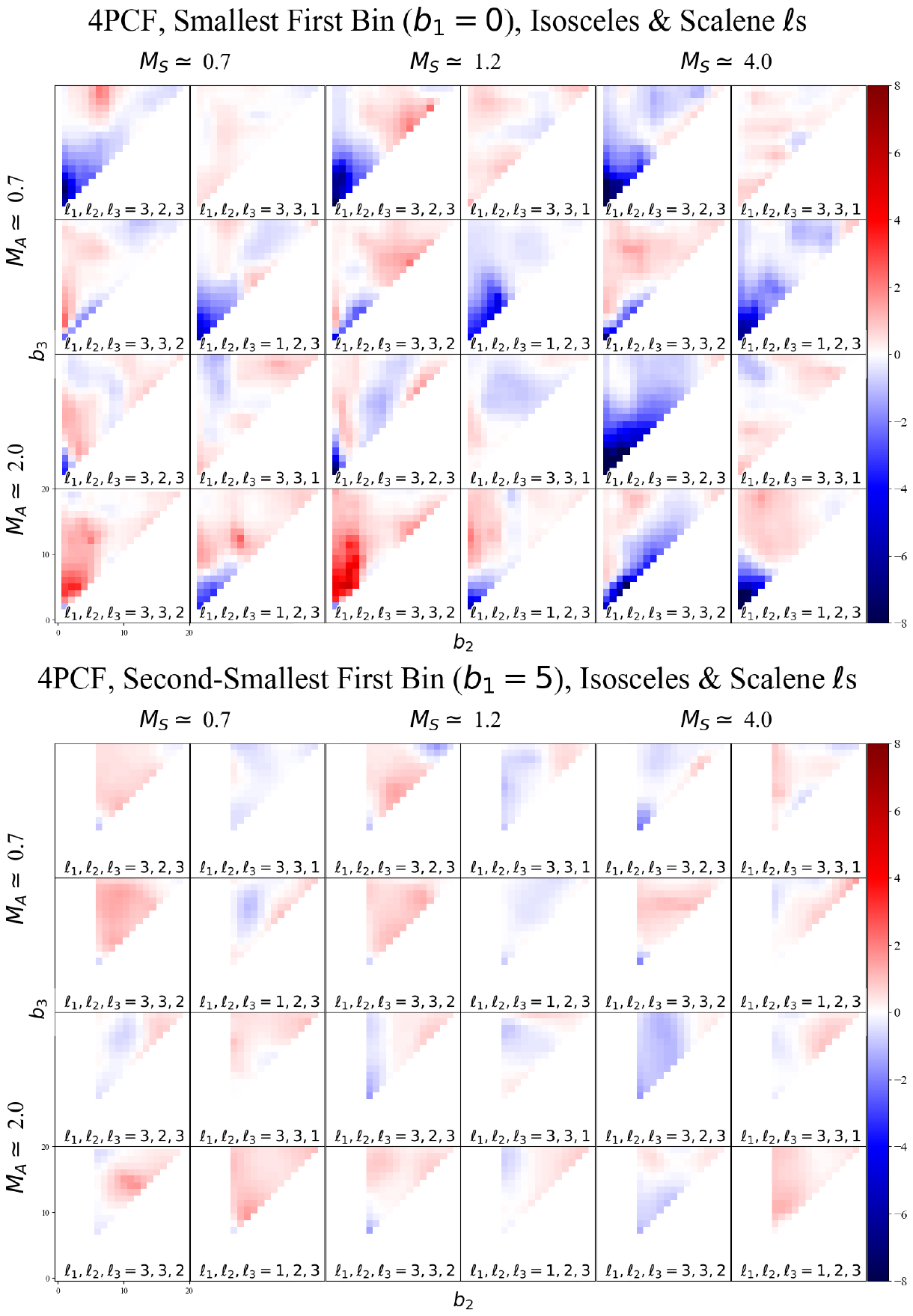}
    \caption{The 3D connected 4PCF signal-to-noise ratio, normalized by the standard deviation across time slices, computed in the isosceles and scalene multipole basis for $\ell_1, \ell_2, \ell_3 = (3, 2, 3)$, $\ell_1, \ell_2, \ell_3 = (3, 3, 1)$, $\ell_1, \ell_2, \ell_3 = (3, 3, 2)$, and $\ell_1, \ell_2, \ell_3 = (1, 2, 3)$. The top panel represents when $b_1 = 0$ and the bottom panel corresponds to when $b_1 = 5$. These plots adhere to the same $b_1, b_2$ limitations and can be interpreted by the same method described for Fig. \ref{fig:equil_start.pdf}. Strong signal intensity is observed at lower $M_A$, with peaks distributed across bins. Symmetry is preserved, showing the balance between correlated and anti-correlated features. Further analysis is offered in Section \ref{sec:Discussion of Qualitative Trends}.}
	\label{fig:i&s1_start.pdf}
\end{figure*}

Scalene configurations show stronger signal retention than both equilateral and isosceles configurations, particularly for parity-odd modes. Signals are more distributed across bin space, even at higher $b_1, b_2, b_3$, with less fragmentation than in equilateral configurations. parity-odd modes dominate at higher parameters in these configurations, with stronger and more localized signals than parity-even modes. We also see peaks in scalene configurations remain more relatively spread out with less suppression observed at higher $M_S$ and $b_2$ (see Fig. \ref{fig:i&s1_start.pdf}). In scalene configurations, parity-even modes are symmetric at smaller parameters but weaken and localize at higher values.

\graphicspath{{Figures/figures/4pcfmergepdf/equil_start.pdf}}
 
\clearpage

\graphicspath{{Figures/figures/4pcfmergepdf/equil_end.pdf}}
 
\clearpage

\graphicspath{{Figures/figures/4pcfmergepdf/squeezed1_start.pdf}}
 
\begin{figure*}
    \centering
    \includegraphics[width=0.95\textwidth, height=0.89\textheight]{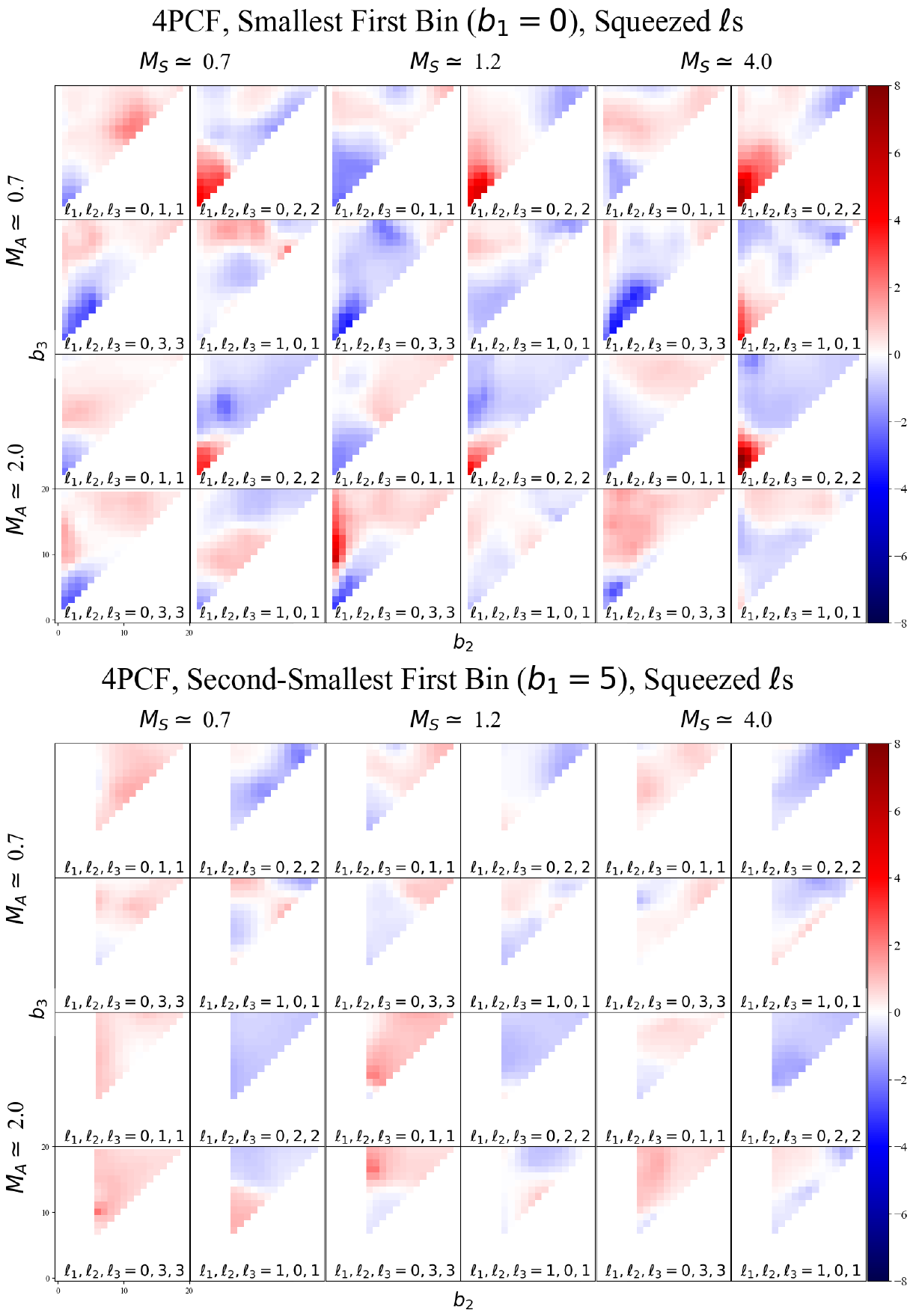}
    \caption{The 3D connected 4PCF signal-to-noise ratio, normalized by the standard deviation across time slices, computed in the squeezed multipole basis for $\ell_1, \ell_2, \ell_3 = (0, 1, 1)$, $\ell_1, \ell_2, \ell_3 = (0, 2, 2)$, $\ell_1, \ell_2, \ell_3 = (0, 3, 3)$, and $\ell_1, \ell_2, \ell_3 = (1, 0, 1)$. The top panel represents when $b_1 = 0$ and the bottom panel corresponds to when $b_1 = 5$. These plots adhere to the same $b_1, b_2$ limitations and can be interpreted by the same method described for Fig. \ref{fig:equil_start.pdf}. Lower $M_A$ produces symmetric patterns, with correlated and anti-correlated features well balanced. Peaks are widely distributed, particularly in parity-odd modes. Further analysis is offered in Section \ref{sec:Discussion of Qualitative Trends}.}
    \label{fig:squeezed1_start.pdf}
\end{figure*}

\clearpage
\graphicspath{{Figures/figures/4pcfmergepdf/squeezed1_end.pdf}}
 
\clearpage

\graphicspath{{Figures/figures/4pcfmergepdf/squeezed2_start.pdf}}
 
\begin{figure*}
    \centering
    \includegraphics[width=0.95\textwidth, height=0.89\textheight]{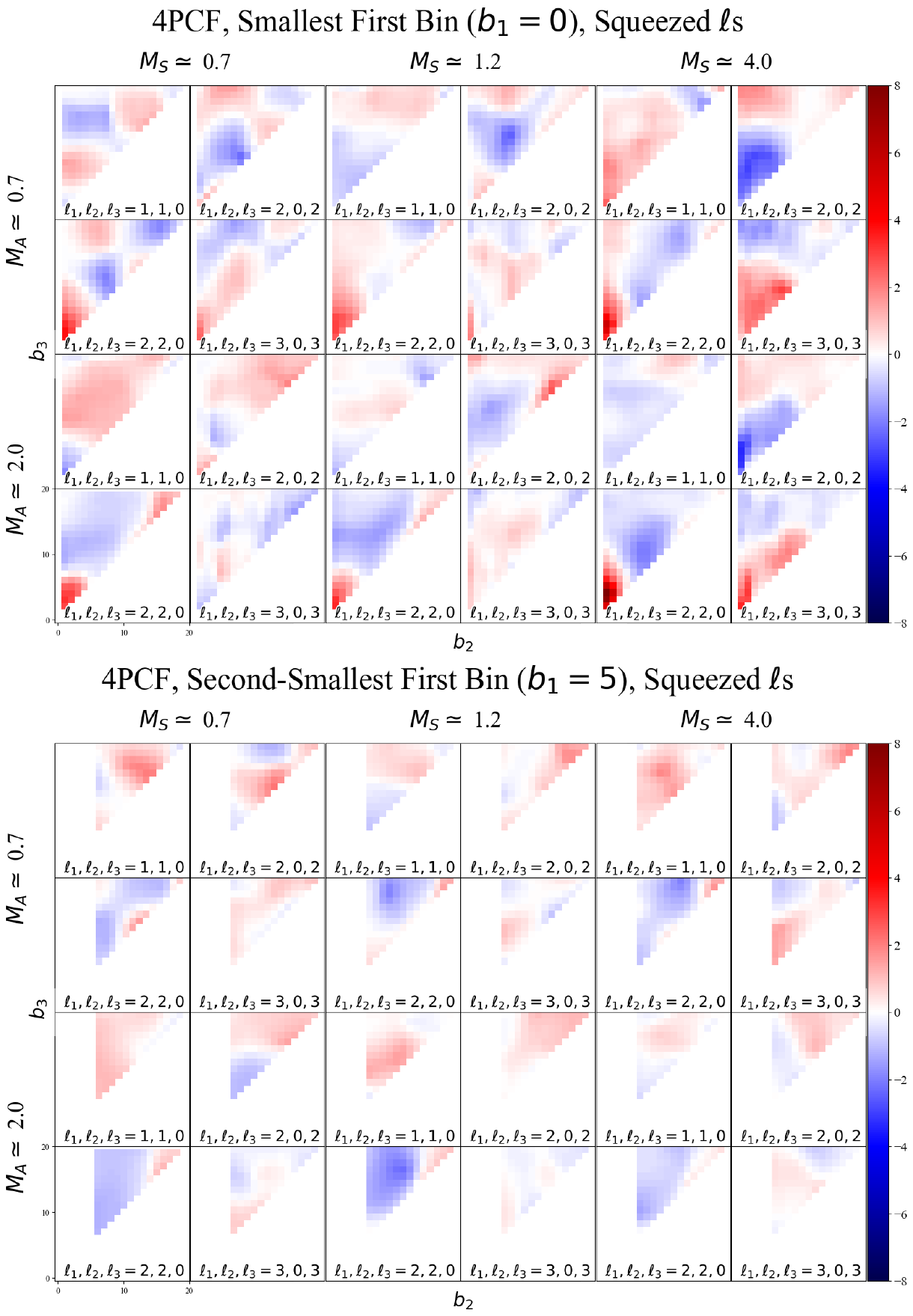}
    \caption{Same as Fig.  \ref{fig:squeezed1_start.pdf} but for $\ell_1, \ell_2, \ell_3= (1, 1, 0)$, $\ell_1, \ell_2, \ell_3 = (2, 0, 2)$, $\ell_1, \ell_2, \ell_3 = (2, 2, 0)$, and $\ell_1, \ell_2, \ell_3 = (3, 0, 3)$. At lower $M_A$, peaks are symmetric and evenly distributed. Signal is evenly distributed across bins and both parity-odd and parity-even modes exhibit strong intensity. Further analysis is offered in Section \ref{sec:Discussion of Qualitative Trends}.}
    \label{fig:squeezed2_start.pdf}
\end{figure*}
\clearpage

\graphicspath{{Figures/figures/4pcfmergepdf/squeezed2_end.pdf}}
 
\begin{figure*}
    \centering
    \includegraphics[width=0.95\textwidth, height=0.89\textheight]{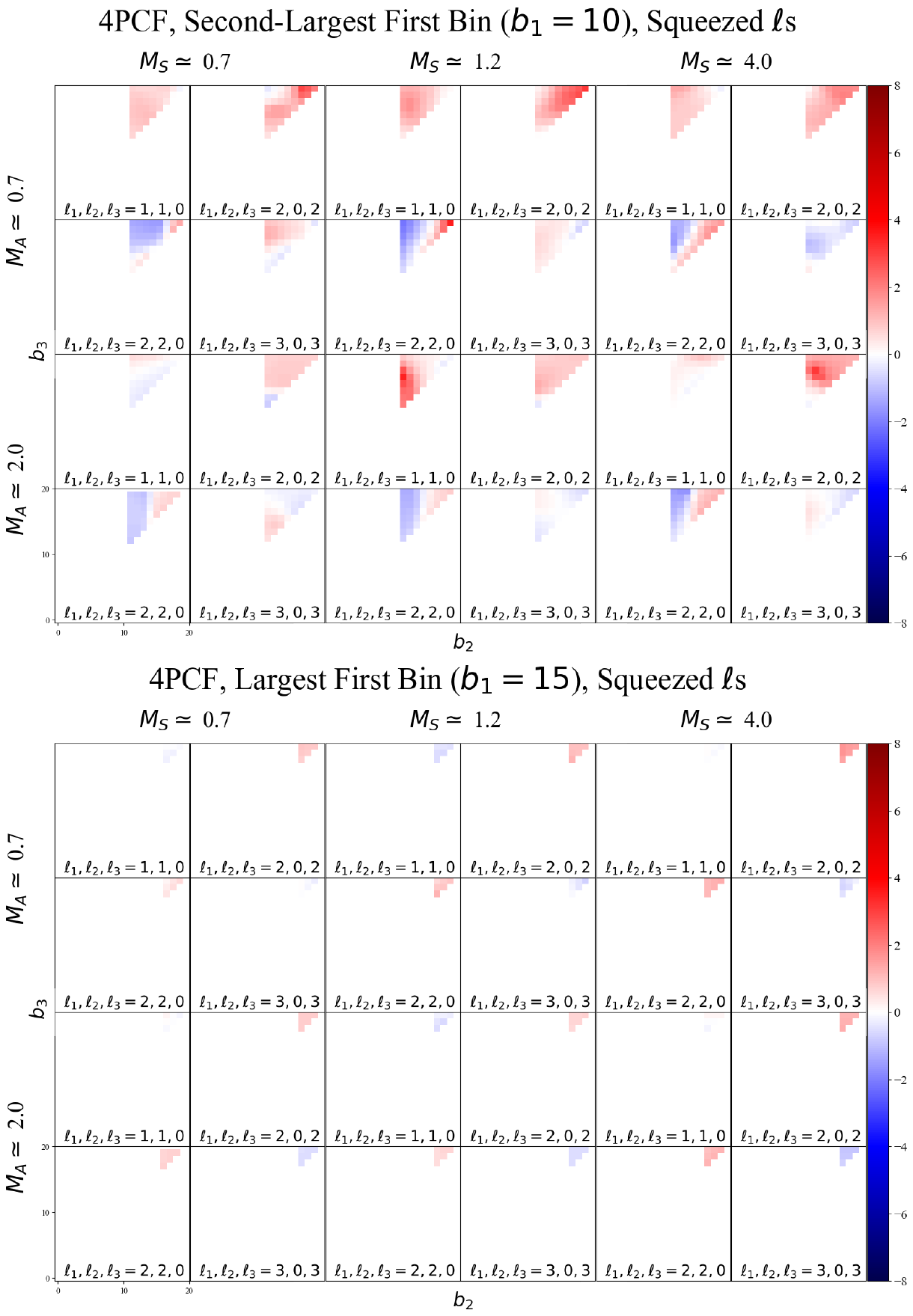}
    \caption{Same as Fig.  \ref{fig:squeezed1_end.pdf} but for $\ell_1, \ell_2, \ell_3= (1, 1, 0)$, $\ell_1, \ell_2, \ell_3 = (2, 0, 2)$, $\ell_1, \ell_2, \ell_3 = (2, 2, 0)$, and $\ell_1, \ell_2, \ell_3 = (3, 0, 3)$. At higher $M_A$, correlated features dominate as anti-correlations fade. Parity-odd modes retain structure, while parity-even modes show greater suppression. Further analysis is offered in Section \ref{sec:Discussion of Qualitative Trends}.}
    \label{fig:squeezed2_end.pdf}
\end{figure*}
\clearpage

\graphicspath{{Figures/figures/4pcfmergepdf/_3,3,0__start.pdf}}

\begin{figure*}
    \centering
    \includegraphics[width=0.95\textwidth, height=0.89\textheight]{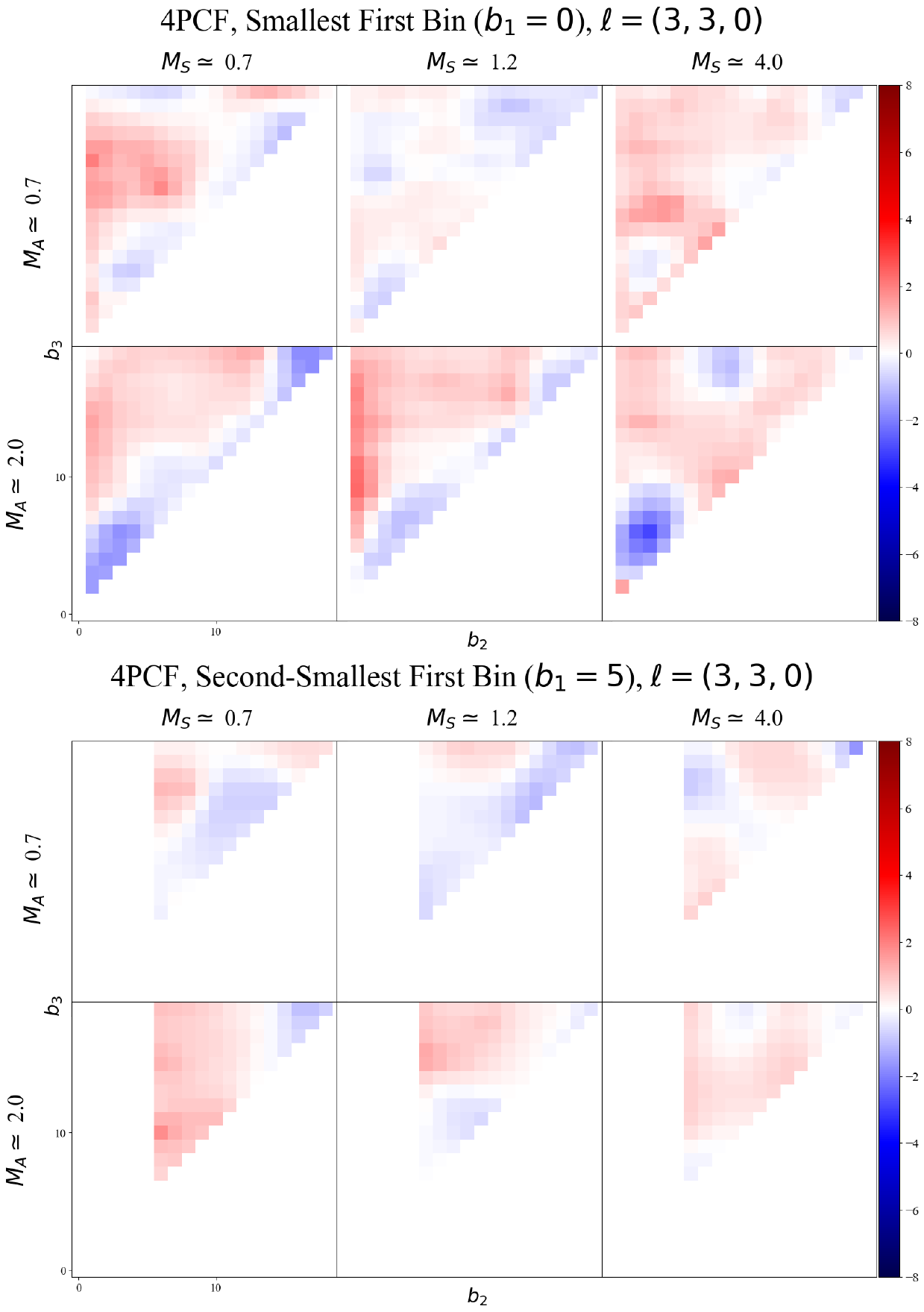}
    \caption{The 3D connected 4PCF signal-to-noise ratio, normalized by the standard deviation across time slices, computed in the squeezed multipole basis for $\ell_1, \ell_2, \ell_3 = (3, 3, 0)$. The top panel represents when $b_1 = 0$ and the bottom panel corresponds to when $b_1 = 5$. These plots adhere to the same $b_1, b_2$ limitations and can be interpreted by the same method described for Fig. \ref{fig:equil_start.pdf}. Parity-even modes show strong intensity at lower $M_A$, with peaks evenly distributed across bins. These broader features reflect smoother transitions between correlated and anti-correlated features. Further analysis is offered in Section \ref{sec:Discussion of Qualitative Trends}.}
	\label{fig:(3,3,0)_start.pdf}
\end{figure*}
\clearpage

\graphicspath{{Figures/figures/4pcfmergepdf/_3,3,0__end.pdf}}

\begin{figure*}
    \centering
    \includegraphics[width=0.95\textwidth, height=0.89\textheight]{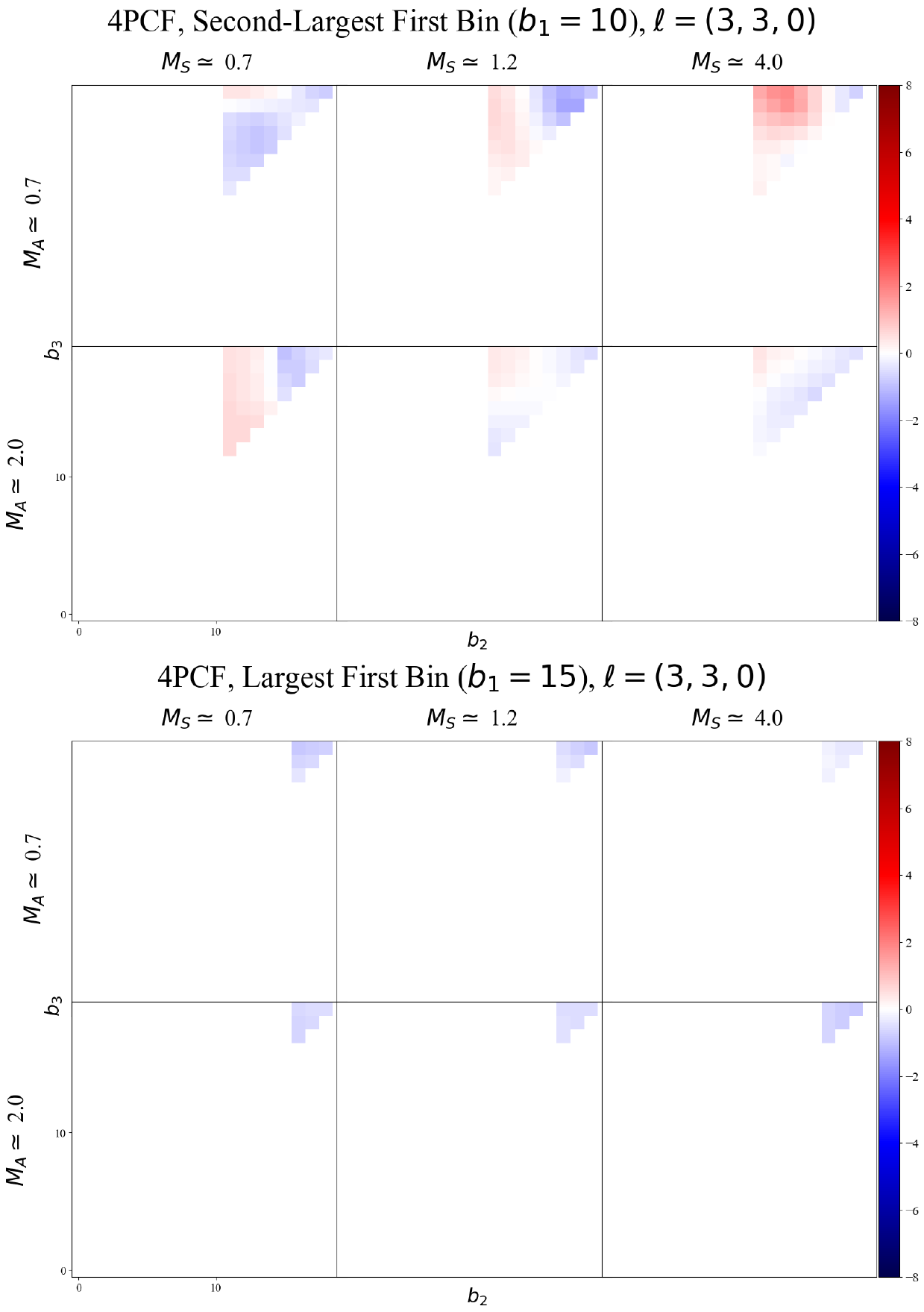}
    \caption{The 3D connected 4PCF signal-to-noise ratio, normalized by the standard deviation across time slices, computed in the squeezed multipole basis for $\ell_1, \ell_2, \ell_3 = (3, 3, 0)$. The top panel represents when $b_1 = 10$ and the bottom panel corresponds to when $b_1 = 15$. These plots adhere to the same $b_1, b_2$ limitations and can be interpreted by the same method described for Fig. \ref{fig:equil_start.pdf}. Signal intensity fades at higher $M_A$, with peaks remaining widespread but weaker. Parity-even modes show uniform reduction in signal compared to lower $M_A$. Further analysis is offered in Section \ref{sec:Discussion of Qualitative Trends}.}
	\label{fig:(3,3,0)_end.pdf}
\end{figure*}
\clearpage

\graphicspath{{Figures/figures/4pcfmergepdf/isosceles1_start.pdf}}

\begin{figure*}
    \centering
    \includegraphics[width=0.95\textwidth, height=0.89\textheight]{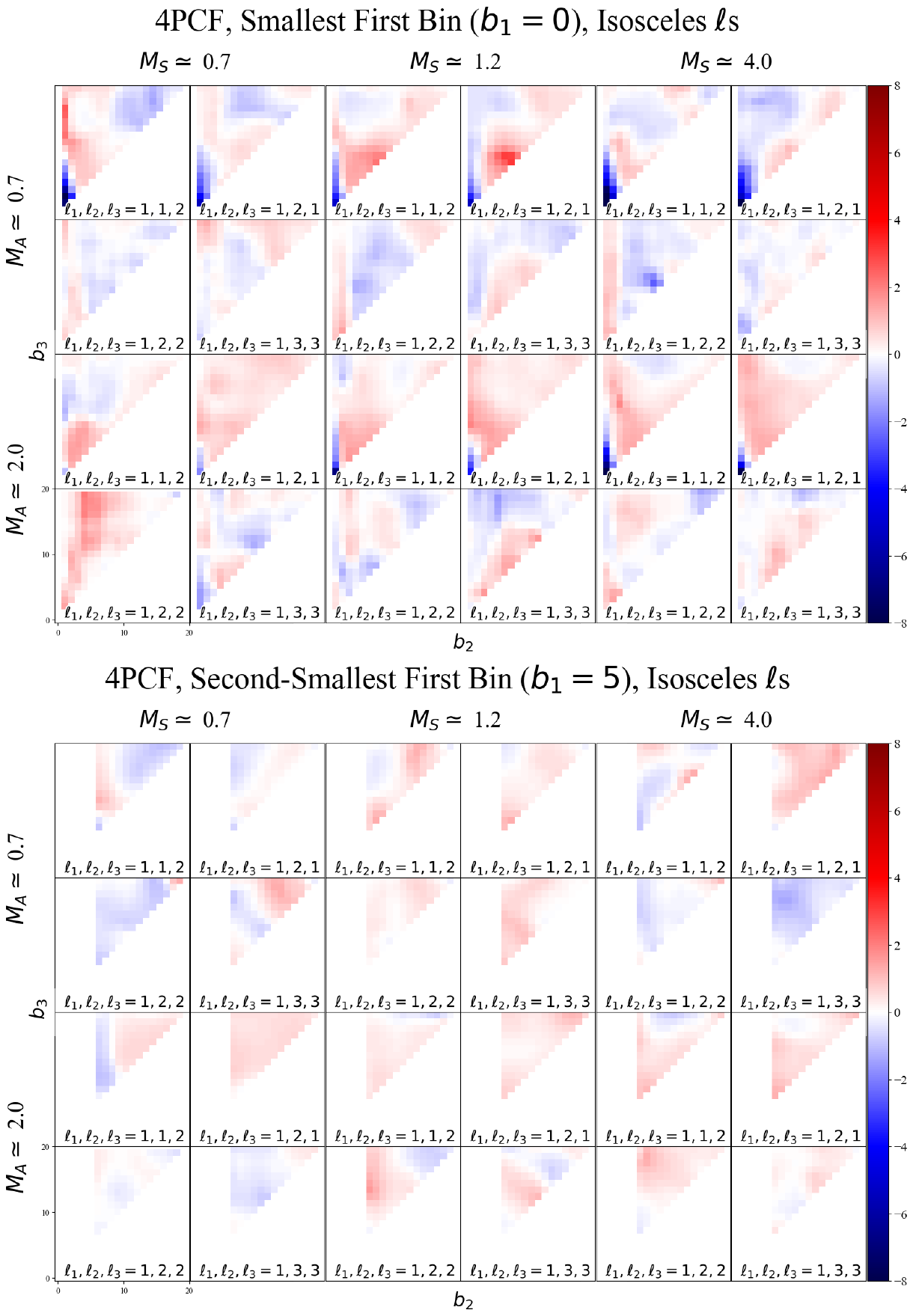}
    \caption{The 3D connected 4PCF signal-to-noise ratio, normalized by the standard deviation across time slices, computed in the isosceles multipole basis for $\ell_1, \ell_2, \ell_3 = (1, 1, 2)$, $\ell_1, \ell_2, \ell_3 = (1, 2, 1)$, $\ell_1, \ell_2, \ell_3 = (1, 2, 2)$, and $\ell_1, \ell_2, \ell_3 = (1, 3, 3)$. The top panel represents when $b_1 = 0$ and the bottom panel corresponds to when $b_1 = 5$. These plots adhere to the same $b_1, b_2$ limitations and can be interpreted by the same method described for Fig. \ref{fig:equil_start.pdf}. Lower $M_A$ produces strong anti-correlations with peaks distributed across bins, particularly in lower $b_3$ regions. Further analysis is offered in Section \ref{sec:Discussion of Qualitative Trends}.}
	\label{fig:isosceles1_start.pdf}
\end{figure*}
\clearpage

\graphicspath{{Figures/figures/4pcfmergepdf/isosceles1_end.pdf}}

\clearpage

\graphicspath{{Figures/figures/4pcfmergepdf/isosceles2_start.pdf}}

\begin{figure*}
    \centering
    \includegraphics[width=0.95\textwidth, height=0.89\textheight]{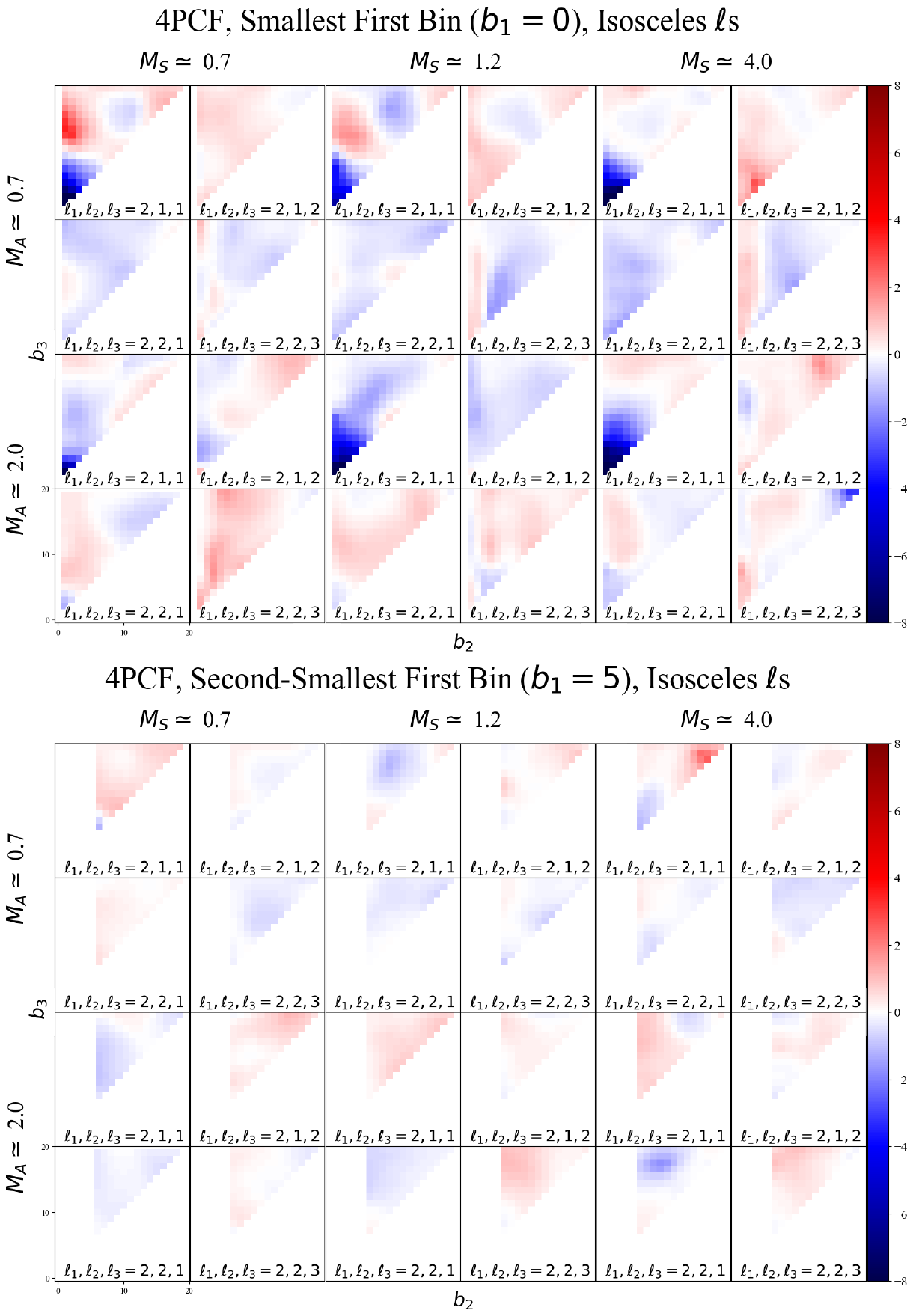}
    \caption{Same as Fig. \ref{fig:isosceles1_start.pdf} but for $\ell_1, \ell_2, \ell_3 = (2,1,1)$, $\ell_1, \ell_2, \ell_3 = (2,1,2)$, $\ell_1, \ell_2, \ell_3 = (2,2,1)$, and $\ell_1, \ell_2, \ell_3 = (2,2,3)$. The first image shows these $\ell$ combinations when $b_1 = 0$  and the second image shows when $b_1 = 5$. At lower $M_A$, signal symmetry is preserved with strong anti-correlation in higher bins. Peaks are broadly distributed, reflecting a balance signal. Further analysis is offered in Section \ref{sec:Discussion of Qualitative Trends}.}
	\label{fig:isosceles2_start.pdf}
\end{figure*}
\clearpage

\graphicspath{{Figures/figures/4pcfmergepdf/isosceles2_end.pdf}}

\clearpage

\graphicspath{{Figures/figures/4pcfmergepdf/isosceles3_start.pdf}}

\begin{figure*}
    \centering
    \includegraphics[width=0.95\textwidth, height=0.89\textheight]{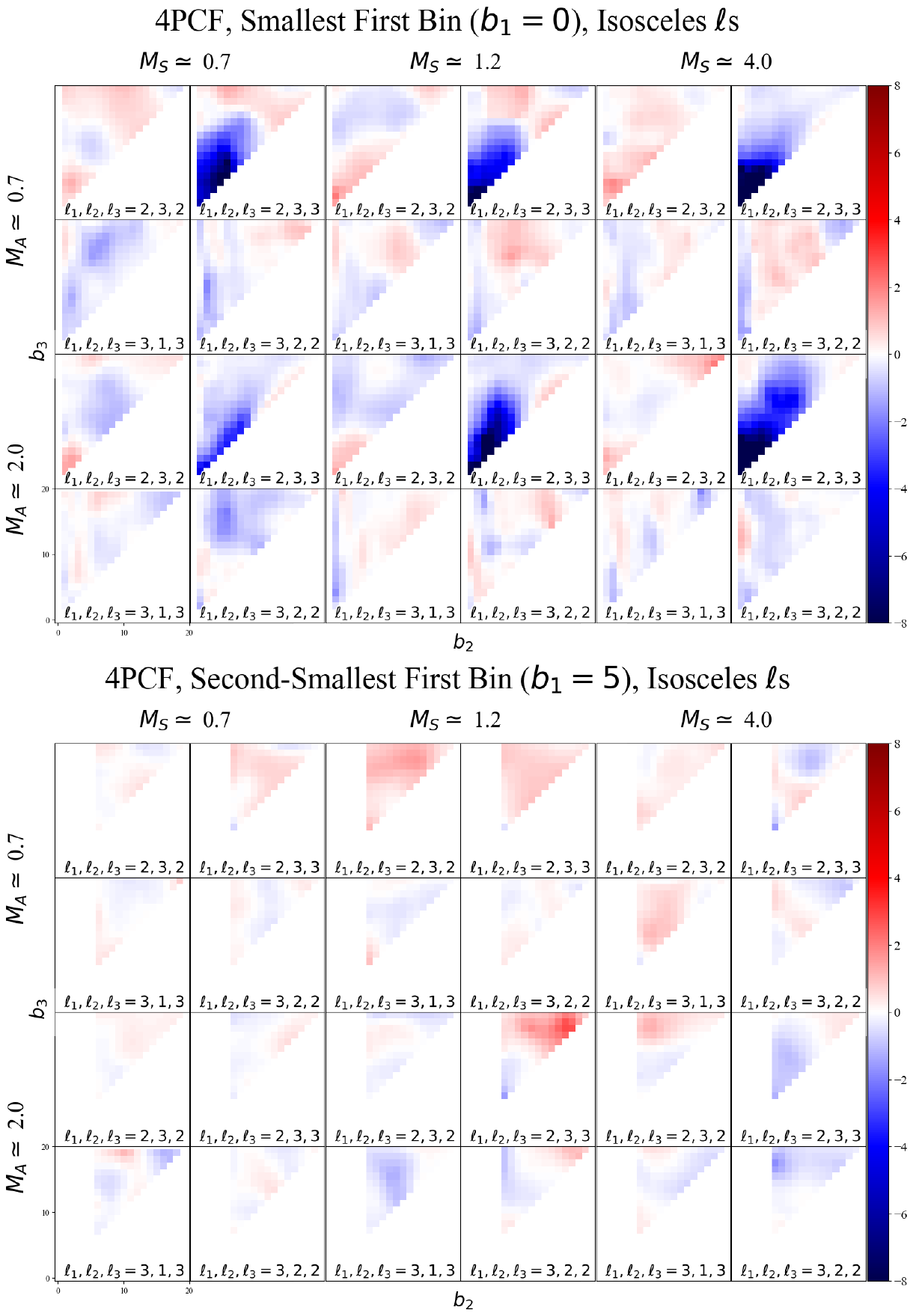}
    \caption{Same as Fig. \ref{fig:isosceles1_start.pdf} but for $\ell_1, \ell_2, \ell_3 = (2, 3, 2)$, $\ell_1, \ell_2, \ell_3 = (2,3,3)$, $\ell_1, \ell_2, \ell_3 = (3, 1, 3)$, and $\ell_1, \ell_2, \ell_3 = (3,2,2)$. The first image shows these $\ell$ combinations when $b_1 = 0$  and the second image shows when $b_1 = 5$. At lower $M_A$, peaks are broadly distributed across bins, with symmetry between correlated and anti-correlated regions. Strong intensity is visible throughout. Further analysis is offered in Section \ref{sec:Discussion of Qualitative Trends}.}
	\label{fig:isosceles3_start.pdf}
\end{figure*}
\clearpage

\graphicspath{{Figures/figures/4pcfmergepdf/isosceles3_end.pdf}}

\clearpage

\graphicspath{{Figures/figures/4pcfmergepdf/i_s1_start.pdf}}

\clearpage

\graphicspath{{Figures/figures/4pcfmergepdf/i_s1_end.pdf}}

\begin{figure*}
    \centering
    \includegraphics[width=0.95\textwidth, height=0.89\textheight]{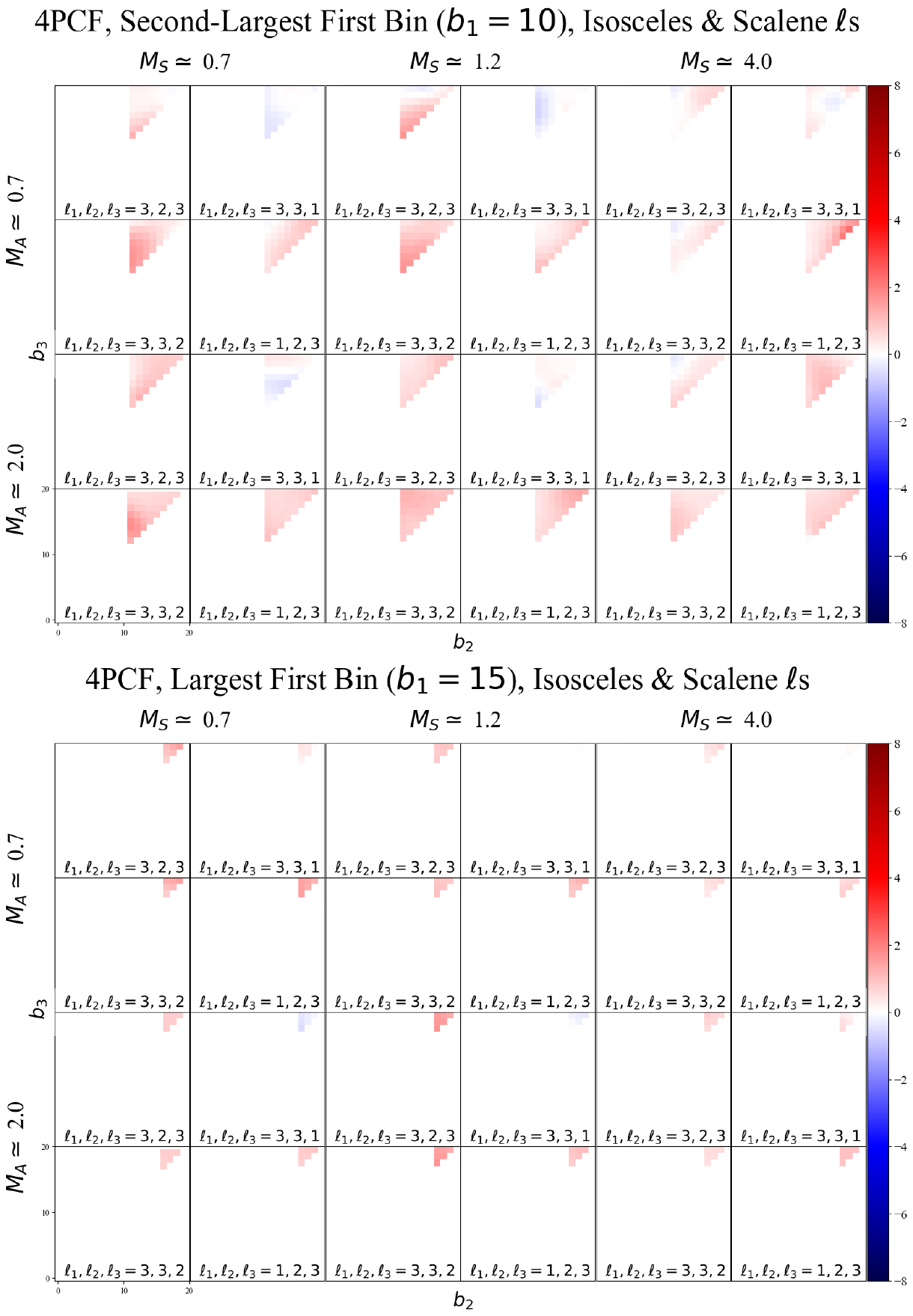}
    \caption{The 3D connected 4PCF signal-to-noise ratio, normalized by the standard deviation across time slices, computed in the isosceles and scalene multipole basis for $\ell_1, \ell_2, \ell_3 = (3, 2, 3)$, $\ell_1, \ell_2, \ell_3 = (3, 3, 1)$, $\ell_1, \ell_2, \ell_3 = (3, 3, 2)$, and $\ell_1, \ell_2, \ell_3 = (1, 2, 3)$. The top panel represents when $b_1 = 10$ and the bottom panel corresponds to when $b_1 = 15$. These plots adhere to the same $b_1, b_2$ limitations and can be interpreted by the same method described for Fig. \ref{fig:equil_start.pdf}. As $M_A$ increases, anti-correlated features diminish faster than correlated ones, disrupting symmetry. Peaks become more localized, reflecting a clustering effect in the 4PCF signal. Further analysis is offered in Section \ref{sec:Discussion of Qualitative Trends}.}
	\label{fig:i&s1_end.pdf}
\end{figure*}
\clearpage

\graphicspath{{Figures/figures/4pcfmergepdf/i_s2_start.pdf}}

\begin{figure*}
    \centering
    \includegraphics[width=0.95\textwidth, height=0.89\textheight]{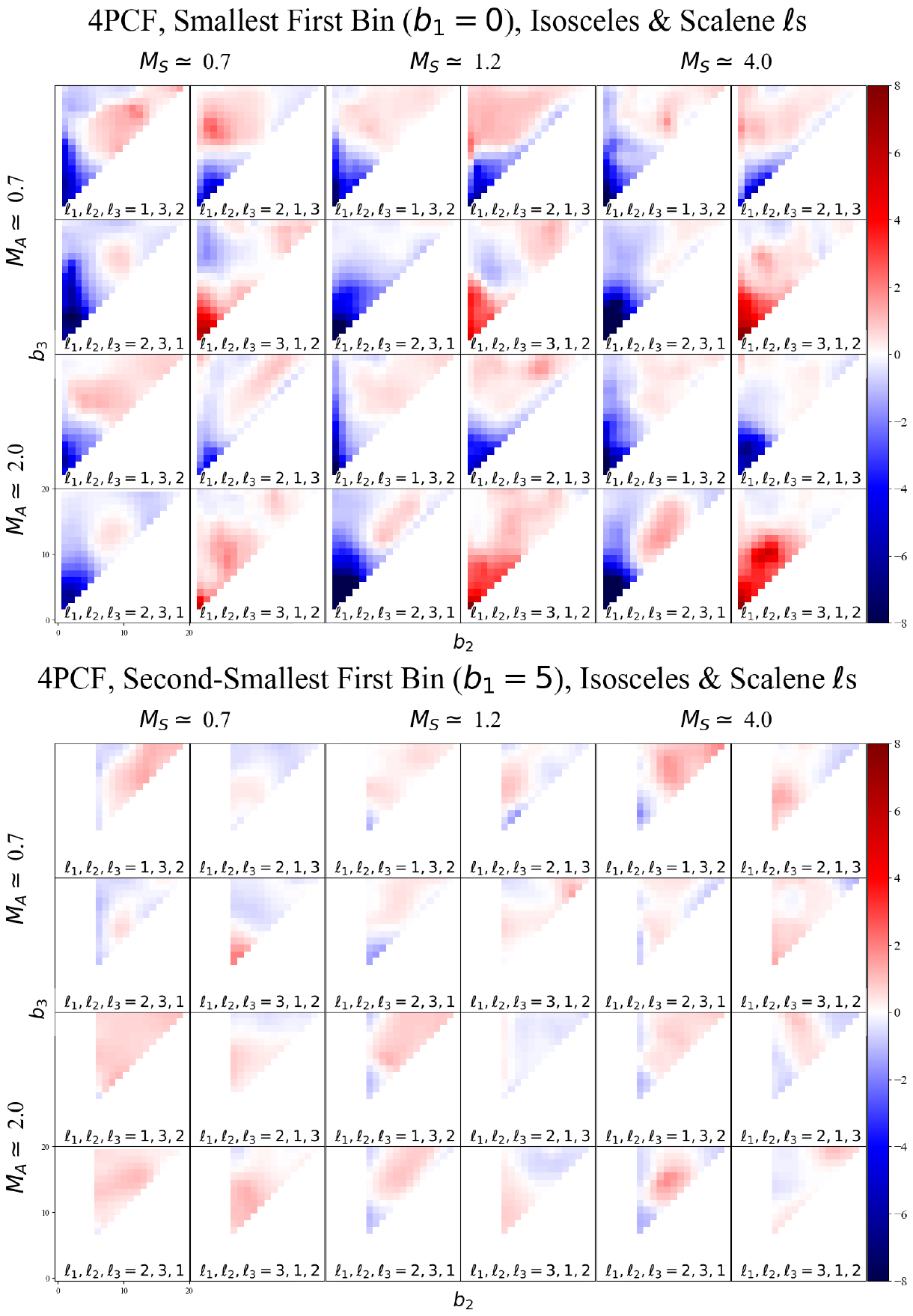}
    \caption{Same as Fig. \ref{fig:i&s1_start.pdf} but for $\ell_1, \ell_2, \ell_3 = (1, 3, 2)$, $\ell_1, \ell_2, \ell_3 = (2, 1, 3)$, $\ell_1, \ell_2, \ell_3 = (2, 3, 1)$, and $\ell_1, \ell_2, \ell_3 = (3, 1, 2)$. The first image shows these $\ell$ combinations when $b_1 = 0$  and the second image shows when $b_1 = 5$. Lower $M_A$ produces broadly distributed signal intensity, with clear peaks in anti-correlated regions. Symmetry between correlated and anti-correlated regions is clear. Further analysis is offered in Section \ref{sec:Discussion of Qualitative Trends}.}
	\label{fig:i&s2_start.pdf}
\end{figure*}
\clearpage

\graphicspath{{Figures/figures/4pcfmergepdf/i_s2_end.pdf}}

\begin{figure*}
    \centering
    \includegraphics[width=0.95\textwidth, height=0.89\textheight]{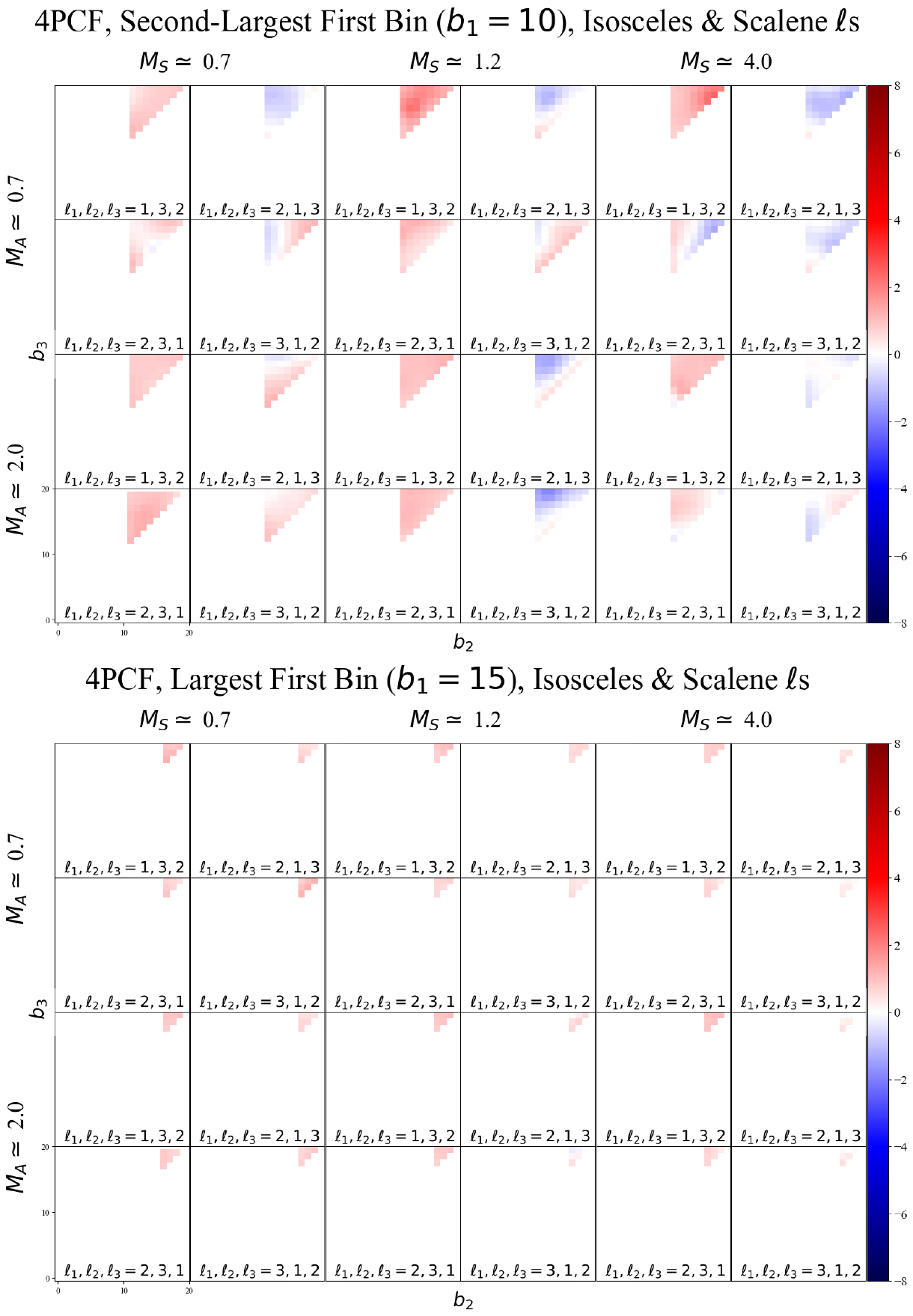}
    \caption{Same as Fig. \ref{fig:i&s1_end.pdf} but for $\ell_1, \ell_2, \ell_3 = (1, 3, 2)$, $\ell_1, \ell_2, \ell_3 = (2, 1, 3)$, $\ell_1, \ell_2, \ell_3 = (2, 3, 1)$, and $\ell_1, \ell_2, \ell_3 = (3, 1, 2)$. The first image shows these $\ell$ combinations when $b_1 = 10$  and the second image shows when $b_1 = 15$. As $M_A$ increases, correlated features dominate as anti-correlations diminish rapidly. Parity-even modes weaken faster, while parity-odd modes reatin more structure. Further analysis is offered in Section \ref{sec:Discussion of Qualitative Trends}.}
	\label{fig:i&s2_end.pdf}
\end{figure*}
\clearpage

\graphicspath{{Figures/figures/4pcfmergepdf/_3,2,1__start.pdf}}

\begin{figure*}
    \centering
    \includegraphics[width=0.95\textwidth, height=0.89\textheight]{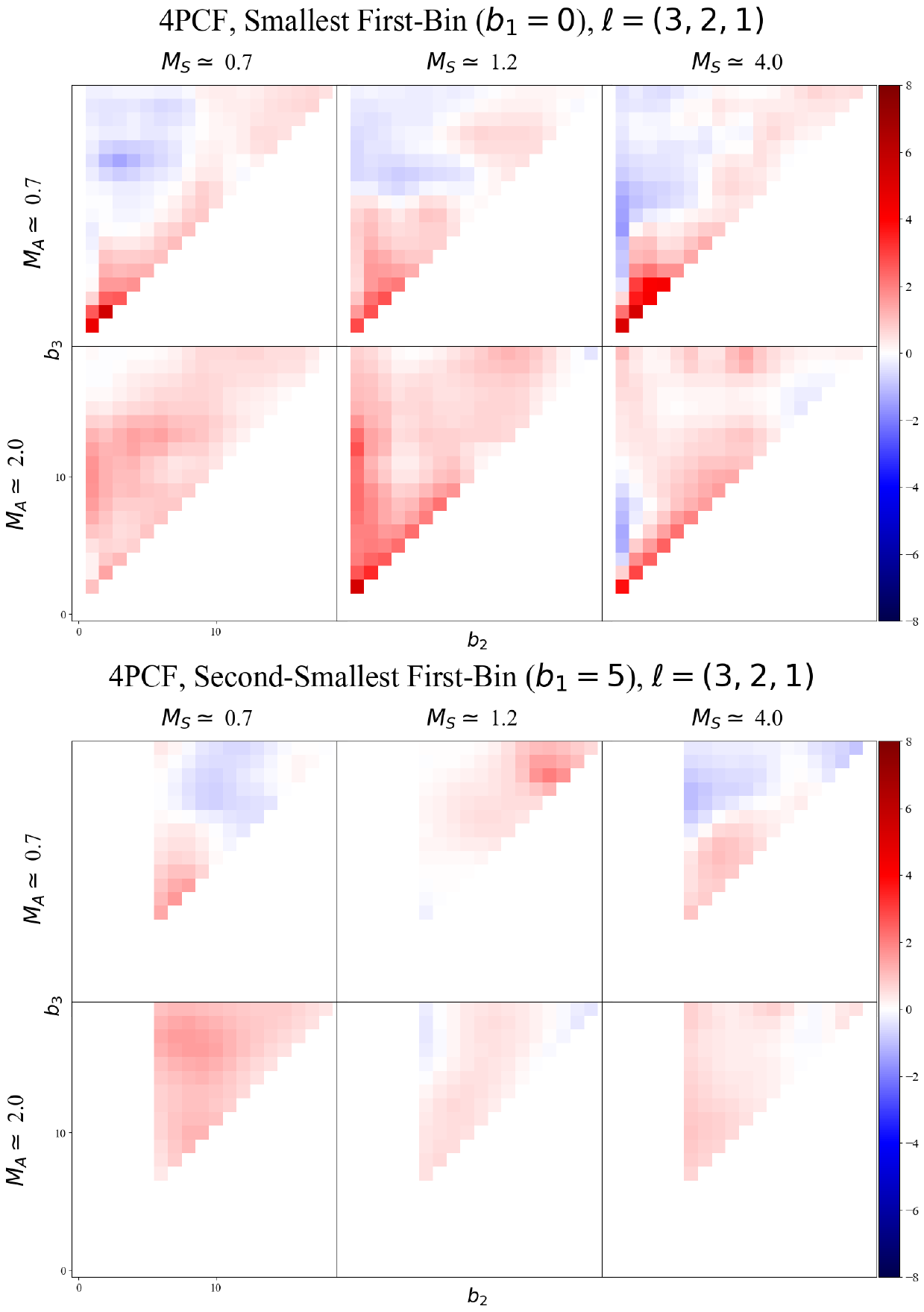}
    \caption{The 3D connected 4PCF signal-to-noise ratio, normalized by the standard deviation across time slices, computed in the scalene multipole basis for $\ell_1, \ell_2, \ell_3 = (3, 2, 1)$. The top panel represents when $b_1 = 0$ and the bottom panel corresponds to when $b_1 = 5$. These plots adhere to the same $b_1, b_2$ limitations and can be interpreted by the same method described for Fig. \ref{fig:equil_start.pdf}. At low $M_A$, the signal shows strong intensity with well-preserved symmetry between correlated and anti-correlated patterns. Peaks are distributed across bins, particularly in lower $\ell$ combinations. Further analysis is offered in Section \ref{sec:Discussion of Qualitative Trends}.}
	\label{fig:(3,2,1)_start.pdf}
\end{figure*}
\clearpage
\graphicspath{{Figures/figures/4pcfmergepdf/_3,2,1__end.pdf}}

\begin{figure*}
    \centering
    \includegraphics[width=0.95\textwidth, height=0.89\textheight]{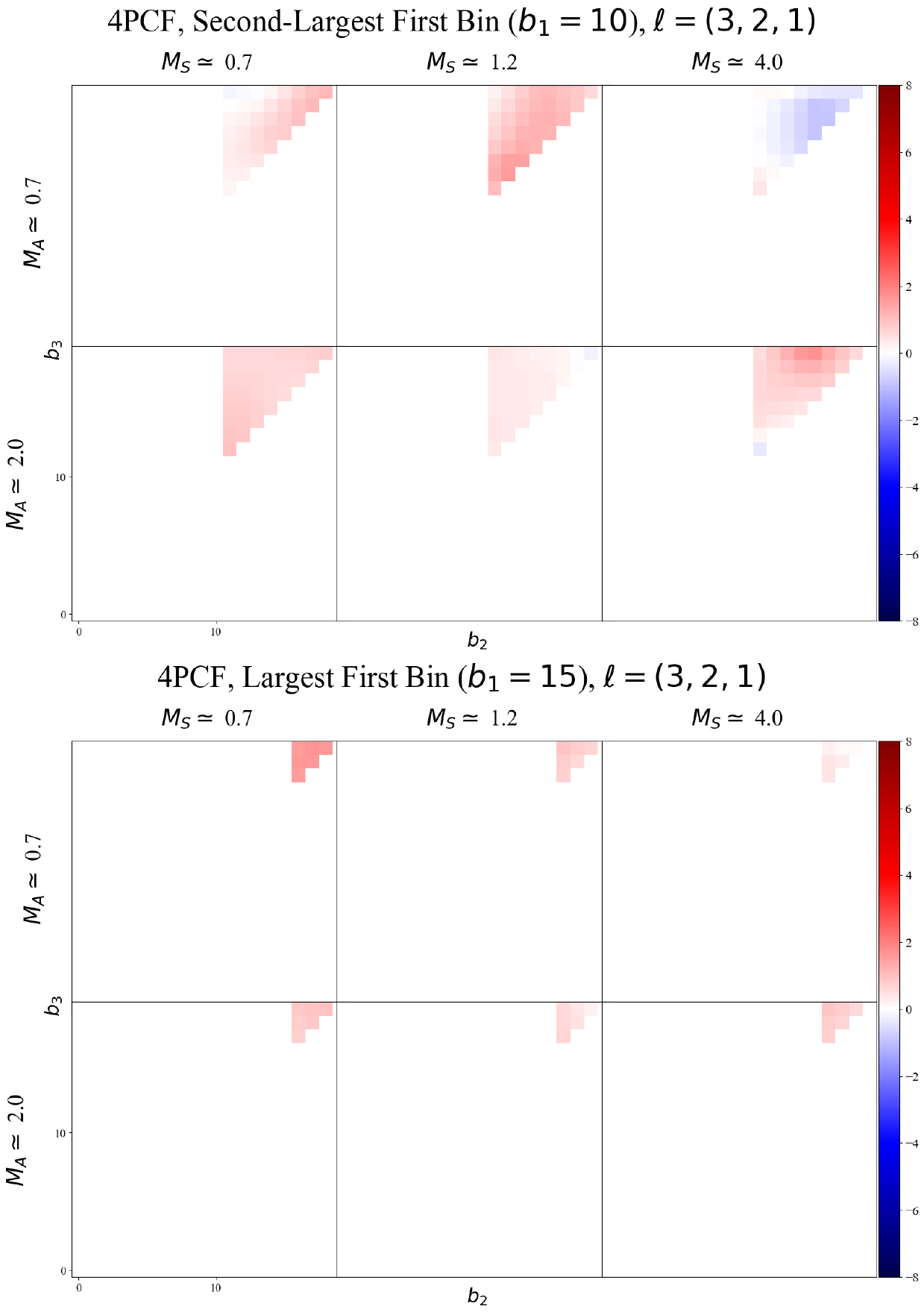}
    \caption{The 3D connected 4PCF signal-to-noise ratio, normalized by the standard deviation across time slices, computed in the scalene multipole basis for $\ell_1, \ell_2, \ell_3 = (3, 2, 1)$. The top panel represents when $b_1 = 10$ and the bottom panel corresponds to when $b_1 = 15$. These plots adhere to the same $b_1, b_2$ limitations and can be interpreted by the same method described for Fig. \ref{fig:equil_start.pdf}. At higher $M_A$, signal intensity decreases significantly, with localization of peaks towards smaller bins. Parity-even modes undergo a more drastic reduction in intensity compared to parity-odd modes. Further analysis is offered in Section \ref{sec:Discussion of Qualitative Trends}.}
	\label{fig:(3,2,1)_end.pdf}
\end{figure*}
\clearpage

\section{Discussion \& Conclusions}
\label{sec:Discussion & Conclusions}
This study represents the first-ever application of the 4-Point Correlation Function (4PCF) to magnetohydrodynamic turbulence simulations of the interstellar medium (ISM). Our high signal-to-noise detection of non-Gaussian correlations in the density field demonstrates the significant value of the 4PCF in probing high-order features of turbulence that cannot be captured by the 2PCF alonne, and argue strongly that the 4PCF also adds information beyond the 3PCF. By isolating the connected component, with the use of the \texttt{python} package \textsc{sarabande}, we were able to efficiently analyze many sets of simulations with varying sonic Mach and Alfv\'enic Mach parameters. This work revealed new patterns and correlations that are crucial to developing a more comprehensive understanding of turbulence and star formation in the ISM.

We found that parity-even components dominated the signal. Our detection of parity-odd components is significant in opening avenues to study coherent anisotropies, such as magnetic fields, that may be present.

The computational framework developed in this work lays the groundwork for future investigations of the complex relationship between turbulence, magnetic fields, and star formation. It establishes the 4PCF as a tool to study complex astrophysical systems, as we plan to expand the range of simulations analyzed and integrate observational data to further enhance our understanding of turbulence in shaping the ISM.

These findings demonstrate the powerful ability of the 4PCF as a probe for non-Gaussian features in turbulent mediums. By isolating the connected component, this methodology is able to efficiently gain access to information that was previously inaccessible with lower-order statistics such as the 2PCF or 3PCF. This work lays the foundation for using the 4PCF to study the ISM, providing insight into the role of turbulence and magnetic fields in star formation. 

\begin{figure*}
    \centering
    \includegraphics[width=\textwidth]{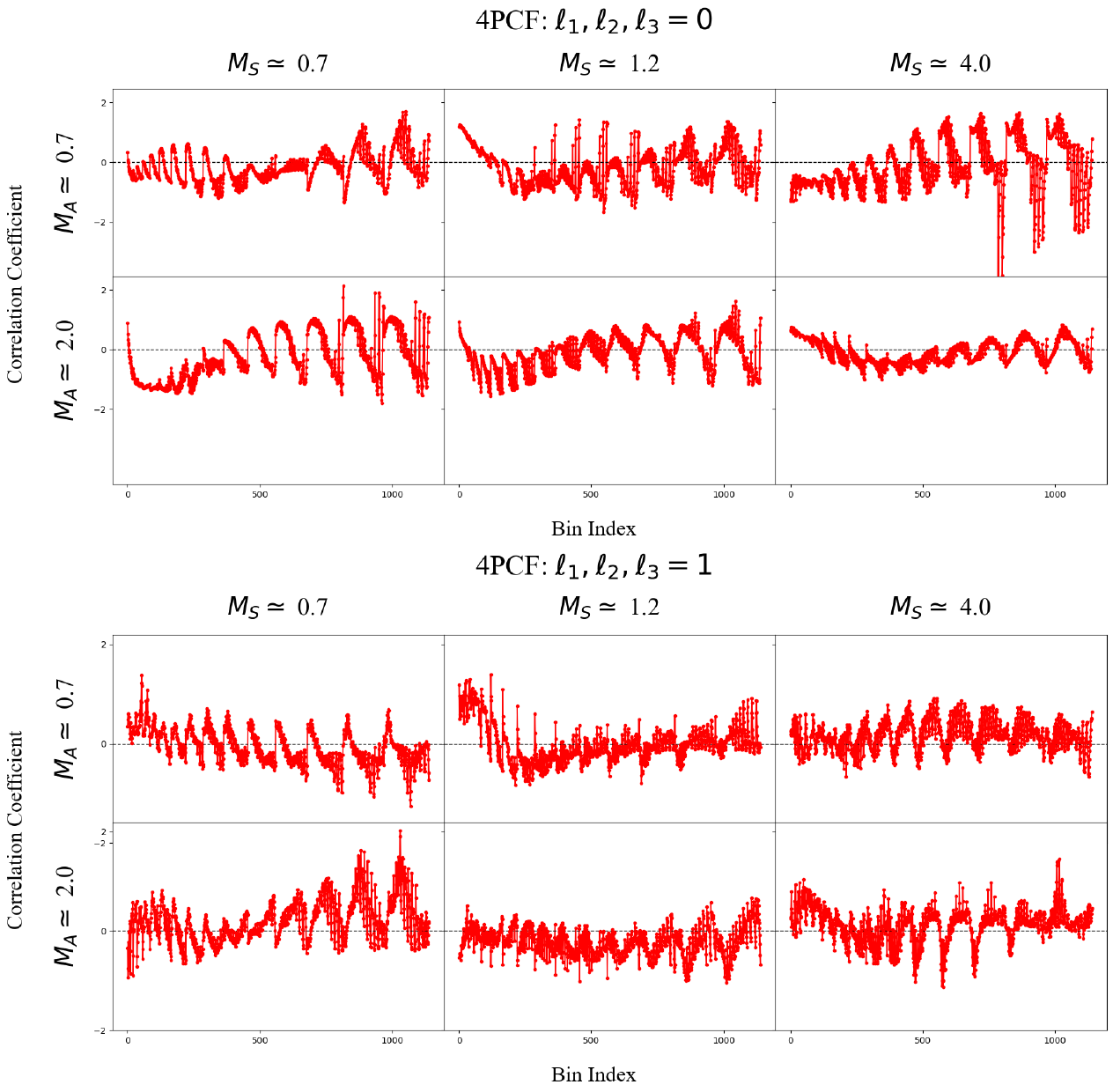}
    \caption{This plot shows the 4PCF coefficients for $\ell_1, \ell_2, \ell_3 = 0$ (top) and $\ell_1, \ell_2, \ell_3 = 1 $(bottom), mapped to 1D by bin combinations. The $\ell = 0$ coefficients exhibit smooth, periodic oscillations with a slight decay at higher bin indices, indicating dominant isotropic correlations that weaken at larger separations. In contrast, the $\ell = 1$ coefficients show more irregular oscillations with pronounced peaks and troughs, capturing complex angular dependencies and anisotropic contributions. These patterns highlight the increasing sensitivity of higher $\ell$ modes to angular clustering details and demonstrate how the 4PCF encodes both isotropic and anisotropic spatial correlations.}
    \label{fig:1d_first.pdf}
\end{figure*}

\begin{figure*}
    \centering
    \includegraphics[width=\textwidth]{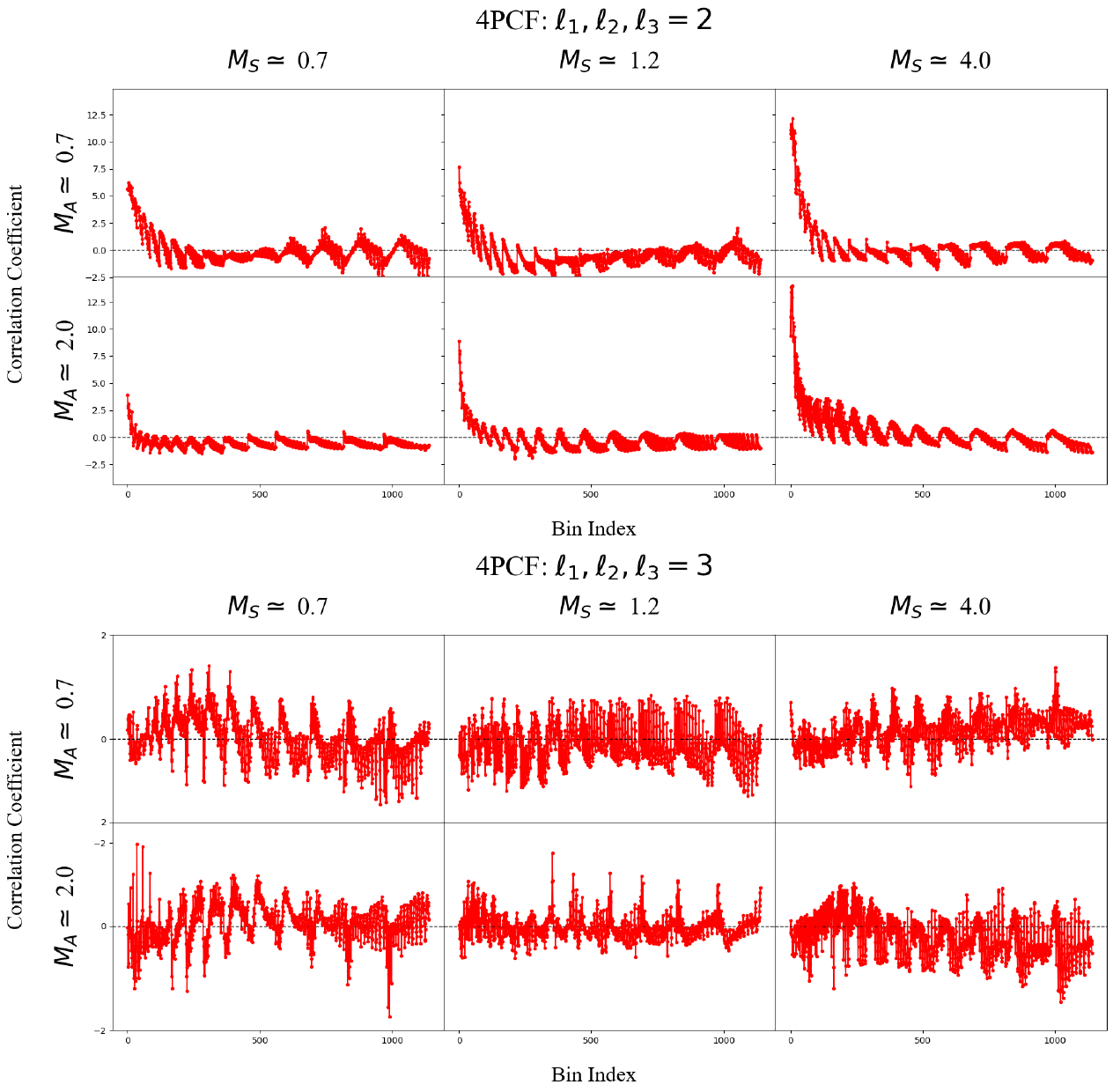}
    \caption{This plot shows the 4PCF coefficients for $\ell_1, \ell_2, \ell_3 = 2 $ (top) and $\ell_1, \ell_2, \ell_3 = 3 $ (bottom), mapped to 1D by bin combinations. The $\ell = 2$ plot shows a rapid decay with smoother oscillations, while the $\ell = 3$ plot exhibits slower decay and more persistent, higher-amplitude oscillations. These trends highlight the increasing angular complexity and slower decay associated with higher multipole moments, offering insights into the scale and angular dependence of the 4PCF.}
    \label{fig:1d_last.pdf}
\end{figure*}

Another promising direction for future work from these results is exploring parity-odd components to better understand the role they play in turbulence. Figures \ref{fig:1d_first.pdf} and \ref{fig:1d_last.pdf} condense the 4PCF coefficients, for a set of $\ell$ combinations, into 1-dimension to allow for further comparison of even and odd parity. This can be replicated for other $\ell$ combinations in future studies to better understand these results. Parity-odd modes could be investigated through the data displayed in Fig. \ref{fig:fullpowerlaw}. This figure depicts the power spectra and power law fit of the connected 4PCF data, which when tested for covariance can determine an above null detection by eliminating the possibility of noise detection in the data and confirming that the results could not have been modeled with 2PCFs or power spectra.

We plan to extend this framework to observational datasets to validate its applicability beyond simulations. By applying \textsc{sarabande} to CO emission maps, we will measure the projected 4PCF of real data (\citealt{kim_2017_TIGRESS}). "Projected" refers to integrating out the line of sight so that the 3D density field becomes 2D, which is crucial for extracting meaningful 4PCF signals from real density fields (\citealt{Sunseri_2022_SARABANDE}). From this we can conduct the first ever-4PCF study of the ISM with feedback and establish a foundation for exploring how feedback shapes galaxy evolution and star formation. Furthermore, with these in hand, we can use the algorithm recently presented in \cite{const_real} to then very cheaply produce a large number of simulated ISM density fields that have matching 2, 3, and 4PCFs to the real universe, and use these to test systematics pipelines in large-scale structure surveys such as DESI or CMB experiments such as Planck, both of which suffer from galactic extinction as a possible systematic.
 
\section*{Acknowledgments}
We acknowledge the University of Florida Research Computing and Princeton University Research Computing for providing computational resources and technical support. We would like to thank Stephen Portillo for development of the 3-Point Correlation Function on MHD turbulence, a helpful starting point for the work of this paper. The University of Florida’s University Scholars Program provided funding towards resources and hourly work. We thank the Slepian group members for useful discussions.

%%%%%%%%%%%%%%%%%%%%%%%%%%%%%%%%%%%%%%%%%%%%%%%%%%
\section*{Data Availability}

The simulations used in this work are publicly available as part of the Catalog for Astrophysical Turbulence Simulations (CATS, \citealt{Cho_2003_MHD}, \citealt{Burkhart_2009_MHDISM}, 
\citealt{Burkart_Lazarian_2016},
\citealt{Portillo_2018_3pcfISM}, \citealt{BialyBurkhart_2020_TurbulentDecorrelationScale}).\footnote{\href{https://www.mhdturbulence.com/cho}{mhdturbulence.com}} All our measured data, along with resources to replicate this work, can be accessed through our project website. \footnote{\href{https://sites.google.com/ufl.edu/4pcfsoftheism}{4pcfsoftheism.com}}
 
% The inclusion of a Data Availability Statement is a requirement for articles published in MNRAS. Data Availability Statements provide a standardised format for readers to understand the availability of data underlying the research results described in the article. The statement may refer to original data generated in the course of the study or to third-party data analysed in the article. The statement should describe and provide means of access, where possible, by linking to the data or providing the required accession numbers for the relevant databases or DOIs.

\graphicspath{{figures/figures/1d_plots/1d_first.pdf}}

\clearpage

\graphicspath{{figures/figures/1d_plots/1d_last.pdf}}
 
\clearpage



\bibliographystyle{mnras}
\bibliography{biblio} % if your bibtex file is called example.bib

%\begin{thebibliography}{99}
%\end{thebibliography}

%%%%%%%%%%%%%%%%%%%%%%%%%%%%%%%%%%%%%%%%%%%%%%%%%%

%%%%%%%%%%%%%%%%% APPENDICES %%%%%%%%%%%%%%%%%%%%%

\appendix

\section{Odd-Parity Modes Only}
\clearpage
\graphicspath{{Figures/figures/oddmergepdf/1_start.pdf}}

\begin{figure*}
    \centering
    \includegraphics[width=0.95\textwidth, height=0.89\textheight]{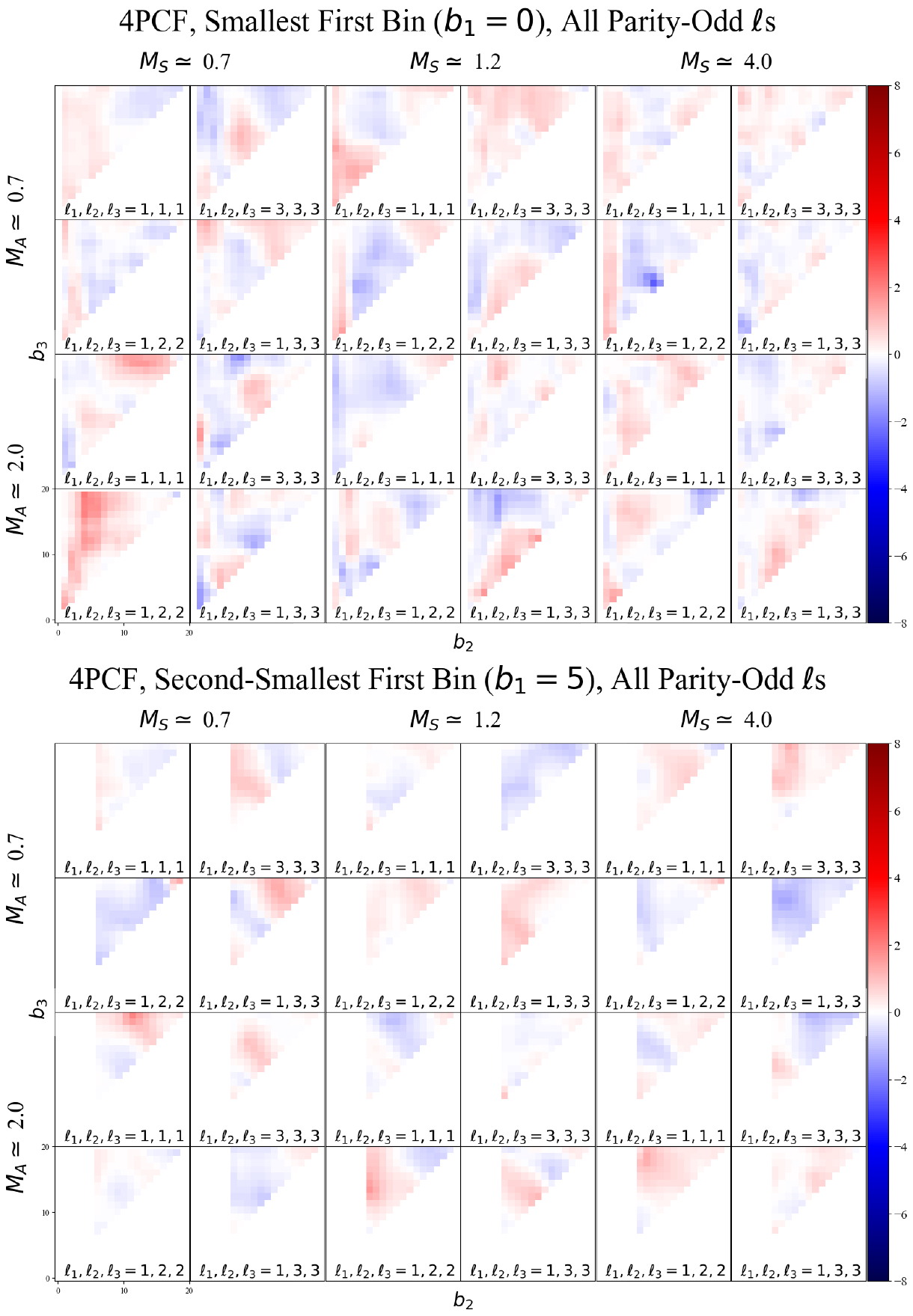}
    \caption{The 3D connected 4PCF signal-to-noise ratio, normalized by the standard deviation across time slices, computed for parity-odd modes only in the multipole basis for $\ell_1, \ell_2, \ell_3 = (1,1,1)$, $\ell_1, \ell_2, \ell_3 = (3,3,3)$, $\ell_1, \ell_2, \ell_3 = (1,2,2)$, and $\ell_1, \ell_2, \ell_3 = (1,3,3)$. The first image shows these $\ell$ combinations when $b_1 = 0$  and the second image shows when $b_1 = 5$.}
	\label{fig:1_start.pdf}
\end{figure*}
\clearpage

\graphicspath{{Figures/figures/oddmergepdf/1_end.pdf}}

\begin{figure*}
    \centering
    \includegraphics[width=0.95\textwidth, height=0.89\textheight]{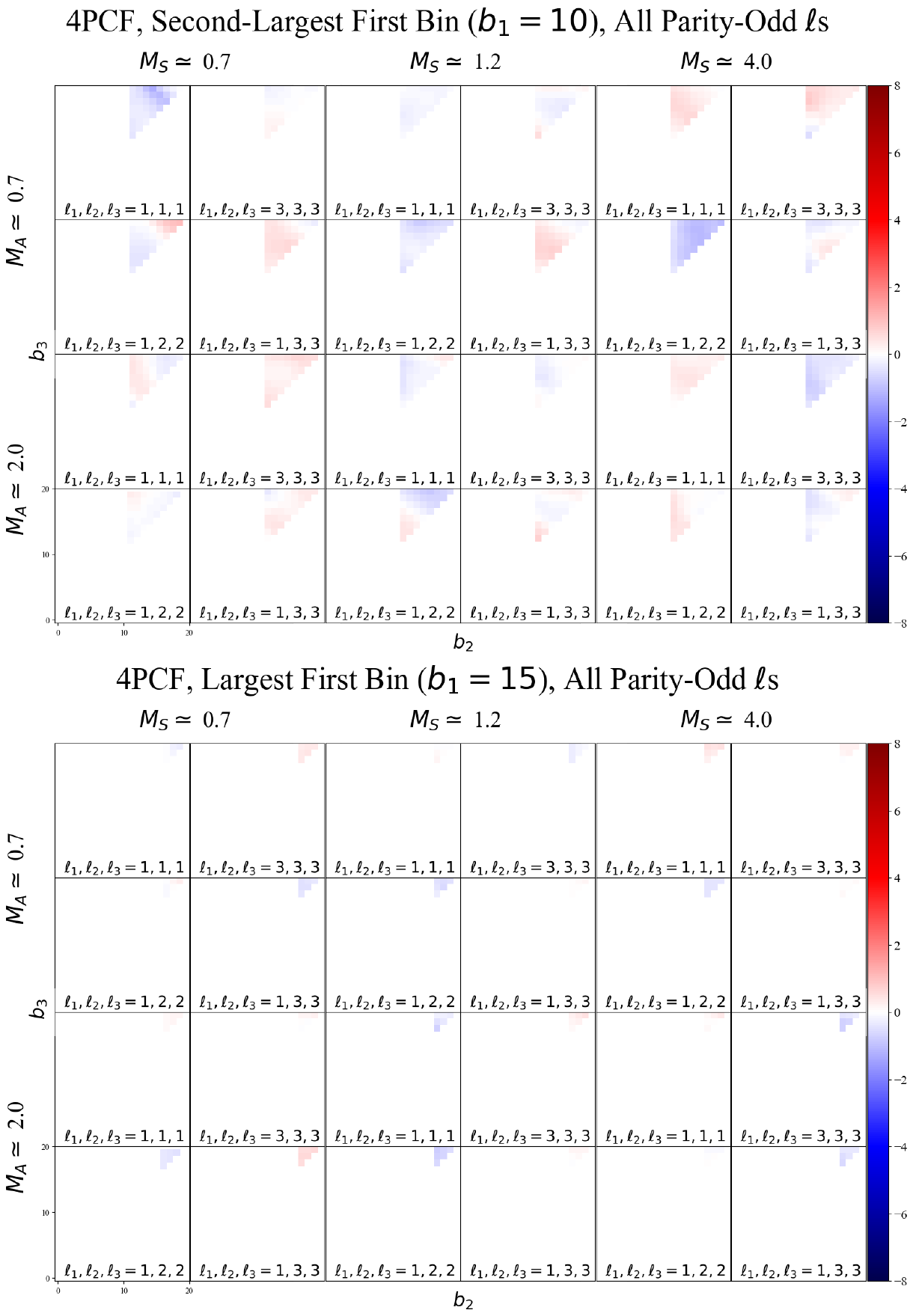}
    \caption{The 3D connected 4PCF signal-to-noise ratio, normalized by the standard deviation across time slices, computed for parity-odd modes only in the multipole basis for $\ell_1, \ell_2, \ell_3 = (1,1,1)$, $\ell_1, \ell_2, \ell_3 = (3,3,3)$, $\ell_1, \ell_2, \ell_3 = (1,2,2)$, and $\ell_1, \ell_2, \ell_3 = (1,3,3)$. The first image shows these $\ell$ combinations when $b_1 = 10$  and the second image shows when $b_1 = 15$.}
	\label{fig:1_end.pdf}
\end{figure*}
\clearpage

\graphicspath{{Figures/figures/oddmergepdf/2_start.pdf}}

\begin{figure*}
    \centering
    \includegraphics[width=0.95\textwidth, height=0.89\textheight]{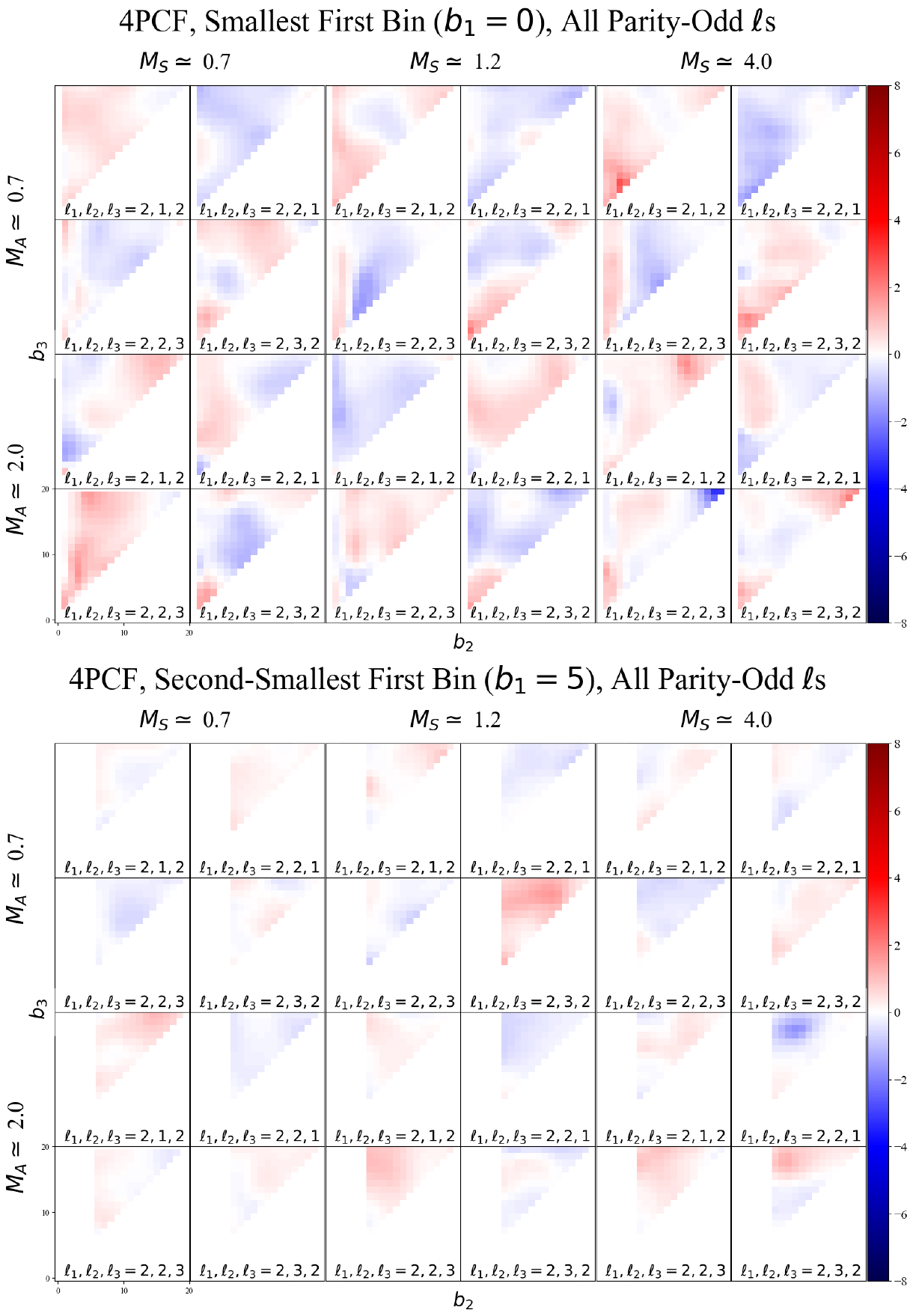}
    \caption{The 3D connected 4PCF signal-to-noise ratio, normalized by the standard deviation across time slices, computed for parity-odd modes only in the multipole basis for $\ell_1, \ell_2, \ell_3 = (2,1,2)$, $\ell_1, \ell_2, \ell_3 = (2,2,1)$, $\ell_1, \ell_2, \ell_3 = (2,2,3)$, and $\ell_1, \ell_2, \ell_3 = (2,3,2)$. The first image shows these $\ell$ combinations when $b_1 = 0$  and the second image shows when $b_1 = 5$.}
	\label{fig:2_start.pdf}
\end{figure*}
\clearpage

\graphicspath{{Figures/figures/oddmergepdf/2_end.pdf}}

\begin{figure*}
    \centering
    \includegraphics[width=0.95\textwidth, height=0.89\textheight]{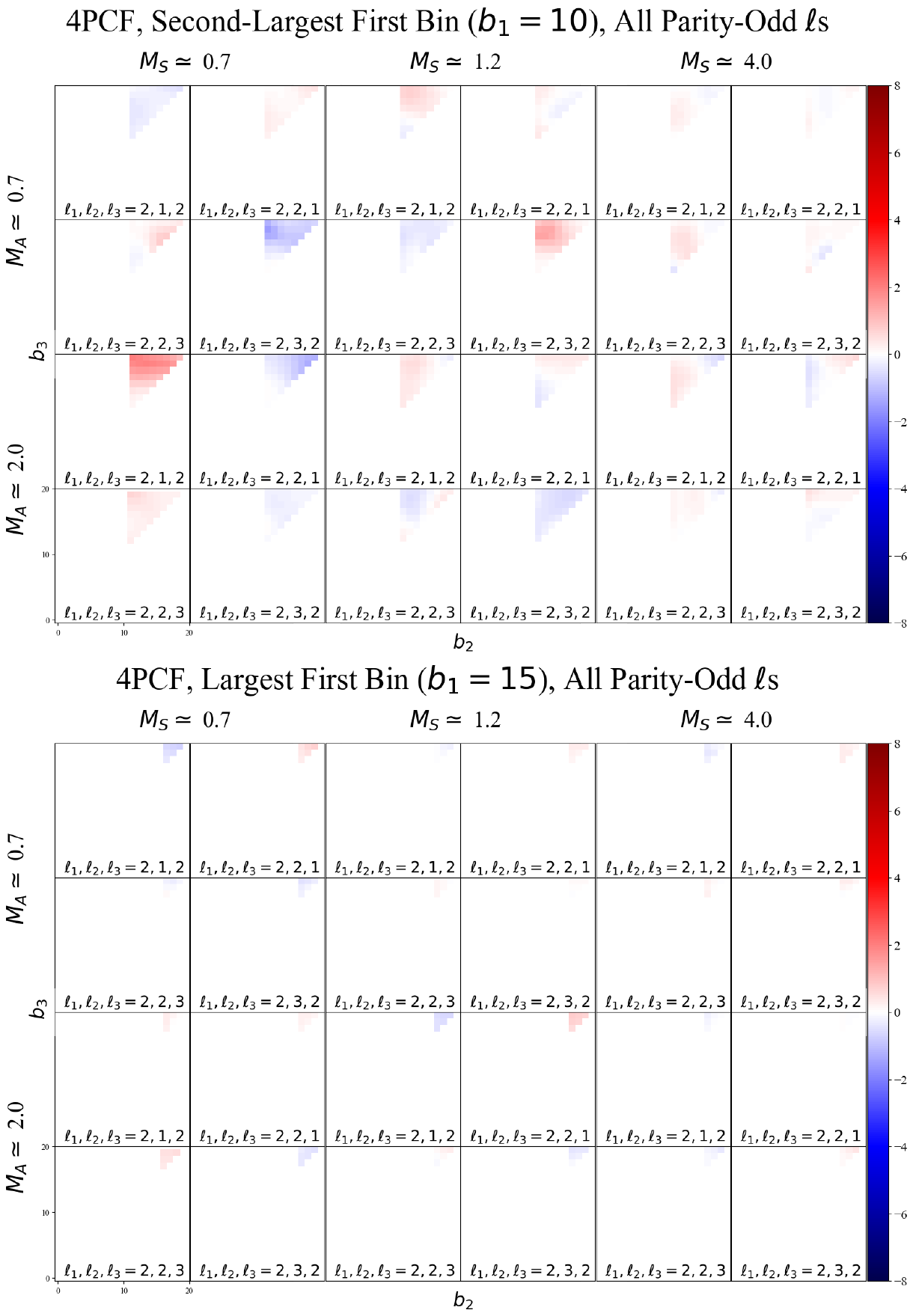}
    \caption{The 3D connected 4PCF signal-to-noise ratio, normalized by the standard deviation across time slices, computed for parity-odd modes only in the multipole basis for $\ell_1, \ell_2, \ell_3 = (2,1,2)$, $\ell_1, \ell_2, \ell_3 = (2,2,1)$, $\ell_1, \ell_2, \ell_3 = (2,2,3)$, and $\ell_1, \ell_2, \ell_3 = (2,3,2)$. The first image shows these $\ell$ combinations when $b_1 = 10$  and the second image shows when $b_1 = 15$.}
	\label{fig:2_end.pdf}
\end{figure*}
\clearpage

\graphicspath{{Figures/figures/oddmergepdf/_3,1,3__start.pdf}}

\begin{figure*}
    \centering
    \includegraphics[width=0.95\textwidth, height=0.89\textheight]{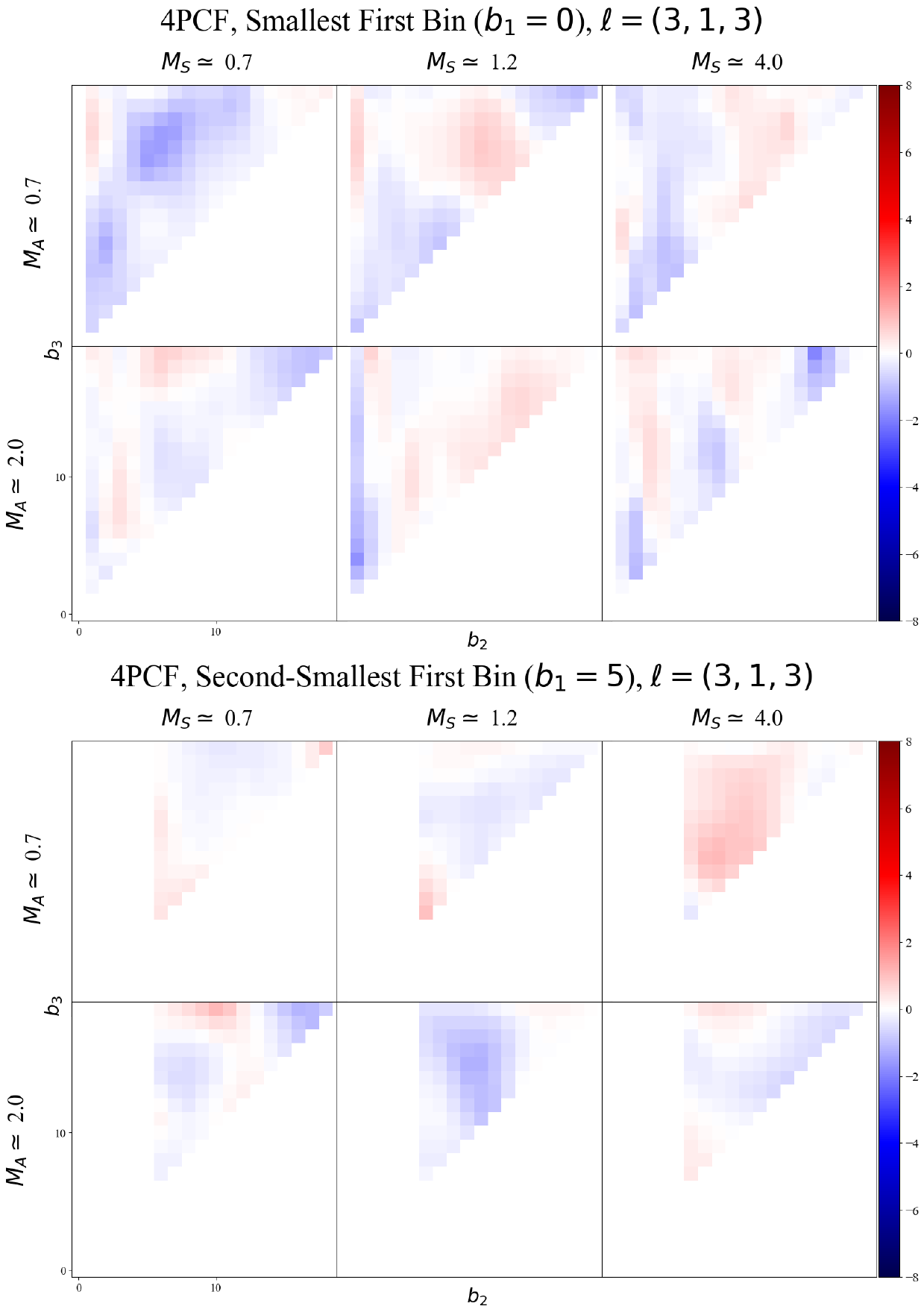}
    \caption{The 3D connected 4PCF signal-to-noise ratio, normalized by the standard deviation across time slices, computed for the parity-odd mode in the multipole basis for $\ell_1, \ell_2, \ell_3 = (3,1,3)$. The first image shows these $\ell$ combinations when $b_1 = 0$  and the second image shows when $b_1 = 5$.}
	\label{fig:(3,1,3)_start.pdf}
\end{figure*}
\clearpage

\graphicspath{{Figures/figures/oddmergepdf/_3,1,3__end.pdf}}

\begin{figure*}
    \centering
    \includegraphics[width=0.95\textwidth, height=0.89\textheight]{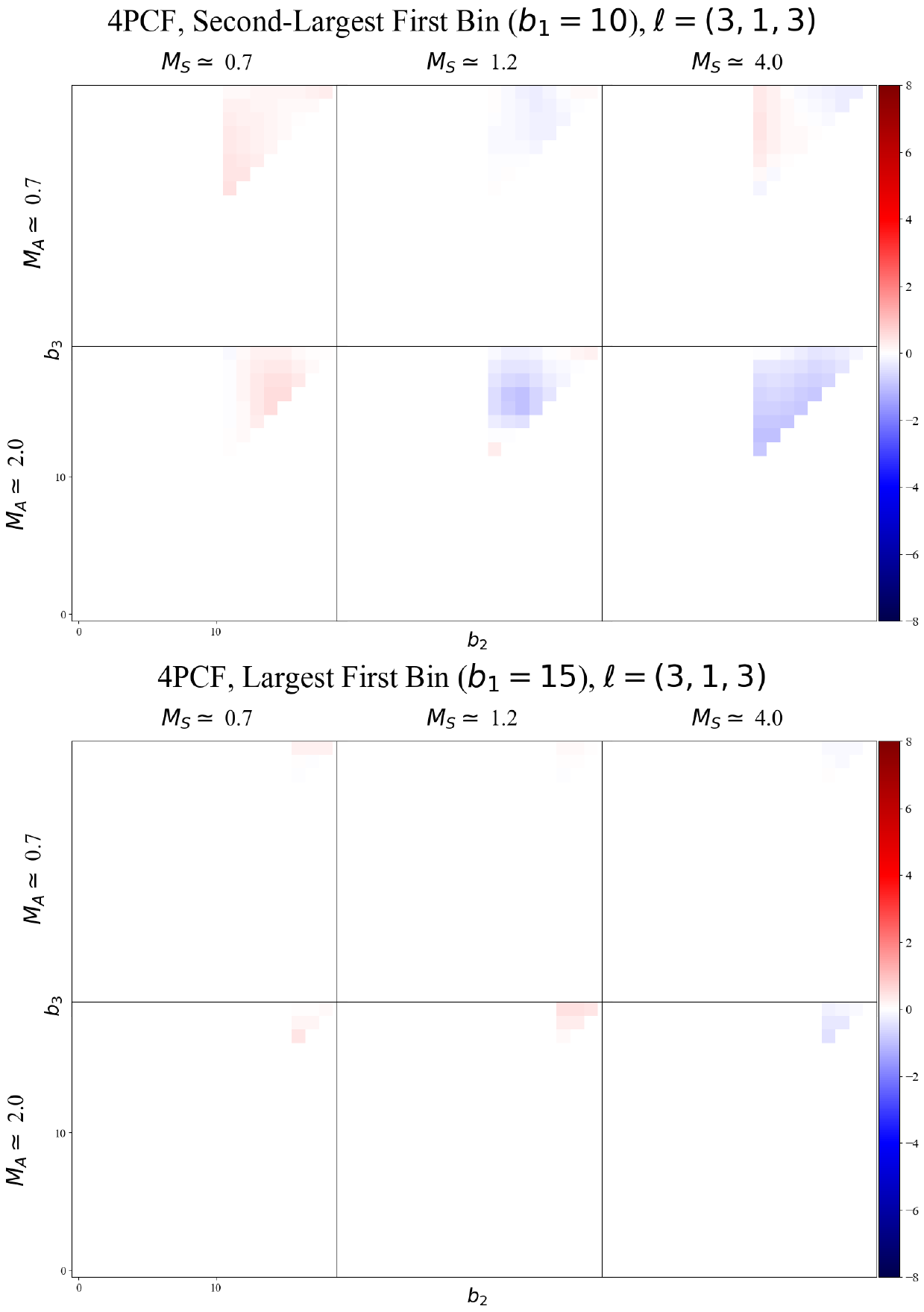}
    \caption{The 3D connected 4PCF signal-to-noise ratio, normalized by the standard deviation across time slices, computed for the parity-odd mode in the multipole basis for $\ell_1, \ell_2, \ell_3 = (3,1,3)$. The first image shows these $\ell$ combinations when $b_1 = 10$  and the second image shows when $b_1 = 15$.}
	\label{fig:(3,1,3)_end.pdf}
\end{figure*}
\clearpage

\graphicspath{{Figures/figures/oddmergepdf/_3,2,2__start.pdf}}

\begin{figure*}
    \centering
    \includegraphics[width=0.95\textwidth, height=0.89\textheight]{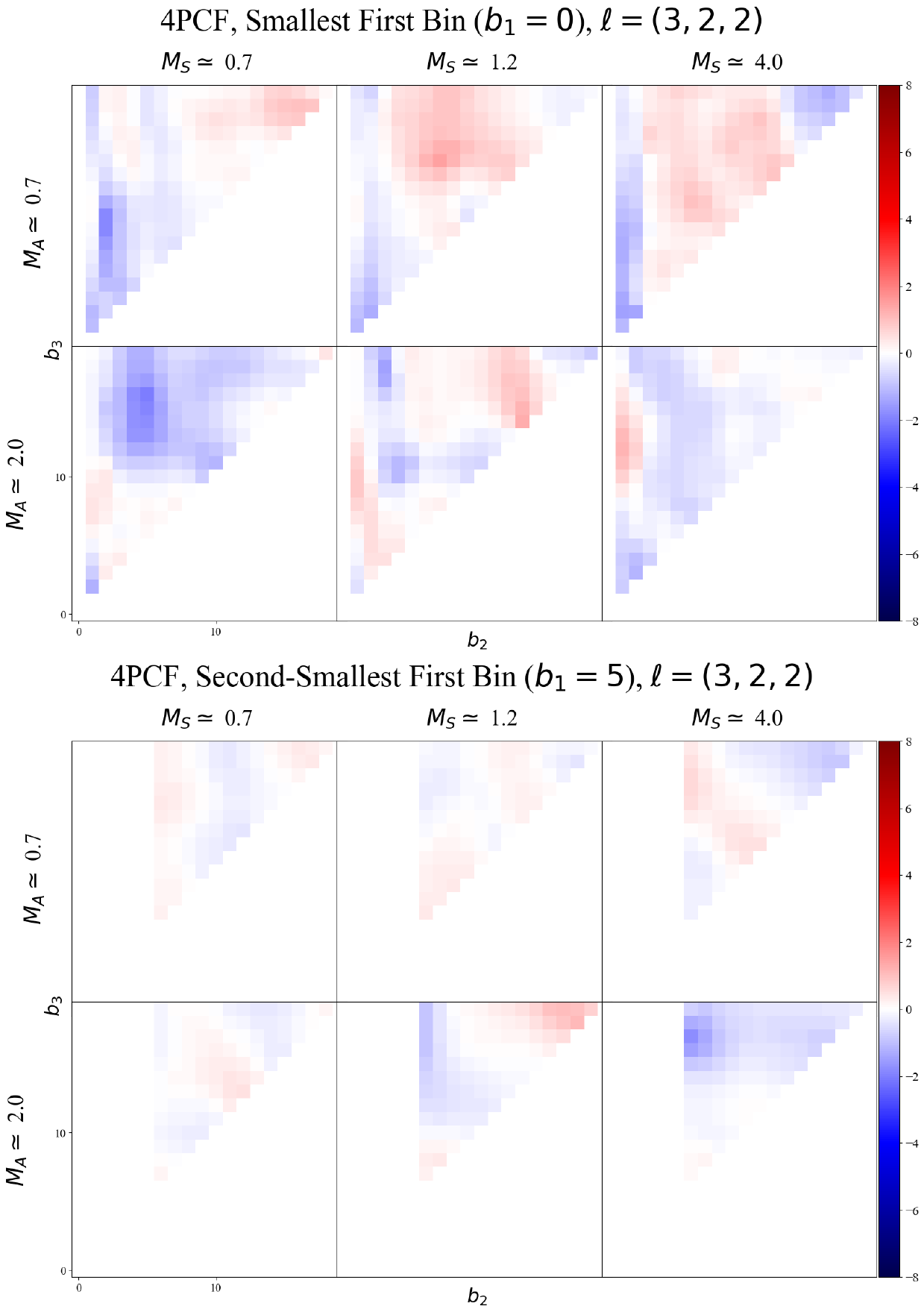}
    \caption{The 3D connected 4PCF signal-to-noise ratio, normalized by the standard deviation across time slices, computed for the parity-odd mode in the multipole basis for $\ell_1, \ell_2, \ell_3 = (3,2,2)$. The first image shows these $\ell$ combinations when $b_1 = 0$  and the second image shows when $b_1 = 5$.}
	\label{fig:(3,2,2)_start.pdf}
\end{figure*}
\clearpage

\graphicspath{{Figures/figures/oddmergepdf/_3,2,2__end.pdf}}

\begin{figure*}
    \centering
    \includegraphics[width=0.95\textwidth, height=0.89\textheight]{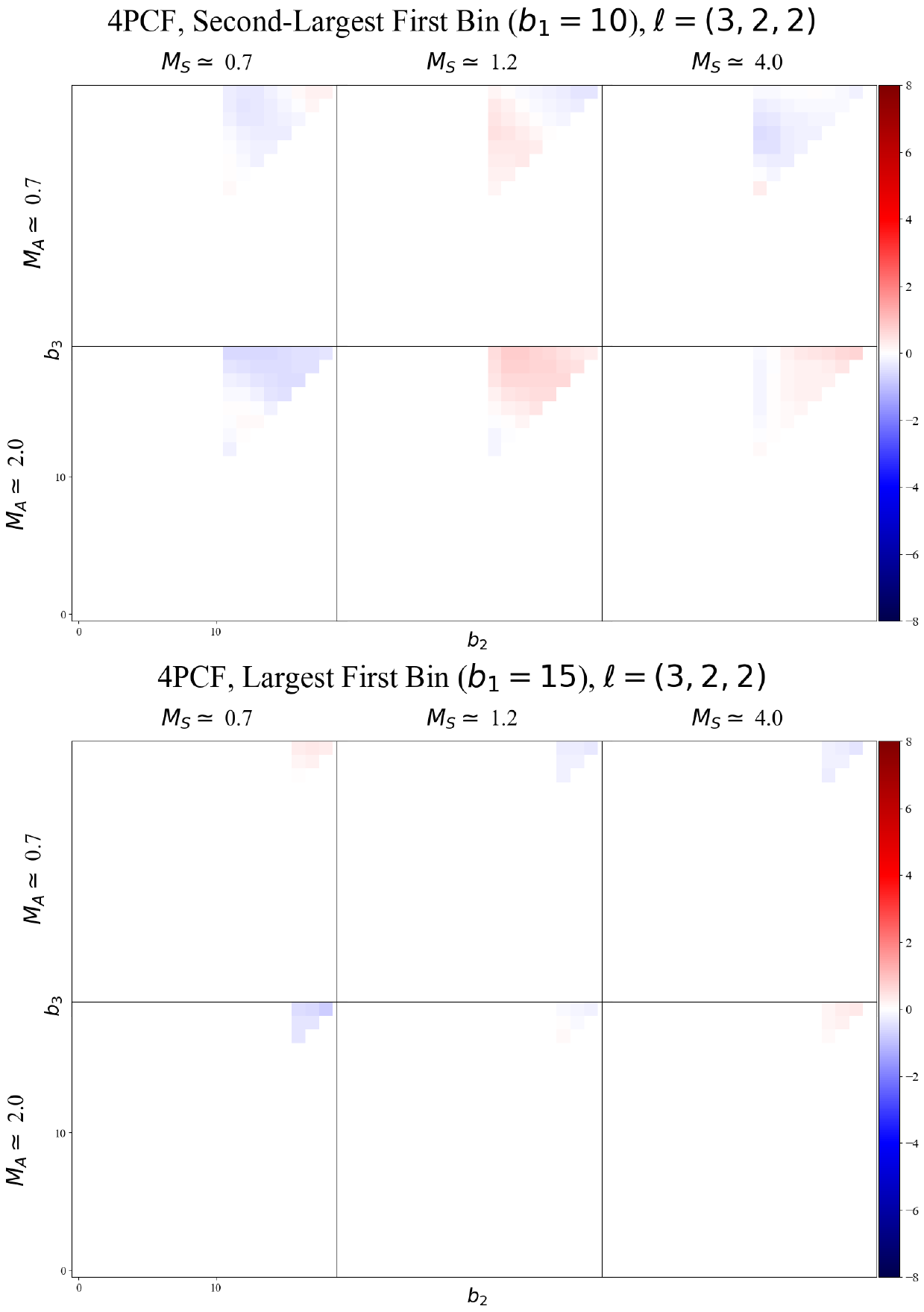}
    \caption{The 3D connected 4PCF signal-to-noise ratio, normalized by the standard deviation across time slices, computed for the parity-odd mode in the multipole basis for $\ell_1, \ell_2, \ell_3 = (3,2,2)$. The first image shows these $\ell$ combinations when $b_1 = 10$  and the second image shows when $b_1 = 15$.}
	\label{fig:(3,2,2)_end.pdf}
\end{figure*}
\clearpage

\graphicspath{{Figures/figures/oddmergepdf/_3,3,1__start.pdf}}

\begin{figure*}
    \centering
    \includegraphics[width=0.95\textwidth, height=0.89\textheight]{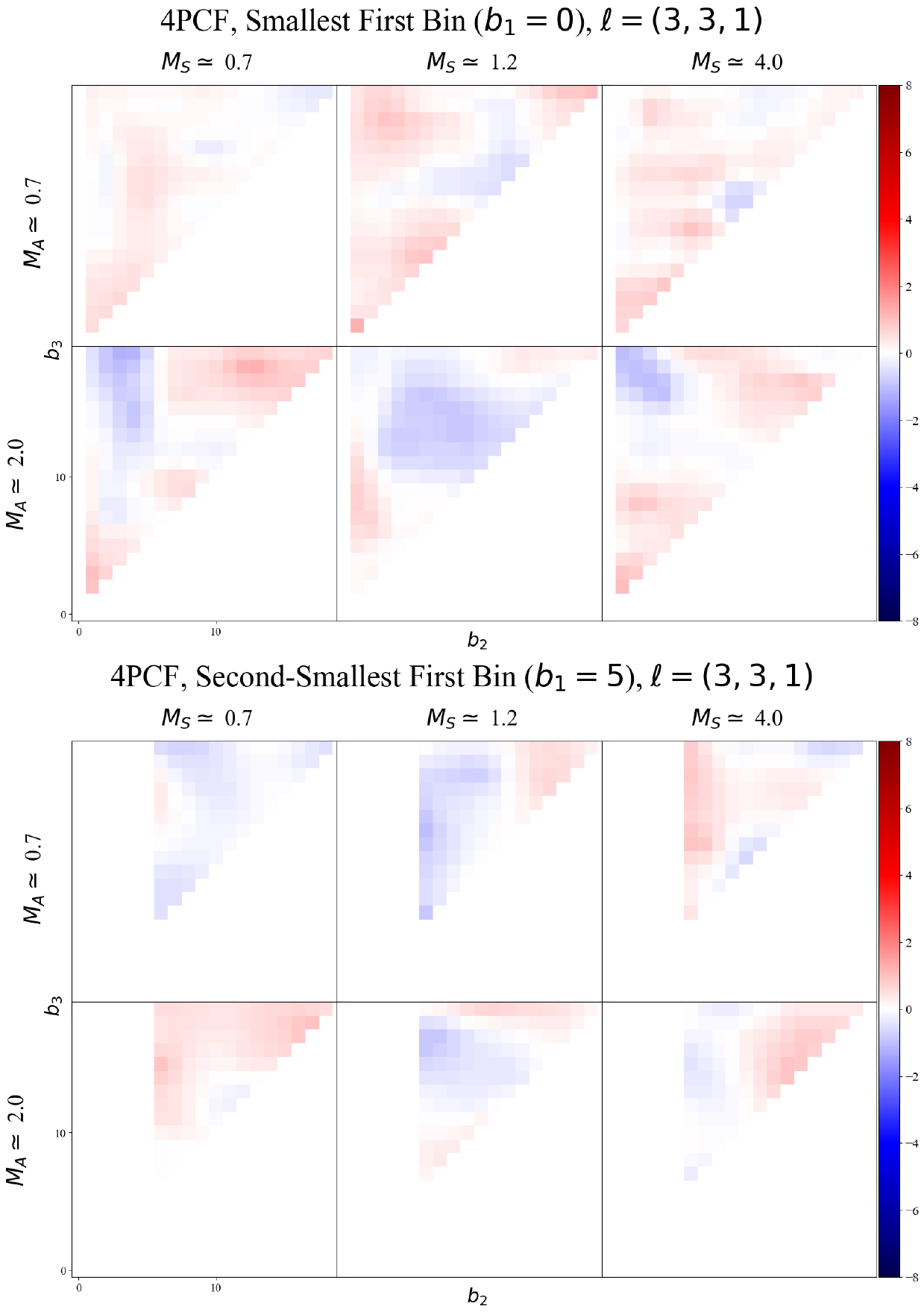}
    \caption{The 3D connected 4PCF signal-to-noise ratio, normalized by the standard deviation across time slices, computed for the parity-odd mode in the multipole basis for $\ell_1, \ell_2, \ell_3 = (3,3,1)$. The first image shows these $\ell$ combinations when $b_1 = 0$  and the second image shows when $b_1 = 5$.}
	\label{fig:(3,3,1)_start.pdf}
\end{figure*}
\clearpage

\graphicspath{{Figures/figures/oddmergepdf/_3,3,1__end.pdf}}

\begin{figure*}
    \centering
    \includegraphics[width=0.95\textwidth, height=0.89\textheight]{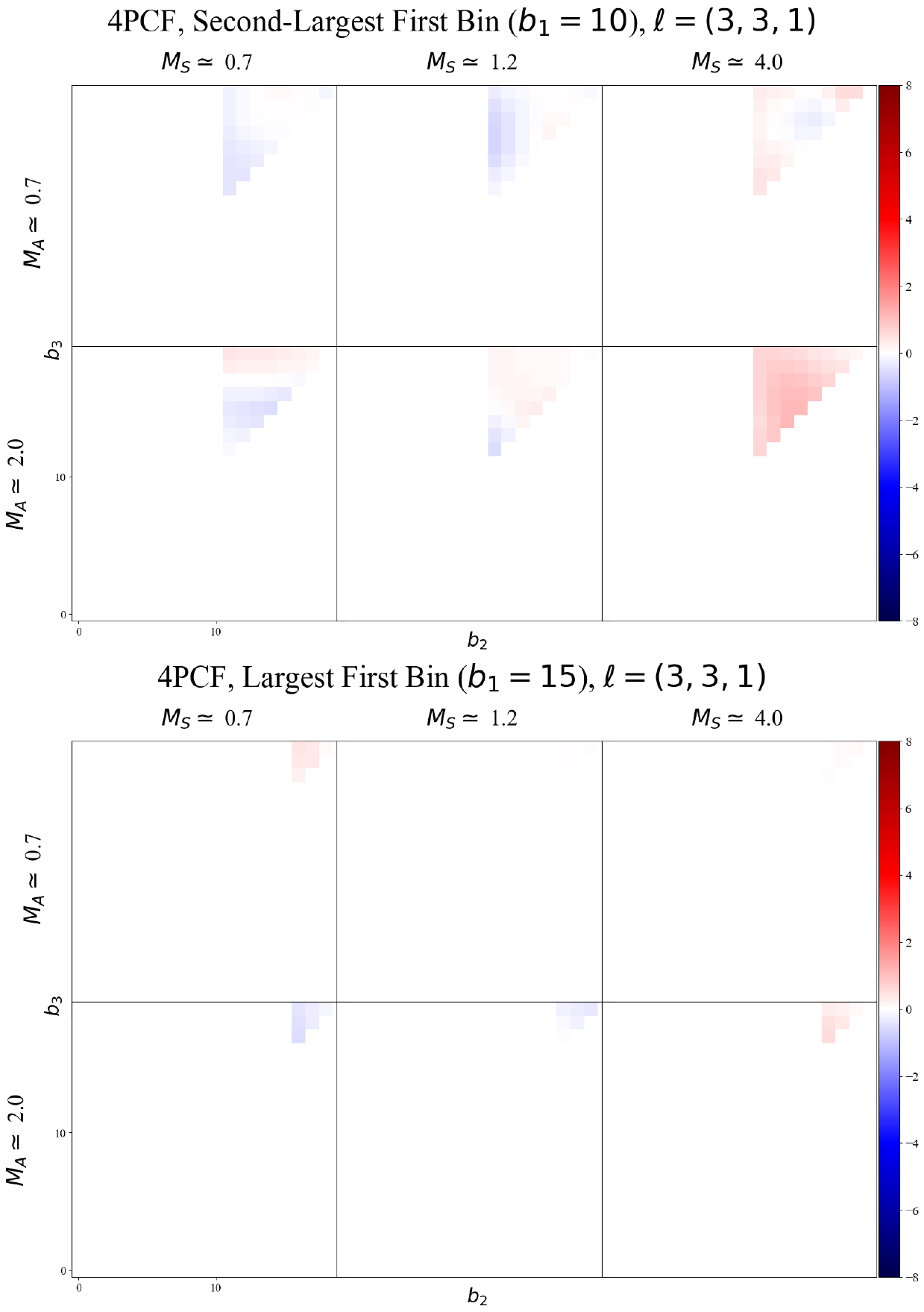}
    \caption{The 3D connected 4PCF signal-to-noise ratio, normalized by the standard deviation across time slices, computed for the parity-odd mode in the multipole basis for $\ell_1, \ell_2, \ell_3 = (3,3,1)$. The first image shows these $\ell$ combinations when $b_1 = 10$  and the second image shows when $b_1 = 15$.}
	\label{fig:(3,3,1)_end.pdf}
\end{figure*}
\clearpage
% If you want to present additional material which would interrupt the flow of the main paper,
% it can be placed in an Appendix which appears after the list of references.

%%%%%%%%%%%%%%%%%%%%%%%%%%%%%%%%%%%%%%%%%%%%%%%%%%

% Don't change these lines
\bsp	% typesetting comment
\label{lastpage}
\end{document}